\documentclass[11pt,oneside,letterpaper]{article}
\usepackage{amssymb}
\usepackage{amsmath}
\usepackage{amsthm}
\usepackage[dvips]{graphicx}
\usepackage{setspace}
\usepackage{amsfonts}
\usepackage{fancyhdr}
\usepackage{xcolor}
\usepackage{graphicx}
\usepackage{rotating}
\usepackage{comment}
\usepackage{color}
\usepackage{cite}
\usepackage{braket}
\usepackage{caption}
\usepackage[utf8]{inputenc}
\usepackage{mathtools}
\usepackage{tikz}
\usepackage{overpic}
\usepackage{stackengine}
\usepackage[normalem]{ulem}
\usepackage{bbold}

\definecolor{darkgreen}{rgb}{0,0.5,0}
\definecolor{darkblue}{rgb}{0,0,0.6}
\definecolor{purple}{rgb}{0.4,.2,0.7}
\newcommand{\p}{\partial}

\newcommand{\be}{\begin{equation}}
\newcommand{\ee}{\end{equation}}

\usepackage[colorlinks=true,citecolor=darkgreen,linkcolor=black,urlcolor=purple]{hyperref}

\usepackage{pdfsync}

\makeatletter
\newcommand*{\defeq}{\mathrel{\rlap{%
                     \raisebox{0.3ex}{$\m@th\cdot$}}%
                     \raisebox{-0.3ex}{$\m@th\cdot$}}%
                     =} 
\makeatother

\def\be{\begin{eqnarray}}
\def\ee{\end{eqnarray}}

\newcommand{\tr}{\textrm{Tr}\,}

\newcommand{\bea}{\begin{eqnarray}}
\newcommand{\eea}{\end{eqnarray}}
\def\ben{\begin{equation}}
\def\een{\end{equation}}
\def\half{{\textstyle{\frac{1}{2}}}}

\let\l=\lambda    \let\p=\phi \let\r=v

\def\be{\begin{equation}}
\def\ee{\end{equation}}
\def\ba{\begin{array}}
\def\ea{\end{array}}

\def\ba#1\ea{\begin{align}#1\end{align}}
\def\bs#1\es{\begin{split}#1\end{split}}

\renewcommand{\p}{\partial}

\newcommand{\bz}{\bar{z}}
\newcommand{\bsigma}{\bar{\sigma}}

\newcommand{\btau}{\bar{\tau}}

\newcommand{\bh}{\bar{h}}
\newcommand{\F}{{\cal F}}
\newcommand{\bx}{\bar{x}}
\newcommand{\bOmega}{\bar{\Omega}}
\newcommand{\Cdozz}{C_{\rm \footnotesize DOZZ}}
\newcommand{\tCdozz}{\widehat{C}_{\rm \footnotesize DOZZ}}

\renewcommand{\O}{{\cal O}}
\newcommand{\blambda}{\bar{\lambda}}

\definecolor{vert}{rgb}{0.1367 0.543 0.1367}

\newcommand{\id}{\mathbb{1}}

\DeclareMathOperator*{\Res}{Res}
\DeclareMathOperator*{\re}{Re}

\allowdisplaybreaks  

\numberwithin{equation}{section}

\def \be {\begin{equation}}
\def \ee {\end{equation}}
\def \half {{1\over 2}}	
\def \JM#1 {{\color{blue}  JM: #1 }}
\def \AAl#1 {{\color{red}  AA: #1 }}

\begin{document}
\onehalfspacing

\begin{center}

{\LARGE  {
Semiclassical 3D gravity as \\
 an average of large-$c$ CFTs
\\
}}

\vskip1cm

Jeevan Chandra,$^1$\ 
Scott Collier,$^2$\ 
Thomas Hartman,$^1$\ 
and Alexander Maloney$^3$

\vskip5mm
$^1${\it Department of Physics, Cornell University, Ithaca, New York, USA}\\
$^2$ {\it Princeton Center for Theoretical Science, Princeton University, Princeton, NJ 08544, USA}\\
$^3${\it Department of Physics, McGill University, Montr{\'e}al, Canada}

\vskip5mm

{\tt jn539@cornell.edu, scott.collier@princeton.edu, hartman@cornell.edu, maloney@physics.mcgill.ca }

\end{center}

\vspace{4mm}

\begin{abstract}
\noindent 

A two-dimensional CFT dual to a semiclassical theory of gravity in three dimensions must have a large central charge $c$ and a sparse low energy spectrum. 
This constrains the OPE coefficients and density of states of the CFT via the conformal bootstrap.  We define an ensemble of CFT data by averaging over OPE coefficients subject to these bootstrap constraints, and show that calculations in this ensemble reproduce semiclassical 3D gravity.  We analyze a wide variety of gravitational solutions, both in pure Einstein gravity and gravity coupled to massive point particles, including Euclidean wormholes with multiple boundaries and higher topology spacetimes with a single boundary.  In all cases we find that the on-shell action of gravity agrees with the ensemble-averaged CFT at large $c$.
The one-loop corrections also match in the cases where they have been computed.  
We also show that the bulk effective theory has random couplings induced by wormholes, providing a controlled, semiclassical realization of the mechanism of Coleman, Giddings, and Strominger.

 \end{abstract}

\pagebreak
\pagestyle{plain}

\setcounter{tocdepth}{2}
{}
\vfill

\ \vspace{-2cm}
\renewcommand{\baselinestretch}{1}\small
\tableofcontents
\renewcommand{\baselinestretch}{1.18}\normalsize

\section{Introduction}

Quantum gravity in two-dimensional anti-de Sitter spacetime is dual to random matrix theory \cite{Saad:2019lba}. This is fundamentally different from earlier examples of the AdS/CFT correspondence, because the boundary theory is not a single quantum theory --- it is an ensemble average of many quantum theories. Thus products of observables do not factorize as they would in quantum mechanics.  For example, the average $\overline{Z(\beta_1) Z(\beta_2)} \neq \overline{Z(\beta_1)} \times \overline{Z(\beta_2)}$ where $Z$ is the partition function \cite{Cotler:2016fpe,Saad:2018bqo}. Connected contributions to these averages encode statistical properties of the underlying ensemble, and are reproduced on the gravity side by geometries with multiple boundaries. Higher topologies with a single boundary also have an ensemble interpretation that beautifully explains the paradoxical behavior of late-time correlation functions \cite{Saad:2019pqd}.

There is ample evidence that averaging must also play a role in more realistic, higher-dimensional theories of quantum gravity. There are many examples of multi-boundary solutions with no known instabilities \cite{Maldacena:2004rf,Marolf:2021kjc}. 
Double black hole topologies, though difficult to study in a controlled manner, seem to correctly reproduce the universal eigenvalue repulsion of chaotic quantum theories \cite{Cotler:2016fpe,Saad:2018bqo,Cotler:2020ugk,Cotler:2021cqa}. Furthermore, all higher-dimensional theories of gravity have extremal black hole solutions with near horizon geometries that can be dimensionally reduced to AdS$_2$. On the other hand, we do not expect string theory (or any other UV complete theory of quantum gravity) to be an ensemble average at the microscopic level, and the reduction to AdS$_2$ includes additional states that might derail the ensemble interpretation. How can these viewpoints be reconciled? There are various proposals (e.g.~\cite{Saad:2018bqo,Saad:2019pqd,Marolf:2020xie,Maxfield:2020ale,Altland:2020ccq,McNamara:2020uza,Verlinde:2021kgt,Saad:2021rcu,Eberhardt:2021jvj,Altland:2021rqn,Heckman:2021vzx,Betzios:2021fnm,Collier:2022emf,Schlenker:2022dyo}), but so far there is no general answer to this question.

One interesting possibility is that ensemble averaging is the result of some sort of coarse graining in the effective theory of the large-$N$ limit.  But what sort of coarse graining?  We will attempt to answer this question in AdS$_3$/CFT$_2$, where we will construct a detailed model which makes this idea precise: semiclassical gravity arises by fixing a limited set of bootstrap constraints and averaging over CFT data subject to these constraints.

This is different from the random matrix interpretation of 2D gravity, because we are not averaging over a family of UV-complete quantum theories, but instead over an ensemble of observables. Quite possibly, most draws from this ensemble lie in the swampland. This is an essential difference between averaging over Hamiltonians in a 1D quantum system, where any spectrum is possible in principle, and averaging over CFTs in higher dimensions, which are subject to bootstrap constraints. We will average over the matrix elements of primary operators $\langle i | \O |j \rangle$, on which we impose some --- but not all --- of the bootstrap conditions which must be obeyed in a conformal field theory. Averaging over matrix elements has also been used to explain the behavior of 2D gravity coupled to matter \cite{Yang:2018gdb,Saad:2019pqd,Blommaert:2020seb,Stanford:2020wkf}, and it has been noted previously in the literature that such averages reproduce qualitative features of gravity in higher dimensions \cite{Pollack:2020gfa,Belin:2020hea,Altland:2021rqn}.

Three-dimensional gravity, unlike its 2D counterpart, has a multitude of classical solutions with higher topology.  Our analysis is restricted to the semiclassical expansion around these solutions. We will study not just smooth solutions of pure Einstein gravity, but also gravity coupled to massive point particles that backreact on the geometry to produce conical defects. We define an ensemble of CFTs (or, more precisely, CFT data), and conjecture that it exactly reproduces every solution of 3D gravity, including loops.  The conjecture is tested extensively.
Many of our conclusions probably also apply to bulk theories with quantum fields, but the restriction to conical defects makes the analysis simpler because we can discuss classical solutions instead of summing Witten diagrams, and conformal blocks instead of conformal Mellin amplitudes.

Off-shell topologies in 3D gravity have been considered in the literature. There is a calculation of the gravitational path integral on the double torus \cite{Cotler:2021cqa} which reproduces expected features of the energy-level distribution in a chaotic CFT.  This result is essentially orthogonal to our analysis: we do not need to specify the energy-level distribution of our CFT ensemble in order to reproduce the semiclassical expansion of 3D gravity. Apparently, the energy-level statistics are more sensitive to assumptions about the UV and therefore require input from the off-shell gravitational path integral,
while the statistics of matrix elements are a classical phenomenon in the bulk. Other off-shell topologies in 3D gravity with a single boundary were considered in \cite{Maxfield:2020ale} by a reduction to two dimensions; there is tantalizing evidence that this may resolve the longstanding puzzle of how to calculate the partition function of pure gravity (see \cite{Maloney:2007ud}). Again, this does not connect to our analysis directly, because the wormholes in \cite{Maxfield:2020ale} are not semiclassical. 

From a bootstrap point of view, solutions of 3D gravity with simple topologies are explained by large-$c$ methods in 2D CFT \cite{Hartman:2013mia}. Examples include the BTZ black hole \cite{Maldacena:1998bw,Hartman:2014oaa}, analogous handlebody manifolds at higher genus \cite{Yin:2007gv}, correlators of heavy operators \cite{Headrick:2010zt,Faulkner:2013yia,Hartman:2013mia,Roberts:2014ifa}, probes of black hole eigenstates \cite{Fitzpatrick:2014vua,Asplund:2014coa,Kraus:2016nwo, Kraus:2017ezw}, collapsing black holes \cite{Anous:2016kss}, and light correlators \cite{Maloney:2016kee, Kraus:2017kyl,Kraus:2018pax}. All of these examples satisfy factorization. One motivation and outcome of our analysis is to extend the tools of large-$c$ conformal field theory to account for the statistical properties of heavy operators, and therefore wormholes and other higher topologies in the bulk. Statistics of heavy operators have also been treated from a purely CFT point of view recently in \cite{Cardy:2017qhl,Collier:2019weq,Belin:2021ryy,Belin:2021ibv,Anous:2021caj} and we will make contact with some of their results. 

In higher dimensions, there is also a large literature on matching solutions of the conformal bootstrap equations to effective field theory in AdS, initiated in \cite{Heemskerk:2009pn}. We will not connect directly to these results because we rely heavily on Virasoro symmetry, but this raises a very interesting question: Where are the wormholes in the $d>2$ conformal bootstrap? See \cite{Hartman:2022zik} for further discussion.

The role of averaging in three dimensional gravity was recently discussed in \cite{Schlenker:2022dyo}, who suggested that only sufficiently heavy operators (with dimension of order $c$) exhibit averaging.   This is completely consistent with our results.   The work of \cite{Schlenker:2022dyo} focused primarily on averaging for states above the black hole threshold, where these effects can be seen without the need to insert operators at the boundary.\footnote{
This harmonizes with a recent discussion in \cite{Benjamin:2021ygh} in the context of CFT$_2$, where it was suggested that the modular completion of light states could be interpreted in terms of a gravitational theory where only heavy states are averaged.} 
In this paper 
we will consider a somewhat wider class of observables where heavy (but sub-threshold) operators are inserted at the boundary. 
This setup will make it clear that operators with dimension ${\cal O}(c)$ will exhibit averaging as well, even if they are somewhat below the black hole threshold.
 It is also quite possible that although we see averaging in the effective low energy theory, these effects are ultimately cancelled by UV contributions, like the half-wormholes of the SYK model \cite{Saad:2021rcu,Saad:2021uzi,Mukhametzhanov:2021nea,Mukhametzhanov:2021hdi} (see also \cite{Blommaert:2021fob}). Even if this is the case, based on two-dimensional examples it seems likely that they encode universal information about the statistics of the UV completion.

In the rest of this introduction, we define the CFT ensemble, and summarize the evidence for the conjecture that this ensemble reproduces semiclassical 3D gravity.

\subsection{Averaging over large-$c$ CFTs}\label{ss:introaveraging}

The high energy spectrum of any 2D CFT is constrained by the Cardy formula \cite{Cardy:1986ie},
\begin{align}\label{introCardy}
\overline{\rho(h,\bh)} = \rho_0(h) \rho_0(\bh) , \quad \rho_0(h) &\approx \exp\left[ 2\pi \sqrt{\frac{c}{6}(h - \frac{c}{24}})\right] \ .
\end{align}
The bar indicates that this is the density of states averaged over some window of conformal weights $(h,\bh)$. Similarly, the bootstrap provides information about average OPE coefficients, summarized by the universal formula \cite{Collier:2019weq}
\begin{align}\label{introC2}
\overline{|c_{pqr}|^2}  = C_0(h_p, h_q, h_r) C_0(\bh_p, \bh_q, \bh_r)
\end{align}
where $p,q,r$ label primary operators and $C_0$ is a known function reproduced in \eqref{C0dozz} below. In any CFT, the Cardy formula holds for asymptotically high energies and the OPE formula applies whenever at least one operator dimension is taken to infinity. 

In holographic CFTs both formulas have a wider range of applicability \cite{Hartman:2014oaa,Michel:2019vnk,Das:2020uax}.  These are theories with a large central charge $c$ and a sparse spectrum of low-lying operators. In a theory of pure gravity, the only primary state below the black hole threshold $h,\bh \sim \frac{c}{24}$ is the vacuum, and the Cardy formula applies for all $h, \bh$ above the threshold. In the dual CFT, the extended range of validity of the Cardy formula follows from modular invariance together with the assumption of a sparse low-energy spectrum \cite{Hartman:2014oaa}.\footnote{Modular invariance has only been shown to extend the Cardy regime down to $h \sim c/12$, not $h \sim c/24$ \cite{Hartman:2014oaa}. We will make the stronger assumption that it extends all the way down to $c/24$. 
This can be justified in chiral CFTs or if the partition function holomorphically factorizes \cite{Witten:2007kt}, which we do not assume.}

With this in mind, we consider an ensemble of CFTs defined as follows. We assume the Cardy formula holds above the black hole threshold, and \eqref{introC2} holds for all nontrivial primaries. Below the threshold, we allow (but do not require) a discrete set of scalar `defect' states with $h/c < 1/24$ held fixed in the large-$c$ limit, plus, when appropriate, multiparticle defects (in the sense of Virasoro mean field theory \cite{Kusuki:2018wpa,Collier:2018exn}). 
The defect states correspond to massive point particles on the gravity side, which backreact to produce conical deficits. In the bulk theory the defect particles are non-interacting, except through the effects of gravity (which include interactions mediated by wormholes), so $\overline{c_{pqr}} = 0$. It is likely that many of our results can be extended to include quantum fields in the bulk but we will not do so here. 

We assume furthermore that the statistics of primary OPE coefficients are Gaussian, so that
\begin{align}\label{introGaussian}
\overline{c_{abc}c_{def}^*} = C_0(h_a, h_b, h_c) C_0(\bh_a, \bh_b, \bh_c)(\delta_{ad}\delta_{be} \delta_{cf} \pm \mbox{permutations})
\end{align}
up to terms suppressed by $e^{-S}$,
and higher moments are computed by Wick contractions.  This assumption is closely related to the eigenstate thermalization hypothesis (ETH) \cite{deutsch1991quantum,Srednicki:1994mfb}.  In the context of averaged holographic duality in AdS$_3$/CFT$_2$, the statistics \eqref{introGaussian} have also been considered in the high energy limit recently in \cite{Belin:2020hea,Belin:2020jxr,Belin:2021ryy} and shown to have the correct scaling in the pinching limit of the genus-2 partition function.\footnote{The evidence for a gravitational interpretation in \cite{Belin:2020hea} is that an ensemble of OPE coefficients with Gaussian statistics leads to a genus-2 partition function with no pinching singularity. This matches the fact that genus-2 wormholes have no pinching singularity. Note that when \cite{Belin:2020hea} discusses the gravitational action of the genus-2 wormhole, the boundary CFT is considered with a hyperbolic metric.  This is enough to test the pinching singularity, but not enough for a detailed comparison between bulk and boundary. We will calculate the full gravitational action for this and various other examples and find a quantitative match.}

The statistics in \eqref{introGaussian} are assumed to hold when all three weights satisfy $h > \frac{c}{32}$, or a combination of the weights is large enough (including a regime with $h_a \ll \frac{c}{24}$); see section \ref{s:ensemble} for details.
For our purposes, this fully defines the ensemble to leading order at large $c$:  Black hole states with a Cardy spectrum, defect states, and Gaussian random OPE coefficients with mean zero and variance \eqref{introC2}. Note that we have said nothing about the energy level statistics of black hole microstates. Level statistics have played a central role in interpreting off-shell contributions to the gravitational path integral \cite{Cotler:2016fpe,Saad:2018bqo,Saad:2019lba,Cotler:2020ugk,Maxfield:2020ale}, but they do not affect any of the observables we consider at classical or 1-loop order.

As noted above, our version of averaging is rather different from the random matrix ensembles dual to Jackiw-Teitelboim gravity \cite{Saad:2019lba}. Random matrix theory is an ensemble of microscopic, UV-complete quantum theories. In our case, we have not defined an ensemble of full-blown CFTs, only an ensemble of certain CFT data. It would be very interesting to define an ensemble of 2d CFTs microscopically (cf. the free toy models in \cite{Afkhami-Jeddi:2020ezh,Maloney:2020nni}) but, absent a complete solution to the conformal bootstrap in two dimensions, this seems difficult.  The ensemble that we have defined above could certainly be refined by the imposition of more bootstrap constraints, which would allow one to compare further subleading terms in  the gravitational path integral.

This discussion allows us to address the following important question: Under what circumstances, and in what approximation, does an individual CFT look like semiclassical Einstein gravity in three dimensions?  Our proposed answer is that the primary operator spectrum must obey \eqref{introCardy} for $h > \frac{c}{24}$ and OPE coefficients must obey \eqref{introGaussian} in the range specified below, where the average is now interpreted in the sense of coarse-graining over an energy window.  Of course, 
(\ref{introCardy}) and (\ref{introC2}) are true in any CFT for operators with $h\gg c$, simply as a consequence of crossing and modular invariance.    The requirement that these hold for $h$ of order $c$ is closely related to the statement that the CFT spectrum  is sparse.
The requirement that these are true upon coarse graining over energies means that the CFT must be chaotic enough that it is self-averaging: an average over states is indistinguishable from an average over coupling constants. 
This definition of chaos differs from what is typically required in discussions of the eigenstate thermalization hypothesis, where only light-heavy-heavy OPE coefficients (i.e. matrix elements of light operators between heavy states) are required to behave like random matrices.

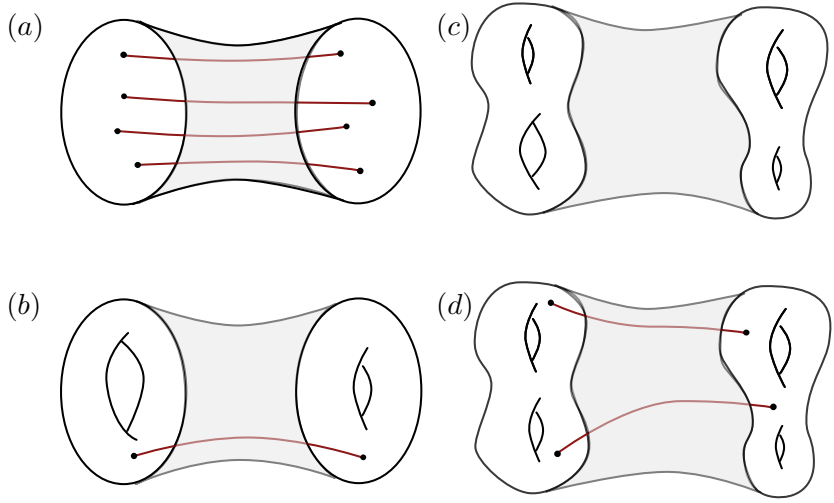
\begin{figure}[t]
\begin{center}
\begin{overpic}[width=300px]{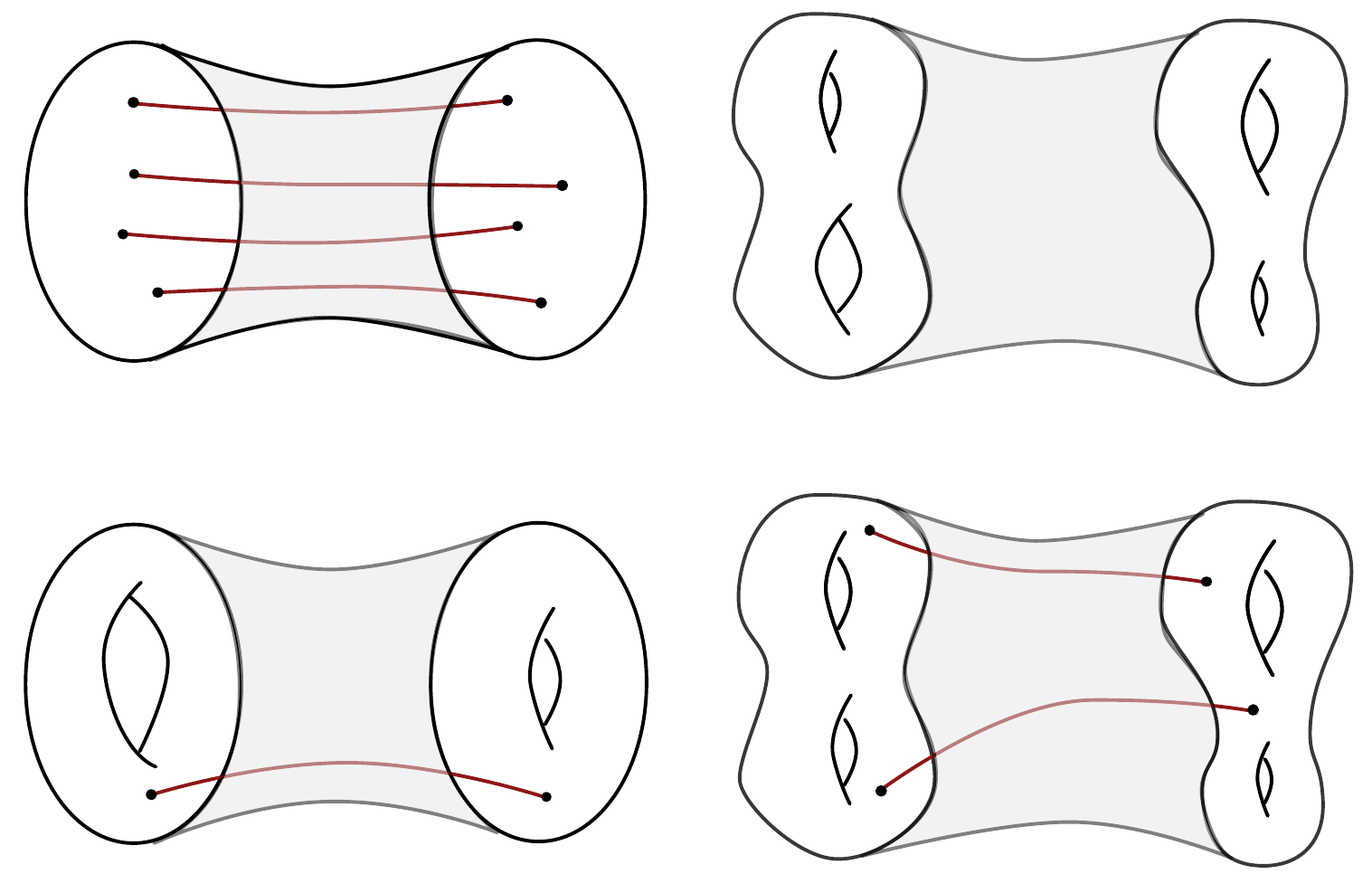}
\put (-5, 60) {$(a)$}
\put (-5, 25) {$(b)$}
\put (49, 60) {$(c)$}
\put (49, 25) {$(d)$}
\end{overpic}
\end{center}
\caption{\small Example 2-boundary (quasi-Fuchsian) wormholes. $(a)$ Spherical topology with 4 massive particles, which contributes to the product of CFT 4-point functions $G(x,\bx)G(x',\bx')$. $(b)$ Torus with 1 massive particle, which contributes to the product of thermal 1-point functions $\langle O \rangle_{\tau} \langle O \rangle_{\tau'}$. $(c)$ Genus-2 wormhole, which contributes to the product of genus-2 partition functions  $Z_{g=2}(\Omega) Z_{g=2}(\Omega')$. $(d)$ A more elaborate wormhole with higher genus and defects.\label{fig:wormholes}}
\end{figure}

The starting point for the comparison of our ensemble with 3D gravity is a computation showing that
the universal asymptotic formula for OPE coefficients, equation (\ref{introC2}), equals the action of a 3-point wormhole:
\begin{align}\label{introC0wormhole}
C_0(h_1,h_2,h_3)^2 \qquad = \qquad \vcenter{\hbox{\includegraphics[width=1.5in]{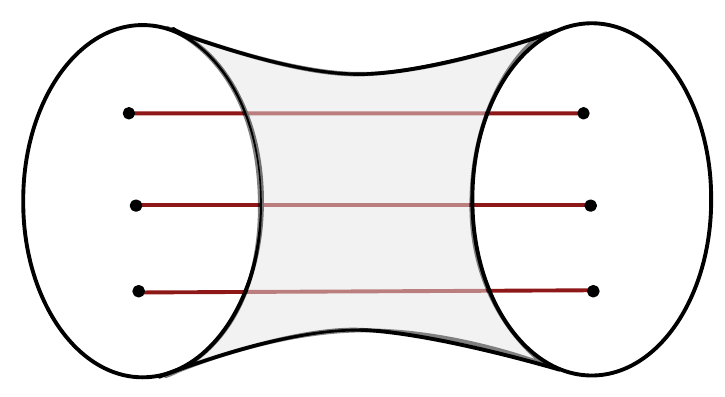}}} \ \quad ,
\end{align}
The picture on the right hand side is of a Euclidean wormhole with two boundaries, each of which is a Riemann sphere with three operators inserted at $0,1,\infty$, and includes massive particles propagating through the bulk. 
Equation (\ref{introC0wormhole}) holds in the semiclassical (large-$c$) limit, where the operator dimensions scale linearly in $c$.

Calculations of more complicated observables in the ensemble of CFTs are straightforward in principle. We simply expand in conformal blocks and apply \eqref{introCardy}, \eqref{introC2}, and \eqref{introGaussian}. For example, consider the product of two CFT 4-point functions of sub-threshold operators,
\begin{align}
G_{1234}G'_{1234} &= \langle O_1(0) O_2(x,\bx) O_3(1) O_4(\infty) \rangle \, \langle O_1(0) O_2(x', \bx') O_3(1) O_4(\infty) \rangle \ .
\end{align}
This has a double expansion in Virasoro conformal blocks,
\begin{align}
G_{1234}G'_{1234} = \sum_{p,q} c_{12p}c_{34p} c_{12q} c_{34q} \left| \F_{1234}(h_p,x) \F_{1234}(h_q, x') \right|^2 \ .
\end{align}
Averaging sets $p=q$, so for the ensemble average we obtain a single correlated sum:
\begin{align}\label{GGintro}
\overline{G_{1234}G'_{1234}} &= \sum_p \overline{c_{12p}^2}\ \overline{c_{34p}^2} \left| \F_{1234}(h_p,x) \F_{1234}(h_p, x') \right|^2
\end{align}
This can be further simplified using \eqref{introCardy} and \eqref{introC2}. We will analyze this particular example in more detail in section \ref{ss:G4G4} (on the CFT side) and section \ref{ss:almostfuchsian} (on the gravity side). The gravity calculation is rather involved for $x \neq x'$. The conclusion is that in the large-$c$ limit, \eqref{GGintro} is reproduced by the 2-boundary wormhole pictured in fig.~\ref{fig:wormholes}a, including both the classical action and 1-loop corrections. The match involves a complicated function of the cross ratios $(x,\bx)$ and $(x', \bx')$ on the two sides of the wormhole that cannot be evaluated in closed form but nonetheless is shown to agree.

Other products of any number of observables in the CFT ensemble can be calculated by similar methods. Generally, the calculation can be done in various OPE channels, and the Gaussian average has multiple contractions among the $c_{ijk}$'s. The sums can also be separated into light and heavy contributions. The picture that emerges is that once we have organized the CFT calculation in this way --- by choosing an OPE channel, doing the OPE contractions, and separating out the light contributions --- each term corresponds to a different wormhole (with the caveat that some OPE channels are deemed equivalent). We will demonstrate this through numerous examples; see section \ref{ss:plan} for a complete list.

\subsection{The Coleman-Giddings-Strominger mechanism}

Coleman \cite{Coleman:1988cy} and Giddings and Strominger \cite{Giddings:1987cg,Giddings:1988cx} proposed that microscopic wormholes at the Planck or string scale would lead to a low energy effective theory with random coupling constants. Integrating out the wormhole gives a bilocal operator at its endpoints, and such a term in the effective action can then be reinterpreted by `integrating in' a random coupling constant. This intriguing idea has never found a home in AdS/CFT, where the random coupling constants would appear in the bulk Lagrangian. (Another aspect of Coleman's approach, namely alpha states, does play a prominent role in recent discussions of AdS$_2$/CFT$_1$ \cite{Marolf:2020xie,Marolf:2020rpm,Goel:2020yxl,Casali:2021ewu,Saad:2021uzi}, but with no clear connection to random bulk coupling constants.)

We will show that the low-energy theory of 3D gravity coupled to sufficiently heavy point particles exhibits a semiclassical version of the Coleman-Giddings-Strominger mechanism: the bulk effective theory has random couplings that can be calculated explicitly by integrating out classical wormholes.  Because the wormholes are on shell, the details are a bit different from \cite{Coleman:1988cy,Giddings:1987cg,Giddings:1988cx}. For a preview, consider the gravity calculation of the CFT correlation function
\begin{align}
G_{1221} = \langle O_1 O_2 O_2 O_1\rangle \ ,
\end{align}
where $O_1$ and $O_2$ are two distinct sub-threshold operators. For concreteness, suppose the bulk theory has three species of particles, 1, 2 and 3. Then the conformal block expansion of this 4-point function has a term
\begin{align}\label{intro123}
c_{123}^2 \left| \F_{1221}(h_3,x) \right|^2 \ .
\end{align}
What is its bulk interpretation? By assumption, our bulk theory has no interactions among the particles, so we might conclude that this term does not appear. However, there is a 2-boundary wormhole that calculates $\overline{c^2_{123}}$, and it is nonzero, so this cannot be correct. The explanation is that there is a single-boundary wormhole responsible for this term:
\begin{align}
\overline{c_{123}^2} \left| \F_{1221}(h_3,x) \right|^2 \quad =  \quad
\vcenter{\hbox{
\begin{overpic}[width=2in]{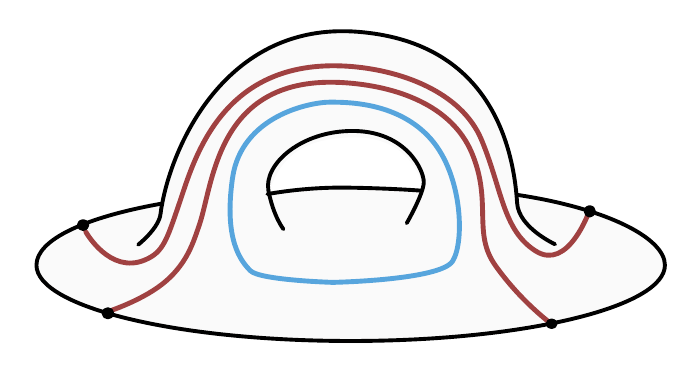}
\put (8,24) {$1$}
\put (11,3) {$2$}
\put (80,2) {$2$}
\put (85,26) {$1$}
\put (47,15) {$3$}
\end{overpic}
}}
\end{align}
This diagram represents a wormhole with the topology obtained by taking a 3-ball, cutting out two smaller 3-balls inside of it, and gluing together their $S^2$ boundaries. 
Particle 3 (in blue) makes a closed loop through the wormhole.
This wormhole is a classical solution to the equations of motion, which can be constructed when the operator dimensions scale linearly with $c$, and the three particle masses are sufficiently large.
The classical action of this wormhole precisely reproduces the left-hand side, including the OPE coefficient, to leading order at large $c$.

Alternatively, we can integrate out the wormholes. In the classical limit this just means writing an effective bulk theory with an explicit 3-point coupling that reproduces \eqref{intro123}. It must also reproduce the correct 3-point function, $\overline{\langle O_1 O_2 O_3 \rangle }= 0$. Therefore the effective theory has a random coupling constant with mean zero and variance $\overline{c^2_{123}}$. In the effective description, the wormhole is replaced by a Witten diagram:
\begin{align}
\vcenter{\hbox{\includegraphics[width=2in]{figures/handle.pdf}}} 
\qquad \Longrightarrow \qquad
\vcenter{\hbox{\includegraphics[width=2in]{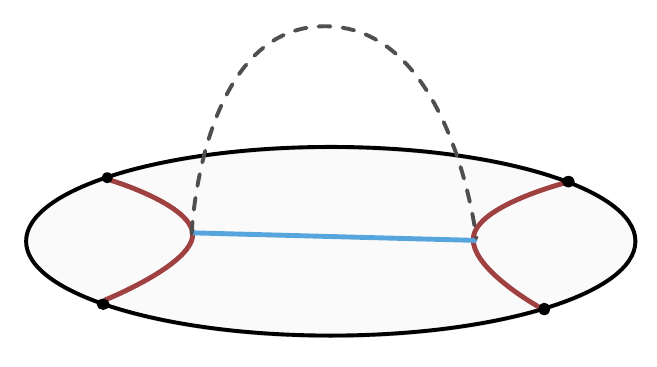}}}
\end{align}
The dashed line indicates that the 3-point couplings in this Witten diagram are correlated by randomness.

One should include either the wormhole or the direct interaction in the bulk theory, not both. The wormhole encodes the average effect of the direct interaction. In a UV-complete theory with fixed couplings, presumably the wormhole does not have a life of its own, but is already included somehow in the sum over microstates (see for example \cite{Eberhardt:2021jvj,Saad:2021rcu}).

\subsection{Multi-boundary wormholes}

We will now describe 3D wormholes in a bit more detail, and state our main results more concretely. Gravity in three dimensions has a wide variety of classical wormholes. The simplest have two asymptotic boundaries with identical topology, as in figure \ref{fig:wormholes}. The topology of such wormholes is $\Sigma_{g,n} \times \mbox{Interval}$, where $\Sigma_{g,n}$ is a genus-$g$ Riemann surface with $n$ conical defects at the locations of the point particles. To find classical solutions we require that $\Sigma_{g,n}$ admits a hyperbolic metric. On the sphere, this imposes $n \geq 3$ and restrictions on the particle masses; on the torus it imposes $n \geq 1$, but allows any mass; on higher genus surfaces, no point particles are required. The 4-point function discussed in section \ref{ss:introaveraging} above was one example in this class. In the special case where the moduli are constant across the wormhole, we refer to this as the Maldacena-Maoz (or Fuchsian) wormhole \cite{Maldacena:2004rf}, and when the moduli vary, it is called a quasi-Fuchsian wormhole.

We denote the period matrix of $\Sigma$ by $\Omega$. The moduli are allowed to vary across the wormhole, so these saddles contribute to CFT observables of the form
\begin{align}
G G' = \langle O_1(x_1) O_2(x_2) \cdots \rangle_{\Sigma(\Omega)}
\langle O_1(x_1') O_2(x_2') \cdots \rangle_{\Sigma(\Omega')} 
 \ ,
\end{align}
where each $G$ is a CFT correlation function of $n$ operators on a surface $\Sigma$. We will show that the semiclassical contribution of the $\Sigma_{g,n} \times \mbox{Interval}$ wormhole to the gravitational path integral matches a term in the CFT average,
\begin{align}\label{introGGW}
\overline{GG'} \quad \supset \quad 
 e^{-S_{\rm wormhole}}  \ Z_{\rm 1-loop}^{\rm gravity} \ .
\end{align}
The wormhole corresponds to a sum over heavy exchanges for a particular OPE channel and contraction. 

For a given OPE channel in the CFT calculation of $\overline{GG'}$, all of the heavy exchanges are accounted for by this class of wormholes. There are also contributions from defect exchanges, considered in section \ref{s:coleman}; these have a natural bulk interpretation as handles added to the $\Sigma_{g,n} \times \mbox{Interval}$ wormhole.

In the process of deriving \eqref{introGGW}, we show that the action of a 2-boundary wormhole with topology $\Sigma_{g,n} \times \mbox{Interval}$ is given by a product of observables in the Liouville CFT. Let $(\sigma, \bsigma)$ and $(\sigma', \bsigma')$ be collectively all of the cross ratios of operators inserted on the left and right boundary, respectively, and denote the Liouville $n$-point correlation function on a surface with period matrix $\Omega$ by $G_L(x,\bx; \Omega,\bOmega)$. The wormhole contribution to the gravitational path integral is found to be
\begin{align}\label{introGL}
&\overline{G(\sigma, \bsigma; \Omega, \bOmega) G(\sigma', \bsigma'; \Omega',\bOmega')} \\
&\qquad \supset e^{-S_{\rm Wormhole}}  \approx G_L(\sigma, \sigma'; \Omega, -\Omega') G_L(\bsigma'; \bsigma; -\bOmega',\bOmega)\notag \ .
\end{align}
The agreement holds at the level of the classical action; for wormholes without defects, we also match the one-loop corrections.
Note that the arguments are mixed up on the two sides of this relation: the first Liouville factor accounts for left-movers in CFT$\times $CFT, and the second Liouville factor accounts for all the right-movers. This pairing can be anticipated from the Chern-Simons/WZW relationship \cite{Witten:1988hf,Elitzur:1989nr} and was also seen in the (off-shell) double-torus calculation of Cotler and Jensen \cite{Cotler:2020ugk} and in a Hamiltonian treatment of certain backgrounds \cite{Henneaux:2019sjx}.

There are also wormhole saddles with $k>2$ asymptotic boundaries. Solutions of 3D gravity are hyperbolic manifolds and therefore quotients of $\mathbb{H}_3$ by subgroups of $SL(2,C)$. The 2-boundary wormholes that we just described correspond to quasi-Fuchsian groups (when the conical defects have finite order), while more generally, multi-boundary wormholes can be constructed as quotients by Kleinian groups. We will discuss them briefly in section \ref{s:kboundary}. The CFT calculation is straightforward, and it leads to a detailed prediction for the gravitational action on various Kleinian manifolds (and conifolds). We have not made a detailed check of this prediction on the gravity side for $k>2$ boundaries, but we will check it in one particular limit and find a quantitative agreement.

\subsection{Average interpretation of simple saddles}

We have emphasized the contributions of higher topologies, but of course the ensemble must also reproduce all of the ordinary saddlepoints of 3D gravity, including all of the examples studied in \cite{Yin:2007gv,Hartman:2013mia,Faulkner:2013yia,Fitzpatrick:2014vua,Asplund:2014coa,Anous:2016kss}. These are the `simple' saddles with a single boundary, including conical defects in global AdS$_3$, the BTZ black hole and its higher-genus analogues known as handlebodies, and conical defects in handlebodies. 

Each of these geometries is dual to a semiclassical Virasoro identity block in some channel \cite{Yin:2007gv,Hartman:2013mia}. We will show that these identity blocks can also be interpreted as an average over heavy states in a crossed channel. This is closely related to the fact that the large-$c$ ensemble is crossing invariant by construction, and certain observables are well approximated by the identity block.

\begin{figure}
\begin{center}
\includegraphics[width=5.5in]{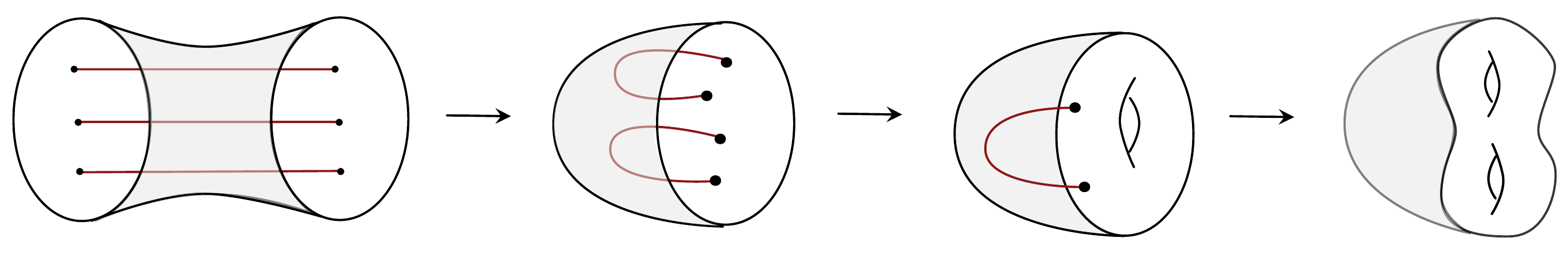}
\end{center}
\caption{If the mass of a conical defect is continued above the black hole threshold, it disappears, and results in two pieces of the boundary being glued together. This relates wormhole amplitudes to analytically continued Virasoro identity blocks. The figure shows how the saddlepoint changes as each additional conical defect is taken above threshold.\label{fig:continuation}}
\end{figure}

There is an important relationship between the Virasoro identity blocks that calculate simple observables and some of the more exotic higher topology saddles. For example, consider the 2-boundary wormhole supported by conical defects propagating across the wormhole, as in fig.~\ref{fig:continuation}. If we increase the mass of one of the defects until it is above the black hole threshold, the topology of this solution changes: the defect disappears, and the two boundary components get glued together.  Thus the two boundary wormhole has become a solution with a single boundary, which is a sphere with four (pairwise identical) operator insertions.  If we continue this process by making another pair of operators heavy the boundary becomes a torus with two identical operator insertions.  Finally, if we make these last two operators heavy the boundary becomes a genus two surface, and the bulk a handlebody which fills in this surface.  At each step in this process, the parameters which label operator dimensions become moduli of the Riemann surface when the operators are made heavy.
This procedure relates our wormhole amplitudes to the analytic continuation of Virasoro identity blocks, and will be described quite explicitly.
A related phenomenon was observed in \cite{Collier:2019weq}, where it was noted that the universal formula for OPE coefficients (\ref{introC2}) applies in three different regimes depending on whether one, two or three operators are taken to be heavy.  In our discussion here we have extended this formula into the regime where all operators are sub-threshold, and the corresponding geometry is a wormhole.  The important distinction is that in \cite{Collier:2019weq} the averaging was interpreted as a coarse-graining over states, while here it is interpreted as an ensemble averaging over CFT data.

\subsection{Plan and list of examples}\label{ss:plan}

In section \ref{s:ensemble} we describe the CFT ensemble in more detail, and review the origins of the Cardy density of states and the universal OPE formula. We also show that the ensemble is crossing invariant to leading order at large $c$.

In section \ref{s:twocopy}, we calculate averaged products of observables in two copies of the CFT. The example of the double 4-point function, $\overline{G_{1234}G_{1234}'}$ was sketched above; this CFT calculation is done in detail in section \ref{ss:G4G4}, and extended to general 2-copy observables in section \ref{ss:general2copy}, with a few more examples in section \ref{ss:examples2copy}, including sphere $n$-point functions, torus 1-point functions, and genus-2 partition functions. For a class of 2-boundary observables  (but not for other cases) both the CFT calculation and the gravity calculation lead to an answer expressed in terms of two copies of the Liouville CFT.

In sections \ref{s:fuchsian} through \ref{s:singleboundary}, we construct various wormholes and other solutions on the gravity side, and compare them to the CFT ensemble. The examples that we check quantitatively are:
\begin{itemize}
\item Single-boundary, handlebody solutions with or without defects (sections \ref{s:ensemble} and \ref{s:singleboundary}). These solutions have been considered previously and matched to CFT identity blocks. We will show that they are also compatible with the large-$c$ ensemble, and that in fact, this interpretation has some advantages because it automatically satisfies crossing.
\item All 2-boundary wormholes with the topology $\Sigma_{g,n} \times \mbox{Interval}$, for any genus $g$ and collection of $n$ defects, with different (or equal) moduli on the two boundaries. 
The case of equal moduli is the Fuchsian wormhole of Maldacena and Maoz \cite{Maldacena:2004rf} (section \ref{s:fuchsian}); the general case is referred to as the quasi-Fuchsian wormhole (section \ref{s:twoboundary}).
\item Single-boundary wormholes with defects propagating through a handle. These are reinterpreted as random couplings in the bulk effective theory, realizing the Coleman-Giddings-Strominger mechanism (section \ref{s:coleman}).
\item $\mathbb{Z}_k$-symmetric, $k$-boundary wormholes. In this case we do the CFT calculation for any $k$, but only compare to gravity for $k$ analytically continued near 2 (section \ref{s:kboundary}).
\end{itemize}
In all of these cases the classical gravity action is matched to the CFT ensemble. Graviton 1-loop corrections are also matched to CFT for the smooth quasi-Fuchsian wormholes (section \ref{s:oneloop}).

\section{The CFT ensemble}\label{s:ensemble}

In this section we describe our ensemble as an average over approximate solutions to the crossing equations in the semiclassical limit.

Before doing so, we briefly establish some notation that will be used throughout the paper. The approximation symbols $\sim$ and $\approx$ are used with the precise meanings
\begin{align}
X \sim  Y &\Rightarrow \lim_{c \to \infty} \frac{X}{Y} = 1 \\
X \approx Y &\Rightarrow \lim_{c \to \infty} \frac{\log X}{\log Y} = 1 \ , \notag
\end{align}
with conformal weights scaling as $h \sim c$ in the large-$c$ limit. Thus `$\approx$' means the classical actions agree, and `$\sim$' means both the classical actions and the 1-loop corrections agree.

Chiral conformal weights are denoted by $(h,\bh)$, with total scaling dimension $\Delta = h+\bh$ and spin $\ell = h - \bh$. We will sometimes adopt the Liouville parameterization
\begin{align}\label{hprelation}
h =  \frac{c-1}{24} + P^2 \ 
\end{align}
that appears naturally in Virasoro representation theory. See appendix \ref{s:liouville} for more details. The semiclassical limit is $c \to \infty$ with $h/c$ and $\bh/c$ held fixed. In this limit we parameterize the weights by $\eta$ or $\gamma$, with 
\begin{align}\label{largecweights}
h &= \frac{c}{6}\eta(1-\eta)  = \frac{c}{24}(1 +  \gamma^2) \ , \qquad \eta = \frac{1}{2}(1 + i \gamma) \ .
\end{align}
The black hole threshold in the large-$c$ limit corresponds to $\eta = \frac{1}{2}$. Below the threshold, $\eta$ is real and falls in the range $[0, \frac{1}{2}]$ (although, for reasons we will discuss shortly, we will often restrict $\eta$ to $[\frac{1}{4},\frac{1}{2}]$). Above the threshold,  $\gamma \in \mathbb{R}$.

\subsection{Definition of the ensemble}\label{ss:ensembleDefinition}

As described in the introduction, we fix the central charge $c \gg 1$ and average over an ensemble of CFT data defined in terms of its leading order spectrum of primary operators and OPE coefficients in the semiclassical limit.  
We will assume the theory has no symmetries beyond conformal invariance. If there are additional symmetries then the ensemble must be modified accordingly.

\bigskip

\noindent \textbf{Spectrum of primaries}\\
We assume the CFT consists of:
\begin{itemize}
\setlength\itemsep{-0.2em}
\item A unique normalizable vacuum state.
\item A finite, discrete list of `defect' scalars with dimensions below the black hole threshold, $ 1 \ll h <\frac{c-1}{24}$.
\item Multi-twist operators built from these.
\item A continuous spectrum of `black hole' states with Cardy density of states, $\rho_0(h) \rho_0(\bh)$, for $h,\bh \geq\frac{c-1}{24}$.
\end{itemize}
We will often make the further assumption that the defect states are all sufficiently heavy such that they do not form additional discrete, infinite towers of multitwist composite primaries. Pairs of defects with $\eta_1+\eta_2 < \half$ have multitrace composites. If there are no primary operators in the spectrum with $h < {c-1\over 32}$, i.e. with $\eta<{1\over 4}$, then the spectrum contains no multitrace composites.
In the bulk, operators with $\eta = \frac{1}{4}$ are dual to particles that backreact on the geometry to produce conical deficits with defect angle $\pi$, which is why they cannot form multiparticle states without becoming a black hole \cite{Collier:2018exn}. Restricting to heavy defects simplifies the CFT analysis, and in some cases, it is necessary in order for averaged observables to receive contributions from real wormhole saddlepoints in the bulk.

Note that although the density of states above the black hole threshold in the ensemble is assumed to be continuous, we have defined an ensemble of \emph{compact} solutions to the crossing equations in the sense that the vacuum state gives a normalizable contribution (with coefficient one) to observables such as correlation functions and partition functions. This distinguishes ensemble-averaged CFT spectra from noncompact solutions to the crossing equations (such as the Liouville CFT or sigma models with noncompact target spaces) in which the density of states is continuous and the identity operator and its descendants are not exchanged in intermediate channels of CFT observables. Indeed, the Virasoro vacuum block that encapsulates the contribution of the identity operator and its descendants to CFT observables will play an important role in what follows.

\bigskip

\noindent \textbf{OPE coefficients}\\
Our ensemble is defined by treating the OPE coefficients of primary operators as Gaussian random variables with zero mean and variance given by the universal asymptotic formula for squared structure constants. For distinct operators, to leading order at large $c$,\footnote{
The absolute value on $c_{ijk}$ is a true absolute value in this equation and throughout the paper, $|c_{ijk}|^2 := c_{ijk} c^*_{ijk}$ with $*$ the complex conjugate. In expressions involving weights, the absolute value symbol is to be interpreted in the standard CFT convention, e.g. $|C_0(h_i, h_j, h_k)|^2 := C_0(h_i, h_j, h_k) C_0(\bh_i, \bh_j, \bh_k) \neq C_0(h_i, h_j, h_k) C_0(h_i, h_j, h_k)^*$ and $|\F(h,x)|^2 = \F(h,x) \bar{\F}(\bh,\bx)$.
}
\begin{align}\label{c2C0}
\overline{|c_{ijk}|^2} &= C_0(h_i, h_j, h_k) C_0(\bh_i, \bh_j, \bh_k) \ .
\end{align}
In a general CFT, this equation is universal for heavy operators  \cite{Collier:2019weq}, with the average interpreted in the microcanonical sense. In our ensemble, it is assumed to hold whenever one of the operators is above the black hole threshold, and also whenever all three operators are below the threshold provided that their dimensions are large enough. Specifically, if all three are below threshold but heavy enough to support a 3-defect wormhole, then \eqref{c2C0} applies; this requires
\begin{align}\label{etaCondition}
\eta_i + \eta_j + \eta_k > 1 \ ,
\end{align}
or equivalently,
\begin{align}
\sqrt{1 - \frac{24h_i}{c}} + \sqrt{1 - \frac{24h_j}{c}} + \sqrt{1 - \frac{24h_k}{c}} <1 \ .
\end{align}
Following the standard conventions of 2d CFT, $c_{ijk}$ is real (imaginary) if $\ell_i + \ell_j+\ell_k$ is even (odd), and the OPE coefficients are conjugated under permutations,  $c_{ijk} = c_{ikj}^*$. Therefore the Gaussian contraction is
\begin{align}\label{contractions}\small
\overline{c_{ijk}c^*_{lmn}} \ \ =\ \  C_0(h_i, h_j, h_k) C_0(\bh_i, \bh_j, \bh_k)\left( 
\delta_{il}\delta_{jm}\delta_{kn} + (-1)^{\ell_i+\ell_j+\ell_k}\delta_{il}\delta_{jn}\delta_{km}
+ \mbox{4 more terms}\right) \ ,
\end{align}
where $i,j,k$ label primary operators. We will not consider observables that involve OPE coefficients of multitrace operators or three light operators violating \eqref{etaCondition}, so whenever $C_0$ appears in this paper, this condition is assumed.

The universal OPE function $C_0$ \cite{Collier:2019weq} is simply related to the DOZZ \cite{Dorn:1994xn,Zamolodchikov:1995aa} structure constants of Liouville CFT. It is given by
\begin{align}\label{C0dozz}
C_0(h_1, h_2, h_3) &\coloneqq {\tCdozz(P_1,P_2,P_3)\over \sqrt{\prod_{k\in\text{heavy}}\rho_0(P_k)}} \ ,
\end{align}
where ``heavy'' refers to the weights above the black hole threshold (which will appear in what follows only as internal operators in OPE decompositions of CFT observables), $\tCdozz$ is the standard definition of the structure constants in the Liouville CFT\footnote{For sufficiently light defects, with $\eta < \frac{1}{4}$, there are extra contributions from analytic continuation discussed in appendix \ref{s:liouville}.} up to a choice of normalization for the operators, and $\rho_0$ is the Cardy density of states defined more precisely below in \eqref{rho0exact}. A brief review of the Liouville CFT and of our conventions is contained in appendix \ref{s:liouville}.\footnote{We are normalizing sub-threshold and above-threshold Liouville operators slightly differently, for reasons explained in the appendix. This is why the denominator in \eqref{C0dozz} only has heavy-operator Cardy factors; the Cardy factors for sub-threshold operators, which are $O(1)$, have been absorbed into the normalization. None of this affects the final answers but this choice of normalization makes for an easier comparison to gravity.}

\bigskip

We pause to emphasize that this data does not define a true CFT. It does, however, define an ensemble of CFT data that one can use to perform meaningful computations of averaged CFT observables. The spectrum is designed to reflect the essential features of semiclassical 3D gravity --- large $c$ and a sparse spectrum of low-dimension operators --- and to obey certain bootstrap constraints. The Cardy spectrum ensures the torus partition function is approximately modular invariant in the semiclassical limit. As we will review shortly, the OPE coefficients are designed to ensure that certain bootstrap conditions on the sphere 4-point functions, torus 2-point functions, and genus-2 partition function are also satisfied approximately. Other bootstrap constraints, with more insertions or higher genus, are also satisfied to leading order. Indeed, we will shortly see that these two characterizations --- that the CFT data capture essential aspects of semiclassical 3D gravity and satisfy bootstrap constraints approximately in the semiclassical limit --- are two sides of the same coin.

Note that although Liouville CFT will appear throughout the paper, our ensemble is certainly not Liouville,\footnote{To name a few obvious differences: the averaged density of states is given by the Cardy spectrum rather than the Liouville density of states (which is flat in the Liouville momentum $P$), the ensemble contains primaries of all spins (indeed the leading semiclassical spectrum does not even see quantization of the spin above the black hole threshold) rather than just scalars, and the identity operator defines a normalizable vacuum state. On the other hand there is a long history of connections between Liouville and 3D gravity starting with \cite{Verlinde:1989ua}; see e.g.~\cite{Jackson:2014nla,Cotler:2018zff} for a recent perspective.} and semiclassical 3D gravity is emphatically not dual to Liouville theory. Liouville quantities are to be regarded only as  auxiliary tools to calculate certain averaged observables, such as $\overline{|c^2_{ijk}|}$, in which $c_{ijk}$ is the OPE coefficient in the dual CFT. 

The definition of the ensemble in terms of $C_0$ and $\rho_0$ is only accurate to leading order in the semiclassical expansion. This is roughly analogous to specifying the spectral curve of a random matrix theory in terms of its genus-zero density of states. In particular, there must be non-Gaussianities in the OPE coefficients at subleading orders due to interactions and the crossing equations \cite{Foini:2018sdb,Belin:2021ryy}. For the most part we consider only the leading terms, but we will touch on these corrections briefly in section \ref{ss:nongaussianities}.

The way we have defined the ensemble implicitly restricts to a distinguished region in Teichm\"{u}ller space. For example, in the density of states we have included only the vacuum and its $S$-transform. 
It would be natural to define an enhanced ensemble where the density of states is obtained by summing the vacuum over all of its $SL(2,\mathbb{Z})$ images, as in \cite{Maloney:2007ud,Keller:2014xba}.
The corrections obtained in this way are exponentially subleading when we restrict $\tau$ to lie within the usual fundamental domain or its image under $\tau\to-\frac{1}{\tau}$, 
but can be dominant for other values of the moduli.\footnote{Similarly, we have also included only the Cardy density of states and not contributions from the $S$-transform of the defect states; including such states would lead to corrections which are exponentially suppressed everywhere in moduli space.  These can be important when analyzing detailed features of the spectrum, as in \cite{Benjamin:2019stq, Benjamin:2020mfz}, but are invisible semi-classically.}  In this paper we will be modest, and consider only the $S$-invariant ensemble, which will be sufficient to reproduce semi-classical gravity in this region of moduli space.
While we could certainly consider the fully $SL(2,\mathbb{Z})$ invariant density of states, 
it is rather more difficult to study the analogous ensemble of OPE coefficients defined by a sum over images. This is technically challenging, and in any case our entire analysis is saddle-by-saddle, so it simpler to work with the ensemble defined above with the understanding that we are restricting to a limited range of moduli. 
To obtain the dominant contribution for other values of the moduli, we would just need to work in the appropriate OPE channel. 
A related fact is that the ensemble as we defined it has continuous, non-integer spins; this is invisible at leading order, but important at subleading orders. Perhaps it can be cured by adding subleading channels.

\subsection{$\rho_0$ and $C_0$ as crossing kernels}

The Cardy formula for the asymptotic density of states can be viewed as the crossing kernel that expresses the Virasoro vacuum character in a complete basis of characters in the cross-channel,
\begin{align}
\chi_0\left(-\frac{1}{\tau}\right) = \int_{(c-1)/24}^\infty dh \, \rho_0(h) \chi_{h}(\tau) =  \int_{\mathbb{R}} \frac{dP}{2} \rho_0(P) \chi_{{c-1\over 24}+P^2}(\tau) \ ,
\end{align}
where $\chi_h$ is the non-degenerate Virasoro character, and we define
\begin{align}\label{measuretrans}
\rho_0(h)dh = \rho_0(P)dP \ .
\end{align}
The exact kernel is given by
\begin{align}\label{rho0exact}
\rho_0(P) = 4\sqrt{2} \sinh(2\pi P/b) \sinh(2\pi P b) \ , 
\end{align}
where $c = 1+6(b + b^{-1})^2$. 
The exponential dependence in the semiclassical approximation is the usual Cardy formula,
\begin{align}
\log \rho_0(P) \sim \frac{\pi c \gamma}{6}  \ , 
\end{align}
with $\gamma$ defined in \eqref{largecweights}.
Note that the Cardy factors appearing in \eqref{C0dozz} are $\rho_0(P)$, which differs from $\rho_0(h)$ by a measure factor $\frac{dh}{dP}$.

Choosing the density of states to be $\rho_0(h) \rho_0(\bh)$ ensures that our ensemble is approximately modular invariant. It is only approximate because we have not included any corrections to the heavy spectrum from the sub-threshold operators, and because the Cardy formula does not impose integrality of spins, so it does not exactly obey $Z(\tau,\btau) = Z(\tau+1,\btau+1)$.

The origin of $C_0$ is similar. The universal OPE formula quoted in the introduction holds in any CFT in the sense of an asymptotic microcanonical average, provided at least one operator is taken to be heavy. It is derived by solving certain bootstrap equations asymptotically in various OPE limits. Since we are assuming that the OPE formula holds for all states, our ensemble satisfies these bootstrap equations for all values of the moduli and cross ratios, at leading order in the large-$c$ limit. In other words, we can use $C_0$ to solve crossing either in the OPE limit or the large-$c$ limit, and here we are choosing to do the latter. This parallels the fact that the Cardy formula generally holds only at very high energies compared to the central charge, but in holographic CFTs, it holds to leading order in $c$ for all energies above the black hole threshold \cite{Hartman:2014oaa}. It is likely that the universal formula for OPE coefficients similarly enjoys an extended regime of validity in holographic CFTs with certain conditions on the light spectrum, but we will not endeavor to quantify the most general set of such conditions (nor the precise sense in which the regime of validity is extended) here. See however \cite{Michel:2019vnk,Das:2020uax}, which explore conditions needed for an extended range of applicability of the universal OPE formula in slightly different contexts (in particular with some subset of the operators kept light in the semiclassical limit).

\begin{figure}
\begin{align*}
\vcenter{\hbox{
\begin{tikzpicture}[scale=0.5]
\draw (0,1) -- (1, 0);
\draw (0, -1) -- (1,0);
\draw (1,0) -- (3,0);
\draw (3,0) -- (4,1);
\draw (3,0) -- (4,-1);
\node at (-0.25,1) {$a$};
\node at (-0.25,-1) {$a$};
\node at (4.25,1) {$b$};
\node at (4.25,-1) {$b$};
\node at (2,0.4) {$\mathbb{1}$};
\end{tikzpicture}
}}
\qquad=\qquad
\int dh \, \rho_0(h) C_0(h_a, h_b, h)
\vcenter{\hbox{
\begin{tikzpicture}[scale=0.5]
\draw (0,0) -- (1,-1);
\draw (1,-1) -- (2,0);
\draw (1,-1) -- (1,-3);
\draw (1,-3) -- (0, -4);
\draw (1,-3) -- (2,-4);
\node at (-0.25,0) {$a$};
\node at (2.25,0) {$b$};
\node at (-0.25, -4) {$a$};
\node at (2.3, -3.9) {$b$};
\node at (1.4,-2) {$h$};
\end{tikzpicture}
}}
\end{align*}
\caption{$C_0$ is the crossing kernel for the identity operator. This equation also applies to subdiagrams inside more complicated blocks.\label{fig:identitycrossing}}
\end{figure}

The defining property of $C_0$ is that it reproduces the identity Virasoro block in the dual channel. That is, 
\begin{align}\label{C0fourpoint}
\int_{(c-1)/24}^{\infty} dh  \, \rho_0(h) C_0(h_1, h_2, h) \F_{1221}(h; x) = \F_{1122}(\id; 1-x) \ ,
\end{align}
where $\F_{1234}(h,x)$ is the chiral Virasoro conformal block on the sphere, with cross-ratio $x$ 
and points ordered as $\O_1(0)\O_2(x,\bx)\O_3(1)\O_4(\infty)$. See fig.~\ref{fig:identitycrossing}. The right-hand side of \eqref{C0fourpoint} is the 4-point identity block. The formula only holds literally for sufficiently heavy operators; in the semiclassical limit, the restriction is  $\eta_1 + \eta_2 > \half$. Otherwise there are are additional contributions from sub-threshold S-channel conformal blocks that must be included on the left-hand side of \eqref{C0fourpoint}.

The same relation applies to internal legs of more complicated conformal blocks. For example, for 2-point Virasoro blocks on the torus we have
\begin{align}\label{eq:torus2ptIdentityCrossing}
\int_{c-1\over 24}^{\infty} dh_2 dh_3\, \rho_0(h_2) \rho_0(h_3) C_0(h_1, h_2, h_3) \,
\vcenter{\hbox{
	\begin{tikzpicture}[scale=0.75]
	\draw[thick] (0,0) circle (1);
	\draw[thick] (-1,0) -- (-2,0);
	\node[above] at (-2,0) {$1$};
	\node[above] at (0,1) {$2$};
	\node[below] at (0,-1) {$3$};
	\draw[thick] (1,0) -- (2,0);
	\node[above] at (2,0) {$1$};
	\node[scale=0.75] at (0,0) {$-1/\tau$};
	\end{tikzpicture}
	}}
=
\vcenter{\hbox{
	\begin{tikzpicture}[scale=0.75]
	\draw[thick] (0,0) circle (1);
	\draw[thick] (0,1) -- (0,2);
	\draw[thick] (0,2) -- (0.866,2+1/2);
	\draw[thick] (0,2) -- (-0.866,2+1/2);
	\node[left] at (0,3/2) {$\id$};
	\node[left] at (-1,0) {$\id$};
	\node[left] at (-0.866,2+1/2) {$1$};
	\node[right] at (0.866,2+1/2) {$1$};
	\node[scale=0.75] at (0,0) {$\tau$};
	\end{tikzpicture}
	}}  
\end{align}
by combining a modular $S$-transformation with the crossing move in figure \ref{fig:identitycrossing}. 
For the genus-2 partition function,
\begin{align}\label{eq:genus2IdentityCrossing}
\int_{c-1\over 24}^{\infty} dh_1dh_2dh_3 \, \rho_0(h_1) \rho_0(h_2) \rho_0(h_3) C_0(h_1,h_2,h_3)\,
\vcenter{\hbox{
	\begin{tikzpicture}[scale=0.75]
	\draw[thick] (0,0) circle (3/2);
	\draw[thick] (-3/2,0) -- (3/2,0);
	\node[above] at (0,3/2) {$1$};
	\node[above] at (0,0) {$2$};
	\node[above] at (0,-3/2) {$3$};
	\end{tikzpicture}
	}} 
	= 
	\vcenter{\hbox{
	\begin{tikzpicture}[scale=0.75]
	\draw[thick] (0,3/2) circle (1);
	\draw[thick] (0,1/2) -- (0,-1/2);
	\draw[thick] (0,-3/2) circle (1);
	\node[left] at (-1,3/2) {$\id$};
	\node[left] at (0,0) {$\id$};
	\node[left] at (-1,-3/2) {$\id$}; 
	\end{tikzpicture}
	}},
\end{align}
where we have combined modular transformations on each torus factor with the fusion move in figure \ref{fig:identitycrossing}. All three of these equations are exact statements about the decompositions of Virasoro vacuum blocks at any value of the central charge, but in this work we will only make use of the semiclassical limits. The semiclassical limit of the OPE function $C_0$ is given by the following, where we take $h_i = {c\over 6}\eta_i(1-\eta_i)$: 
\begin{equation}\label{semiclassicalC0}
\begin{aligned}
\log C_0(h_1,h_2,h_3) \approx& \, \frac{c}{6}\bigg[  \sum_{i=1}^3F(2\eta_i) - F(\eta_1+\eta_2+\eta_3-1) - \left(F(\eta_1+\eta_2-\eta_3) + \text{(2 permutations)}\right)\\
& \, + F(0) + 2\left(\eta_1+\eta_2+\eta_3-1\right) + \sum_{i=1}^3(1-2\eta_i)\log(1-2\eta_i) \bigg] + O(c^0),
\end{aligned}
\end{equation}
where
\begin{equation}
	F(x) \coloneqq \int_{\half}^x dy\log\left({\Gamma(y)\over\Gamma(1-y)}\right).
\end{equation}

It is instructive to compare the crossing integrals to the spectral integrals that appear in the calculations of observables in the Liouville CFT. For example, in terms of the DOZZ structure constant, the left-hand side of \eqref{C0fourpoint} is proportional to  
\begin{align}
 \int_{\mathbb{R}} dP \sqrt{\rho_0(P)} \tCdozz(P_1, P_2, P) \F_{1221}(h_P; x) \ .
\end{align}
The factor of $\sqrt{\rho_0}$ is crucial. Without this factor, for real $x$ in the semiclassical limit this integral would have the same saddlepoint as a Liouville 4-point function (see \eqref{GLhat4}). The extra factor of $\sqrt{\rho_0}$ shifts the saddlepoint so that instead of producing a Liouville correlator, it gives a Virasoro identity block.  

\subsection{Crossing in the large-$c$ ensemble}\label{ss:ensemblecrossing}

Let us now consider observables in our ensemble. Consider the 4-point function of sub-threshold CFT operators, identical in pairs. The conformal block expansion is 
\begin{align}\label{expandG4}
G_4 := \langle \O_1(0) \O_2(x,\bx) \O_2(1) \O_1(\infty) \rangle &= 
\sum_p |c_{12p}|^2 |\F_{1221}(h_p; x) |^2 \ .
\end{align}
The ensemble average is therefore
\begin{align}\label{G4av1}
\overline{G_4} &= \sum_p \overline{|c^2_{12p}|} |\F_{1221}(h_p; x) |^2\\
&\approx  \left| \int_{\frac{c-1}{24}}^\infty dh \, \rho_0(h) C_0(h_1, h_2, h) \F_{1221}(h; x)\right|^2 
\end{align}
We have used the definition of the ensemble to rewrite the sum over black hole states as an integral,
\begin{align}
\sum_{p| h_p \geq \frac{c-1}{24}}  \to \int_{c-1\over 24}^\infty dh d\bh \, \rho_0(h) \rho_0(\bh) \ ,
\end{align}
and to evaluate $\overline{|c^2_{12p}|}$. 
There are also contributions to \eqref{G4av1} from sub-threshold states in the sum, but they are subleading. (They will be discussed in section \ref{s:coleman}). 
Using the $C_0$ crossing relation \eqref{C0fourpoint}, the averaged correlator is therefore an identity block:
\begin{align}\label{G4ansS}
\overline{G_4(x, \bx)} &\approx \left| \F_{1122}(\id; 1-x)\right|^2  \ .
\end{align}

In the first step of this calculation \eqref{expandG4} we expanded in the OPE channel $12 \to p \to 12$. Instead, we can expand in the channel $11 \to p \to 22$. In this case we immediately get the Virasoro identity block, because $\overline{c_{11p}c_{22p}} = 0$ unless $p$ is the identity (we are assuming $\O_1 \neq \O_2$). 

The conclusion is that the ensemble has a crossing-invariant 4-point function to leading order in the semiclassical limit.\footnote{\label{footnote:euclideanwinding} 
More precisely, our ensemble is crossing invariant at leading order in $c$ as long as we restrict $x$ to be sufficiently close to the unit interval $0<x<1$.
Just as with our earlier discussion, the four point function will not satisfy other bootstrap constraints which become important elsewhere in the complex $x$ plane. 
Consider, for example, a cross-ratio with Re $x>1$. The conformal blocks can be calculated by analytically continuing from $x<1$ through the upper or lower complex plane, and the two answers do not agree. Thus the identity block is not unique; it has ambiguities from branch cuts in Euclidean signature. On the gravity side, this reflects the existence of two distinct saddlepoints that differ by braiding the two conical defects \cite{Asplund:2014coa}. 
This leads to subleading corrections to the ensemble of OPE coefficients.  We will not need to consider these corrections here, because we only consider one saddlepoint at a time.  It would be interesting to consider a more refined ensemble which is invariant under the full set of crossing symmetries; the resulting four point function would presumably take the form of a modular sum, as in \cite{Maloney:2016kee}.}

The identity block is exactly what is expected from 3D gravity. It matches the saddlepoint with two non-interacting conical defects in AdS$_3$ \cite{Hartman:2013mia,Faulkner:2013yia},
\begin{align}\label{G4identity}
\overline{G_4} \qquad &=  \qquad \vcenter{\hbox{\includegraphics[width=1.5in]{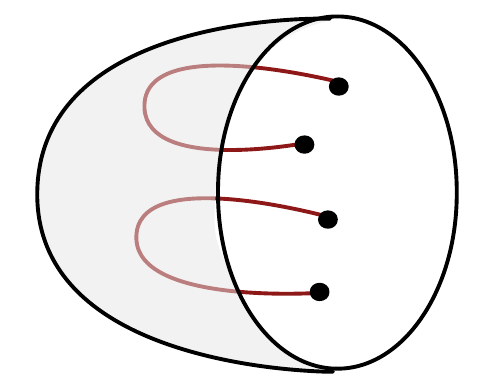}}} 
\qquad =\qquad  
\left|\vcenter{\hbox{
\begin{tikzpicture}[scale=0.5]
\draw (0,1) -- (1, 0);
\draw (0, -1) -- (1,0);
\draw (1,0) -- (3,0);
\draw (3,0) -- (4,1);
\draw (3,0) -- (4,-1);
\node at (2,0.4) {$\mathbb{1}$};
\end{tikzpicture}
}}\right|^2
\ .
\end{align}
This fact is not new, and does not require an ensemble interpretation. We would obtain the same approximate 4-point function if we just assumed that the structure constants $|c_{ijk}|^2$ are well-approximated by the universal OPE formula, with microcanonical rather than ensemble averaging. The point is that the gravity result is also compatible with the ensemble, and when we consider non-factorizing saddles with multiple boundaries, the ensemble interpretation becomes important. 

Similar comments apply to torus 2-point functions and the genus-2 partition function. For example, the averaged genus-2 partition function calculated in the sunset channel agrees with the calculation in the dumbbell channel:
\begin{align}
	\overline{Z_{g=2}} = \sum_{p,q,r}\overline{|c^2_{pqr}|}\left|\vcenter{\hbox{
	\begin{tikzpicture}[scale=0.5]
	\draw[thick] (0,0) circle (3/2);
	\draw[thick] (-3/2,0) -- (3/2,0);
	\node[above] at (0,3/2) {$p$};
	\node[above] at (0,0) {$q$};
	\node[above] at (0,-3/2) {$r$};
	\end{tikzpicture}
	}}\right|^2
	=
	\sum_{p,q,r} \overline{c_{ppq}c_{rrq}}
	\left|\vcenter{\hbox{
	\begin{tikzpicture}[scale=0.5]
	\draw[thick] (0,3/2) circle (1);
	\draw[thick] (0,1/2) -- (0,-1/2);
	\draw[thick] (0,-3/2) circle (1);
	\node[left] at (-1,3/2) {$p$};
	\node[left] at (0,0) {$q$};
	\node[left] at (-1,-3/2) {$r$}; 
	\end{tikzpicture}
	}}\right|^2
\end{align}
and to leading order, these can be recast as a sum over identity blocks in different channels. The ensemble average of the torus 2-point function reproduces the action of a conical defect in the BTZ black hole, and the average genus-2 partition function matches the gravitational contribution of a genus-2 handlebody. We will expand on this in section \ref{s:singleboundary}.

\section{Two-copy observables in the CFT ensemble}\label{s:twocopy}

In this section we calculate averaged products of observables in two copies of the CFT.  These will match wormhole contributions on the gravity side.
We start with the product of 4-point functions, which illustrates the main features. We then describe the result for general observables, and sketch a few other examples.

\subsection{4-point functions}\label{ss:G4G4}

Consider the product of two defect 4-point functions,
\begin{align}
G_{1234} G'_{1234} &:= \langle \O_1(0) \O_2(x,\bx) \O_3(1) \O_4(\infty) \rangle
\langle \O_1(0) \O_2(x', \bx') \O_3(1) \O_4(\infty) \rangle
\end{align}
with the four operators distinct. This is the example sketched in the introduction (section \ref{ss:introaveraging}); we will now discuss the details of the CFT calculation. In a fixed CFT, obviously the two terms factorize, but after averaging they are correlated.   By expanding both correlators in the $12 \to p \to 34$ OPE channel  and taking the average, we find
\begin{align}
\overline{G_{1234} G'_{1234}} &= 
\sum_{p,q} \overline{c_{12p}c_{34p} c_{12q} c_{34q} }\left|
\F_{1234}(h_p;x) \F_{1234}(h_q;x')\right|^2 \\
&=
\sum_p \overline{|c^2_{12p}|} \ \overline{|c^2_{34p}|} \left| \F_{1234}(h_p;x) \F_{1234}(h_p; x') \right|^2
\\
&\approx \left| \int dh \rho_0(h) C_0(h_1, h_2, h) C_0(h_3, h_4, h) \F_{1234}(h;x) \F_{1234}(h;x') \right|^2 \ ,
\end{align}
where we have dropped the subleading contribution from sub-threshold states in the intermediate channel (see section \ref{s:coleman} for a discussion of these corrections).  We also used the fact that the external operators are scalars to remove a factor of $(-1)^{\sum_i \ell_i}$ that arises when permuting indices to write the OPE coefficients as absolute values.  Now we rewrite $C_0$ in terms of the DOZZ formula using \eqref{C0dozz}, and change the integration variable $h \to P$ using \eqref{measuretrans}. This gives
\begin{align}
\overline{G_{1234} G'_{1234}} &\approx 
 \left| \frac{1}{2} \int_{\mathbb{R}}dP\, \tCdozz(P_1, P_2, P)\tCdozz(P_3, P_4, P)
\F_{1234}(h_P;x) \F_{1234}(h_P;x')  \right|^2
\end{align}
All of the factors of $\rho_0$ have cancelled, and the quantity inside the $|\cdot |^2$ is manifestly a Liouville correlation function (see \eqref{GLhat4}). Note, however, that the $P$ integral pairs $x \leftrightarrow x'$ and the $\bar{P}$ integral pairs $\bx \leftrightarrow \bx'$. Therefore we find 
\begin{align}\label{eq:G4equalsGLiouville}
\overline{G_{1234}(x,\bx) G_{1234}(x', \bx')} &\approx 
G^L_{1234}(x,x') G^L_{1234}(\bx', \bx) \ ,
\end{align}
where $G^L_{1234}$ is the Liouville 4-point function.\footnote{We write all Liouville observables with the `hatted' normalization of external operators; see the discussion in appendix \ref{ss:liouvilleBasics}.} The weights in the Liouville correlator are identical to the weights in the averaged CFT correlator, but the cross-ratios on the two sides are paired differently. 
It is an important fact -- and one which will arise in a variety of settings -- that the average of the square of an observable is a product of Liouville observables, but with cross-ratios and moduli permuted as in (\ref{eq:G4equalsGLiouville}).
We note that (\ref{eq:G4equalsGLiouville}) is invariant under crossing transformations that act simultaneously on $(x,\bar x)$ and $(x',\bar x')$. 

Interestingly, even when the CFT kinematics are Euclidean, the Liouville kinematics can be effectively Lorentzian, because the two arguments of $G^L(x,x')$ are not complex conjugates. This means that the Liouville correlation function has branch cuts at kinematics where no such branch cuts are allowed in the exact CFT. These branch cuts have a natural interpretation as coming from the braiding of conical defects on the gravity side. From a CFT point of view they reflect the fact that the ensemble must have subleading corrections that become important when the branch cuts come into play, as in the discussion of footnote \ref{footnote:euclideanwinding}. These corrections can perhaps be included by summing over saddles, but there could also be off-shell topologies that are important.

\subsection{General 2-copy observables}\label{ss:general2copy}
This result immediately generalizes to any product of CFT observables of the form
\begin{align}
G(\sigma,\bsigma; \Omega, \bOmega) G(\sigma',\bsigma'; \Omega', \bOmega') \ ,
\end{align}
where $G(\sigma,\bsigma; \Omega, \bOmega)$ is a CFT $n$-point correlation function of sub-threshold operators on a Riemann surface with period matrix $\Omega$. The operator insertion points are denoted collectively by
\begin{align}
\sigma = (x_1,x_2, x_3, \dots) , \qquad
\bsigma = (\bx_1, \bx_2, \bx_3, \dots ) \ .
\end{align}
We first expand both observables in the same OPE channel, then take the average. The average of a product of many OPE coefficients is calculated by Wick contractions. There is one particular contraction, call it the `paired' term, that sets all of the internal weights equal in the two copies. 
This term is therefore a sum of correlated conformal blocks. Schematically, it leads to an expression like 
\begin{align}
\overline{GG'}|_{\rm paired} =  \sum  
(\overline{|c^2_{pqr}|} \ \overline{|c^2_{rst}|}\ \overline{|c^2_{tuv}|} \cdots)\left|
\vcenter{\hbox{
\begin{tikzpicture}[scale=0.5]
\draw (-.5,.5) -- (1,-1); 
\draw (1,-1) -- (2,0); 
\draw (1,-1) -- (1,-3); 
\draw (1,-3) -- (0, -4); 
\draw (1,-3) -- (2,-4); 
\draw[dashed] (0,-4) -- (-1, -5);
\draw[dashed] (2,-4) -- (3,-5);
\draw[dashed] (2,0) -- (3,1);
\draw (-2, 0.5) arc (-123:30:1.4);
\draw[dashed] (-2,0.5) -- (-3,1.5);
\draw[dashed] (0,2.4) -- (-1,3.4);
\node at (0,-0.35) {$r$};
\node at (2,-0.35) {$s$};
\node at (-0.25, -3.8) {$u$};
\node at (2.3, -3.8) {$v$};
\node at (1.4,-2) {$t$};
\node at (-1.3, -0.15) {$p$};
\node at (0.35, 1) {$q$};
\node at (1,-6) {$(\sigma)$};
\end{tikzpicture}
}}\quad \quad
\vcenter{\hbox{
\begin{tikzpicture}[scale=0.5]
\draw (-.5,.5) -- (1,-1); 
\draw (1,-1) -- (2,0); 
\draw (1,-1) -- (1,-3); 
\draw (1,-3) -- (0, -4); 
\draw (1,-3) -- (2,-4); 
\draw[dashed] (0,-4) -- (-1, -5);
\draw[dashed] (2,-4) -- (3,-5);
\draw[dashed] (2,0) -- (3,1);
\draw (-2, 0.5) arc (-123:30:1.4);
\draw[dashed] (-2,0.5) -- (-3,1.5);
\draw[dashed] (0,2.4) -- (-1,3.4);
\node at (0,-0.35) {$r$};
\node at (2,-0.35) {$s$};
\node at (-0.25, -3.8) {$u$};
\node at (2.3, -3.8) {$v$};
\node at (1.4,-2) {$t$};
\node at (-1.3, -0.15) {$p$};
\node at (0.35, 1) {$q$};
\node at (1,-6) {$(\sigma')$};
\end{tikzpicture}
}}
 \right|^2
\end{align}
where we have drawn just one internal piece of the conformal blocks appropriate to an $n$-point function on a genus-$g$ Riemann surface. Next we replace each sum by an integral over the Cardy spectrum and plug in the OPE formula. All of the $\rho_0$ factors cancel between the density of states and $C_0$, so we are left with a product of Liouville observables,
\begin{align}\label{GGgeneral}
\overline{G(\sigma,\bsigma; \Omega, \bOmega) G(\sigma',\bsigma'; \Omega', \bOmega')}|_{\rm paired}
= G^L(\sigma,\sigma'; \Omega, -\Omega') G^L(\bsigma', \bsigma; -\bOmega',\bOmega) \ .
\end{align}
Both the positions and the period matrices are mixed up on the right-hand side: the first Liouville factor accounts for left-movers in both copies of the CFT and the second accounts for all the right-movers.

The other contributions to $\overline{G G'}$ that are important in the semiclassical limit come from other Wick contractions of the OPE coefficients. For example, the product of two genus-2 partition functions $\overline{Z_{g=2} Z_{g=2}}$ has a factorizing Wick contraction that produces $\overline{Z_{g=2}} \ \overline{Z_{g=2}}$.  There can also be partially-connected Wick contractions, where some contractions occur within a single copy and others connect the copies, and there are contributions from the additional contractions in \eqref{contractions} when some of the intermediate operators are identical, some of which are discussed in the examples below.

If there are identical external operators inserted in $G$, then an additional complication is that there are different ways to pair the operators in $G$ with the operators in $G'$. Each such pairing leads to a product of Liouville correlators, and each of these will come from a different bulk saddle. 

We can also consider products of observables that are different on the two sides, such as the product of a 4-point function and a 6-point function. It is straightforward to do the ensemble CFT calculation. It is always possible to write the answer as integrals of products of $\rho_0$'s, $\tCdozz$'s, and conformal blocks, but generally the factors of $\rho_0$ do not cancel so the result cannot be rewritten as Liouville observables. We will see examples like this below. Similarly, for observables with a number of replicas other than two, the computation in the CFT ensemble will generally not lead to a cancellation of the factors of $\rho_0$ and so the result cannot be rewritten in terms of a product of observables in the Liouville CFT. For instance, $k$-copy products of correlation functions are considered in section \ref{s:kboundary}, and we will see in section \ref{s:singleboundary} that the ensemble-averaged one-copy observables are given by a suitable Virasoro vacuum block to leading order in the semiclassical limit.

\subsection{Explicit examples of two-copy observables}\label{ss:examples2copy}
We have argued that the connected part of the average of two-copy observables in our CFT ensemble is given by a product of corresponding observables in Liouville CFT that couple the left- (and right-)movers on each boundary. Here we will consider a few more examples in explicit detail.

\subsubsection{Sphere $n$-point functions}
We will start by considering the ensemble average of the product of two sphere $n$-point functions of distinct defects
\begin{equation}
    G_{1\cdots n}(x_i,\bar x_i) G_{1\cdots n}(x'_i,\bar x'_i) \coloneqq \langle\mathcal{O}_1(x_1,\bar x_1) \cdots \mathcal{O}_n(x_n,\bar x_n)\rangle \langle \mathcal{O}_1(x'_1,\bar x'_1)\cdots \mathcal{O}_n(x'_n,\bar x'_n)\rangle.
\end{equation}
We expand each correlator in the comb channel, for which we represent the conformal blocks pictorially as\footnote{Here and in what follows we use numbers to label external primaries and $p_i$ to label internal operators.}
\begin{align}
\F^{\rm comb}_{12\cdots n}(h_{p_i};x_i) &=
\vcenter{\hbox{
\begin{tikzpicture}[scale=1]
    \draw[thick] (0,0) -- (1,0) -- (1,1) -- (1,0) -- (2,0) -- (2,1) -- (2,0) -- (3,0);
    \node at (7/2,0) {$\ldots$};
    \draw[thick] (4,0) -- (5,0) -- (5,1) -- (5,0) -- (6,0) -- (6,1) -- (6,0) -- (7,0);
    \node[left] at (0,0) {$1$};
    \node[above] at (1,1) {$2$};
    \node[above] at (2,1) {$3$};
    \node[above] at (5,1) {$n-2$};
    \node[above] at (6,1) {$n-1$};
    \node[right] at (7,0) {$n$};
    \node[below] at (3/2,0) {$p_1$};
    \node[below] at (5/2,0) {$p_2$};
    \node[below] at (4+1/2,0) {$p_{n-4}$};
    \node[below] at (5+1/2,0) {$p_{n-3}$};
\end{tikzpicture}
}}
\end{align}
Performing the Gaussian ensemble average leads to\footnote{We have implicitly assumed that the $n$ external operators are distinct defects. As previously discussed, in the case that some of the defects are identical there would be additional contractions that would contribute. In the case that the operators are pairwise identical this includes fully disconnected contributions.}$^,$\footnote{Our convention throughout the paper is to include the complete position dependence in the blocks $\F$.} 
\begin{equation}
\begin{aligned}
    & \, \overline{G_{1\cdots n}(x_i,\bar x_i) G_{1\cdots n}(x'_i,\bar x'_i)}\\
    =& \, \sum_{p_1,\ldots,p_{n-3}} \overline{|c^2_{12p_1}|}\, \overline{|c^2_{p_13p_2}|}\cdots \overline{|c^2_{p_{n-4}(n-2)p_{n-3}}|}\, \overline{|c^2_{p_{n-3}(n-1)(n)}|}\left|\mathcal{F}^{\rm comb}_{12\cdots n}(h_{p_i};x_i)\mathcal{F}^{\rm comb}_{12\cdots n}(h_{p_i};x'_i)\right|^2\\
    \approx&\, \Bigg|\int_{c-1\over 24}^\infty {dh_{p_1}\rho_0(h_{p_1})}\cdots {dh_{p_{n-3}}\rho_0(h_{p_{n-3}})}C_0(h_1,h_2,h_{p_1})C_0(h_{p_1},h_3,h_{p_2})\cdots C_0(h_{p_{n-3}},h_{n-1},h_n)\\
    & \,~\times \mathcal{F}^{\rm comb}_{12\cdots n}(h_{p_i};x_i)\mathcal{F}^{\rm comb}_{12\cdots n}(h_{p_i};x'_i)\Bigg|^2.
\end{aligned}
\end{equation}
As before, the contributions of exchanges from intermediate defect states are suppressed in the semiclassical limit and the ensemble average pairs the dependence on the positions $x_i$ with their counterparts $x'_i$ (and likewise for $\bar x_i$ and $\bar x'_i$). Assembling the factors of $\rho_0$ and $C_0$, the average can then be re-expressed as a product of sphere $n$-point correlators in the Liouville CFT 
\begin{equation}
    \overline{G_{1\cdots n}(x_i,\bar x_i)  G_{1\cdots n}(x'_i,\bar x'_i) }= 
    G^L_{1\cdots n}(x_i,x'_i)G^L_{1\cdots n}(\bar x'_i,\bar x_i).
\end{equation}

\subsubsection{Thermal one-point functions}\label{sss:thermalOnePointWormhole}

The discussion 
extends to two-copy observables with higher-genus boundaries. As a simple first example, consider the product of torus one-point functions of an external sub-threshold (scalar) defect operator $\mathcal{O}_1$,
\begin{equation}
    G_1(\tau,\bar\tau)G_1(\tau',\bar\tau') = \langle \mathcal{O}_1\rangle_{T^2(\tau,\bar\tau)}\langle\mathcal{O}_1\rangle_{T^2(\tau',\bar\tau')}.
\end{equation}
The averaged product is given by the following in the CFT ensemble
\begin{equation}\label{eq:torusOnePtWormhole}
\begin{aligned}
    \overline{G_1(\tau,\bar\tau)G_1(\tau',\bar\tau')}  
    &= \sum_p\overline{c^2_{1pp}}\,\mathcal{F}^{g=1}_1(h_p;\tau)\mathcal{F}^{g=1}_1(h_p;\tau')\overline{\mathcal{F}}^{g=1}_1(\bar h_p;\bar\tau)\overline{\mathcal{F}}^{g=1}_1(\bar h_p;\bar\tau')\\
    &\approx 2\left|\int dh_p\, \rho_0(h_p)C_0(h_1,h_p,h_p)\mathcal{F}^{g=1}_1(h_p;\tau)\overline{\mathcal{F}}^{g=1}_1(h_p;-\tau')\right|^2\\
    &= 2
    \left|\int_{\mathbb{R}}{dP\over 2}\, \tCdozz(P_1,P,P)\mathcal{F}^{g=1}_1(h_p;\tau)\overline{\mathcal{F}}^{g=1}_1(h_p;-\tau')\right|^2.
\end{aligned}
\end{equation}
The overall factor of 2 comes from the two Wick contractions of $\overline{c_{1pp} c_{1qq}}$. (Note that the `paired' term in \eqref{GGgeneral} was defined to include only one of these contractions.) 
In the bulk, each of the two contractions in \eqref{eq:torusOnePtWormhole} is dual to a two-boundary wormhole whose boundaries are tori with a single operator insertion. 
Thus the averaged product of torus one-point functions is given by a product of one-point functions in Liouville theory in the semiclassical limit, 
\begin{equation}\label{eq:torusOnePtProductLiouville}
     \overline{G_1(\tau,\bar\tau)G_1(\tau',\bar\tau')} \approx 2
     G^L_1(\tau,-\tau')G^L_1(-\bar\tau',\bar\tau).
\end{equation}
The result is covariant under modular transformations that act simultaneously on $\tau$ and $\tau'$
\begin{align}
    \tau \mapsto \gamma\tau\ , \quad
    \tau'  \mapsto (M\gamma M)\tau',\quad \gamma\in PSL(2,\mathbb{Z}), \quad M = \begin{pmatrix} -1 & 0 \\ 0 & 1 \end{pmatrix}.
\end{align}
The need for the factors of $M$, which flip the sign of $\tau$, in the modular transformation that acts on $\tau'$ is due to the orientation reversal on that boundary so that the averaged product (\ref{eq:torusOnePtWormhole}) assembles into a product of Liouville correlators. In order to achieve a result fully covariant under independent modular transformations acting on $\tau$ and $\tau'$, one would need to sum (\ref{eq:torusOnePtProductLiouville}) over relative modular transformations.

The general structure of this result is similar to the off-shell calculation of the double-torus amplitude in 3D gravity by Cotler and Jensen \cite{Cotler:2020ugk}.\footnote{
Cf. $\tilde{Z}(\tau_1,\tau_2)$ defined in their equation (3.54). The result \eqref{eq:torusOnePtProductLiouville} and the double torus amplitude have the same modular properties and similar spectral integrals. The double torus is a contribution to the product of partition functions, i.e. $\langle \id \rangle_{T^2} \langle \id \rangle_{T_2}$, so it is tempting to try to continue $h_1 \to 0$ and compare the two expressions quantitatively. However, this is a 
singular limit of the Liouville torus one-point correlator, as $\tCdozz(P_1,P,P)$ (equivalently $C_0(P_1,P,P)$) diverges in the limit that $h_1$ is taken to zero, while the torus one-point blocks simply reduce to the non-degenerate Virasoro characters in this limit. One may view this as reflecting the fact that the torus partition function of Liouville theory gives a formally divergent volume factor due to its noncompact spectrum of local primary operators.}

\subsubsection{Genus-two partition functions}
As our final example we consider the product of genus two-partition functions in the CFT ensemble. The computation proceeds as in the previous examples. We consider the product of genus-two partition functions, expanded in (for instance) the sunset channel (as in \eqref{genus2sunset}), and average according to the rules of the CFT ensemble. A novel feature compared to the previously considered examples is the presence of a Wick contraction corresponding to a purely disconnected contribution, the product of means $\overline{Z_{g=2}} \ \overline{Z_{g=2}}$. The reason for this is that unlike the previous examples, the ensemble average of the single-boundary observable is non-vanishing for the genus-two partition function.\footnote{If the external defects are pairwise identical, then the sphere $n$-point functions can also have a non-trivial ensemble average. We explicitly consider the cases of the averaged sphere six-point functions and higher-genus partition functions in section \ref{s:singleboundary}.} The paired contraction in the averaged product of genus-two partition functions is 
\begin{align}
    &\left. \overline{Z_{g=2}(\Omega,\overline{\Omega})Z_{g=2}(\Omega',\overline{\Omega'})}\right|_{\rm paired}\notag \\
    &\qquad= \sum_{p_1,p_2,p_3}\left(\overline{|c^2_{p_1p_2p_3}|}\right)^2_{\rm paired} \left|\mathcal{F}^{g=2}_{\rm sunset}(h_{p_i};\Omega)\mathcal{F}^{g=2}_{\rm sunset}(h_{p_i};\Omega') \right|^2 \notag\\
    &\qquad\approx \left|\int_{c-1\over 24}^\infty \left(\prod_{k=1}^3dh_{p_k}\rho_0(h_{p_k})\right)C_0(h_{p_1},h_{p_2},h_{p_3})^2\mathcal{F}^{g=2}_{\rm sunset}(h_{p_i};\Omega)\overline{\mathcal{F}}^{g=2}_{\rm sunset}(h_{p_i};-\Omega')\right|^2.
\end{align}
All the factors of $\rho_0$ combine with the factors of $C_0$ to produce the product of genus-two partition functions in the Liouville CFT
\begin{equation}
     \left. \overline{Z_{g=2}(\Omega,\overline{\Omega})Z_{g=2}(\Omega',\overline{\Omega'})}\right|_{\rm paired} \approx 
     Z^L_{g=2}(\Omega,-\Omega')Z^L_{g=2}(-\overline{\Omega}',\overline{\Omega}).
\end{equation}
There are additional contractions in $\overline{|c^2_{p_1p_2p_3}|}^2$ that lead to similar expressions, but with an element of $Sp(4,\mathbb{Z})$ acting on $\Omega'$.

\section{The Maldacena-Maoz Wormhole}\label{s:fuchsian}

In this section we consider classical wormhole solutions with two identical boundaries, where it is easy to write down explicit bulk saddle points which solve the equations of motion. These are the wormholes discussed in detail by Maldacena and Maoz \cite{Maldacena:2004rf}, generalized slightly to include conical defects. 
We will first explain the geometry of these manifolds, then turn to the calculation of the action.
This action will match the ensemble average of the products of identical observables; non-identical observables, and the corresponding wormholes, will be discussed in the next section.

\subsection{Bulk theory}
The action of 3D gravity coupled to massive point particles is
\begin{align}\label{gravityaction}
S = -\frac{1}{16\pi G} \int_M \sqrt{g}(R+2)  - \frac{1}{8\pi G} \int \sqrt{h}(K-1) + \sum_i m_i \int dl_i \ ,
\end{align}
where the last integral is over the particle wordlines. There are some subtleties near the particles to remove divergences that are addressed in section \ref{ss:fuchsianaction}. The parameter $m_i$, with $0 < m_i < \frac{1}{4G}$, is referred to as the local mass of a particle; due to backreaction, it is not equal to the physical ADM mass. The ADM mass of a particle, or equivalently the total scaling dimension of the dual CFT operator, is 
\begin{align}
\Delta_i = m_i(1-2G m_i) \ .
\end{align}
Therefore with conformal weights parameterized as $h = \frac{c}{6}\eta(1-\eta)$, 
\begin{align}
m_i = \frac{\eta_i}{2G} \ .
\end{align}
The point particles backreact on the geometry to produce conical defects of total angle $2\pi(1-2\eta_i)$. 
The metric of the Maldacena-Maoz wormhole is
\begin{equation}\label{mm}
ds^2 = d\rho^2 + \cosh^2 \rho d\Sigma^2
\end{equation}
where $d\Sigma^2$ is the constant negative curvature metric on a surface $\Sigma$, potentially with conical defects at the locations of the particles. The boundary consists of two disjoint copies of $\Sigma$ at $\rho \to \pm \infty$. It is easy to check that \eqref{mm} solves the equations of motion, i.e. is locally ${\mathbb H}_3$ away from conical defects.

The two boundaries have the same moduli up to an orientation reversal, so denoting the moduli of the left boundary by $(\lambda, \blambda)$, these wormholes contribute to the average of 
\begin{align}
|G(\lambda, \blambda)|^2 = G(\lambda, \blambda) G(\blambda,\lambda)
\end{align}
in the dual CFT.

\subsection{Quotient construction}
\label{ss:quotient}

In order to understand the geometry of this wormhole, let us consider as a warmup the example where $\Sigma$ is the upper half plane, with the usual hyperbolic metric
\begin{equation}\label{h2}
d\Sigma^2 = \frac{dy^2 + dx^2}{y^2}
\end{equation}
and $y>0$.  
Making the change of coordinates
\begin{equation}
{\tilde y}= y \cos \theta,~~~~w= y \sin \theta,~~~~~\theta\equiv\cos^{-1}\left({\tanh}\rho\right)  
\end{equation}
the wormhole (\ref{mm}) becomes
\begin{equation}\label{h3}
ds^2 = \frac{dw^2+d{\tilde y}^2+dx^2}{w^2}~.
\end{equation}
This is the usual hyperbolic metric on ${\mathbb H}_3$, represented as the upper half 3-space with $w>0$.
This is not a surprise, but it is instructive: $\rho$ has become an angular coordinate $\theta$ which rotates around the axis $y=0$.  The full geometry can be thought of as the surface of rotation about the $y=0$ axis, with the boundaries at $\rho=\pm \infty$ now identified as the surfaces $\theta=0,\pi$ where $w=0$.  The boundary at $w=0$ is the full $(x,{\tilde y})$ plane, with ${\tilde y}>0$ the boundary at $\rho\to\infty$ and ${\tilde y}<0$ the boundary at $\rho\to-\infty$.  In particular, when $\Sigma={\mathbb H}_2$ our geometry is not a wormhole at all: the boundary is indeed two copies of ${\mathbb H}_2$, but they have been glued together to form a single copy of the plane at $w=0$.\footnote{Of course, the boundary is actually a sphere once we add the point at infinity.}  

This is a general feature of the Maldacena-Maoz wormhole: when $\Sigma$ itself has a boundary the geometry (\ref{mm}) only has a single boundary, since the boundaries of the two copies of $\Sigma$ at $\rho\to\pm\infty$ get glued together to form a single connected geometry. It is only when $\Sigma$ is compact that the wormhole geometry (\ref{mm}) is a genuine wormhole with two disconnected boundaries.  

To see this, let us consider (\ref{mm}) where $d\Sigma^2$ is the constant negative curvature metric on a general Riemann surface $\Sigma$, allowing for conical singularities. It is a bit simpler to discuss conical defects of finite order with total angle $2\pi/\mathbb{N}$, so we will consider this case first, and generalize to continuous defect angles below. 
Such a negative curvature metric can be obtained by identifying points on the upper half plane (\ref{h2}) by some discrete subgroup $\Gamma$ of the $PSL(2,{\mathbb R})$ isometry group of ${\mathbb H}_2$.  
This $PSL(2,{\mathbb R})$ isometry group acts in the usual way by fractional linear transformations: 
\begin{equation}\label{eq:fractionalLinear}
\gamma:z\to\gamma z\equiv \frac{az+b}{cz+d}, ~~~~~\gamma=\left({a~b\atop c~d}\right)\in PSL(2,{\mathbb R})
\end{equation}
where $z=x+iy$ is the coordinate on the upper half plane.  The discrete subgroup $\Gamma$ is known as a Fuchsian group, and the representation of $\Sigma$ as the quotient $\Sigma = {\mathbb H}_2/\Gamma$ is known as the Fuchsian model of the hyperbolic metric on $\Sigma$.  Since ${\mathbb H}_2$ is simply connected, $\Gamma$ is isomorphic to the fundamental group $\pi_1(\Sigma)$, and $\Gamma$ can be thought of as an embedding $\pi_1(\Sigma)$ into $PSL(2,{\mathbb R})$.  Of course, for a given surface $\Sigma$ there are many such embeddings, which correspond to different choices of hyperbolic metric on $\Sigma$.  Alternately, by the uniformization theorem, they can be regarded as different choices of complex structure on $\Sigma$. 
One way to think about this quotient is to chose a fundamental domain for the action of $\Gamma$ on ${\mathbb H}_2$, and to regard $\Sigma$ as this fundamental domain with its sides glued together by the action of some of the elements of $\Gamma$.  Since isometries map geodesics to geodesics, we can take the boundary of the fundamental domain to be a collection of geodesics in ${\mathbb H}_2$. In other words, the boundary of the fundamental domain will be a collection of arcs of circles which intersect the boundary $y=0$ transversely.

We can now embed this in ${\mathbb H}_3$ to obtain a picture of our Euclidean wormhole.  The wormhole is now interpreted as the quotient ${\mathbb H}_3/\Gamma$, where the same Fuchsian group $\Gamma$ is interpreted as a subgroup of the $SL(2,{\mathbb C})$ isometry group of ${\mathbb H}_3$.  Of course it is a very special kind of subgroup, since it sits inside a $PSL(2,{\mathbb R})$ subgroup of the full isometry group.\footnote{We will consider generalizations later, where $\Gamma$ does not sit inside  $PSL(2,{\mathbb R})$.}
We can think of the full geometry as obtained by starting with the boundary at $\theta=0$ and rotating about the $y=0$ axis until we reach the other boundary at $\theta=\pi$.   Just as above, our wormhole can be thought of as a fundamental domain in ${\mathbb H}_3$ with its sides identified by the action of $\Gamma$.  The fundamental domain is obtained by rotating the original fundamental domain about the $y=0$ axis.  The action of $\Gamma$ now identifies geodesic surfaces with geodesic surfaces, i.e. it identifies hemispheres with hemispheres in ${\mathbb H}_3$.  Our wormhole is obtained by gluing together the sides of this fundamental domain in ${\mathbb H}_3$. 

Some examples will help make this picture more clear.  One is the case where we take $\Gamma={\mathbb Z}$ to be generated by a single element of $PSL(2, {\mathbb R})$, say the element 
\begin{equation}
\gamma = 
\begin{pmatrix} q^{1/2} & 0 \\ 0 & q^{-1/2} \end{pmatrix}
\end{equation} 
for some positive real $q$.  This maps $z\to q z$, so we can take our fundamental domain to be the region between the two arcs $|z|=1$ and $|z|=q$.  In this case $\Sigma={\mathbb H}_2/{\mathbb Z}$ is the annulus, obtained by identifying the two arcs.  The two boundaries of the annulus are line segments $(1,q)$ and $(-1,-q)$ on the $y=0$ axis, each of which is a circle because of the identification $z\sim qz$.  The full wormhole is now the geometry ${\mathbb H}_3/{\mathbb Z}$, which is obtained by rotating this annulus about the $y=0$ axis.  The fundamental domain is the region between the two hemispheres $1\le \sqrt{w^2 + {\tilde y}^2+x^2} < q$, whose boundaries are identified under the action of the group $\Gamma$ which takes $(w,{\tilde y},x)\to q (w, {\tilde y}, x)$.  In this case the ``wormhole" is not really a wormhole at all: the boundary is the torus found by identifying $z \sim zq$.  We can put this in a more familiar form by letting $z=e^{2\pi i {\hat z}}$, so that the identifications are the usual identifications ${\hat z}\sim {\hat z}+1 \sim {\hat z}+\tau$ of a torus with modular parameter $\tau= \frac{1}{2\pi i} \log (q)$.  The bulk  is topologically a solid donut (a handlebody) that fills in this boundary.  This handlebody has a single non-contractible cycle with geodesic length $\log q$.  This geometry can be interpreted as either Thermal AdS or the Euclidean BTZ black hole.\footnote{We have taken $\tau$ to be purely imaginary because we wish to write our geometries in the form (\ref{mm}).
One can easily consider quotients with ${\rm Re}\,\tau\ne 0$, which are the Euclidean continuation of rotating black holes (or thermal AdS with non-zero angular potential); in this case the action $z\to qz$ does not preserve the upper half plane, but the quotient still leads to a bulk solution with the topology of a solid donut.}

In both of the examples above $\Sigma$ had a boundary, and as a result the corresponding ``wormhole" geometry was not a wormhole, but instead had a connected boundary.  When $\Sigma$ is compact, however, the resulting wormhole has the topology of $\Sigma\times I$, and its boundary has two disconnected components. For example, we can realize a smooth genus 2 surface as a quotient of the upper half plane in which the fundamental domain is an octagon and the action of $\Gamma$ identifies the opposite sides of this octagon together. 
This wormhole leads to a connected contribution to the variance of the genus 2 partition function, $\overline{ |Z_{g=2}|^2}$. 

In the above example the surface $\Sigma$ was smooth, and as a result all of the elements of $\Gamma$ were hyperbolic.\footnote{Elements of $PSL(2,{\mathbb R})$ are elliptic, parabolic, or hyperbolic depending on whether $|{\rm Tr}\,\gamma|$ is less than, equal to, or greater than 2.  An elliptic element has a single fixed point, which is in the interior of ${\mathbb H}_2$.  A parabolic element has a single fixed point on the boundary of ${\mathbb H}_2$.  A hyperbolic element has a pair of fixed points on the boundary of ${\mathbb H}_2$.}  If we allow our group $\Gamma$ to contain an elliptic element $\gamma$, then $\Sigma$ will no longer be smooth: it will have a singularity at the fixed point of $\gamma$.  In particular, the hyperbolic metric on ${\mathbb H}_2$ has a conical singularity, with total angle $2\phi$ given by the formula
\begin{equation}
\cos\phi=\frac{1}{2}{\rm Tr}\left(\gamma\right).
\end{equation}
When $\Gamma$ has a parabolic element $\gamma$, on the other hand, the surface $\Sigma$ is smooth but non-compact and the hyperbolic metric has a cusp at the fixed point of $\gamma$.  

We emphasize that, although they are singular, these examples have an important physical interpretation: they are the geometries which arise when we have operator insertions at the boundary. 
To see this, note that when $\Sigma$ has singularities, the corresponding bulk wormhole (\ref{mm}) will be singular as well.  It has a dimension one orbifold singularity which traces out a path through the bulk, connecting the two boundaries. This path is a geodesic, and is the trajectory of a particle through the bulk.  Indeed, this is just the usual situation where a massive particle in AdS$_3$ back-reacts on the metric to produce a conical singularity.  The endpoints of the geodesic on the two boundaries are the locations of operator insertions which source this bulk particle. 
If we denote by $\Delta$ the dimension of the operator inserted at the boundary, then\footnote{Note that we have taken spinless particles here, for the same reason that we took $q$ to be real above.  It is straightforward to generalize to spinning particles be considering more general types of elliptic elements, but the resulting groups $\Gamma$ will not preserve the upper half plane.} 
\begin{equation}\label{defect}
\phi=\pi \sqrt{1-\frac{12 \Delta}{c}} = \pi(1-2\eta)~.
\end{equation}
A few comments are in order.  First, this is an expression for the classical back-reaction of a massive particle on our geometry, so can be trusted only at leading order in $c$.  Second, if the operator is too light ($\Delta \ll c$) then it will not backreact on the geometry at all and one would instead consider a particle moving in a fixed background.  Third, the existence of a constant negative curvature metric on $\Sigma$ (and therefore of a Fuchsian group $\Gamma$) requires that 
\begin{equation}
2g+\sum_i \left(1-\sqrt{1-\frac{12 \Delta_i}{c}}\right) >2
\end{equation}
where $g$ is the genus of the surface and the sum is over all conical defects.

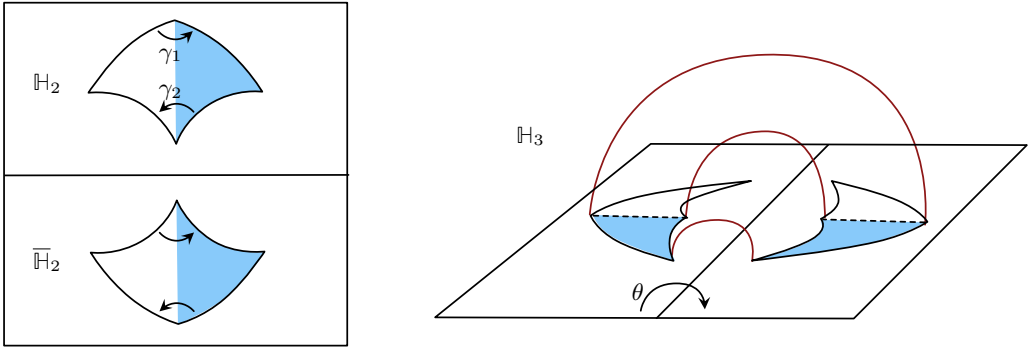
\begin{figure}
\begin{center}
{\footnotesize
\begin{overpic}[width=5.5in]{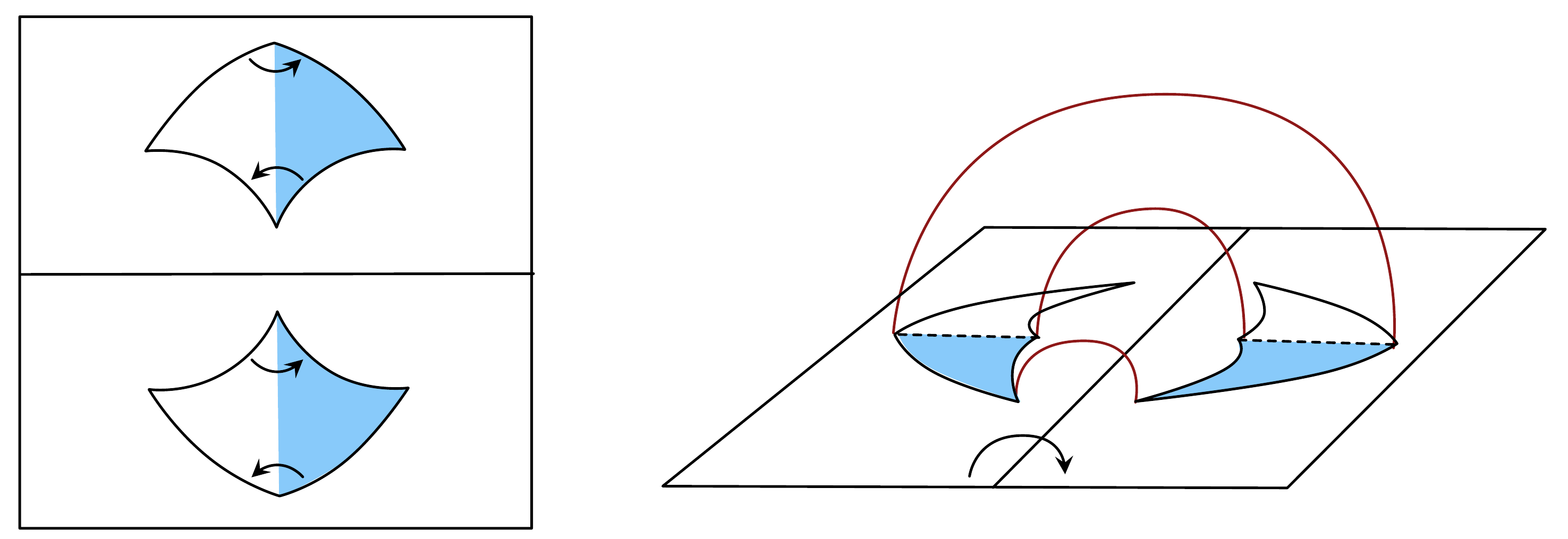}
\put (16,28) {$\gamma_1$}
\put (16,24.5) {$\gamma_2$}
\put (61,5) {$\theta$}
\put (4,25) {$\mathbb{H}_2$}
\put (4,8) {$\overline{\mathbb{H}}_2$}
\put (50,20) {$\mathbb{H}_3$}
\end{overpic}
}
\end{center}
\caption{A Maldacena-Maoz wormhole with genus $g=0$ and $n=3$ conical defects. Left:  The quotient $\mathbb{H}_2/\Gamma$ and its reflection in the lower half-plane. Right: The wormhole is the quotient $\mathbb{H}_3/\Gamma$. The shaded region is a hyperbolic triangle, and two copies of the triangle form the fundamental domain for $\Gamma$ acting on $\mathbb{H}_2$. The red curves are the conical defects.
\label{fig:batman}
}
\end{figure}

When $\Sigma$ has singularities, the corresponding wormholes contribute to the variance of partition functions with operator insertions.  For example, if we wish to compute the variance of the three point function then we simply need to consider a wormhole geometry where $\Sigma$ is a sphere with three conical defects. An example of this is depicted in Fig \ref{fig:batman}. This shows the geometry of the wormhole with two genus-0 boundaries and three conical defects, which we write schematically as
\begin{align}
\vcenter{\hbox{\includegraphics[width=1.5in]{figures/G3G3.pdf}}} \ . 
\end{align}
It is instructive to work out this example in detail.
We will consider the identification under the action of the $SL(2,R)$ group elements
\begin{equation}
\gamma_1 = 
\begin{pmatrix}\cos \phi_1 & -\sin\phi_1 \\ \sin\phi_1 & \cos\phi_1 \end{pmatrix}
,~~~~~
\gamma_2 = 
\begin{pmatrix} \cos \phi_2 & e^{-\alpha} \sin\phi_2 \\ -e^{\alpha}\sin\phi_2 & \cos\phi_2 \end{pmatrix}
\end{equation}
The two generators $\gamma_1$ and $\gamma_2$ have fixed points at $z=i$ and $z=ie^{-\alpha}$, which sit at the boundary of the fundamental domain.   Computing the trace of these generators, we see that these lead to conical defects with angles $\phi_1$ and $\phi_2$, respectively. The product $\gamma_1 \gamma_2$ also has a fixed point, with corresponding total angle $2\phi_3$ with  
\begin{equation}\label{phi3}
\cos \phi_3=\frac{1}{2}{\rm Tr}\,\left( \gamma_1 \gamma_2 \right)= \cos \phi_1\cos\phi_2+ \cosh\alpha \sin\phi_1 \sin\phi_2~.
\end{equation}
This is a 3-parameter family of Fuchsian groups labelled by $(\phi_1, \phi_2, \alpha)$, and by varying these parameters we can vary the dimensions of the operators inserted at the conical defects.\footnote{The only constraint is that $\cosh\alpha>1$, which means that the angles must satisfy the triangle inequality $\cos \phi_3> \cos (\phi_1-\phi_2)$.}
In this discussion we have assumed that the elliptic elements $\gamma_1$ and $\gamma_2$ are of finite order, so that they generate a discrete group. However this assumption was not really necessary. The defect geometry makes sense for any positive real parameters $(\phi_1, \phi_2, \alpha)$ and thus for continuous conformal dimensions, though it can no longer be expressed as $\mathbb{H}_2/\Gamma$.

This geometry contributes to the variance of the three point function $\langle {\cal O}_1 {\cal O}_2 {\cal O}_3 \rangle$ where the operators have dimensions $\Delta_i$ given in terms of the $\phi_i$ by equation (\ref{defect}).  This is therefore an explicit geometry which contributes to the variance of the OPE coefficients. We need only remember that the operators here are inserted at the points $i, ie^{-\alpha}$ and $z$ on the upper half plane, where $z$ is the unique solution of $(\gamma_1 \gamma_2) z=z$\footnote{Recall that the $PSL(2,\mathbb{R})$ element $\gamma_1\gamma_2$ acts on $z$ by fractional linear transformation as in (\ref{eq:fractionalLinear}).}, whereas in the literature it is standard to put the operators at $0$, $1$ and $\infty$ in the complex plane.

\begin{figure}
\begin{center}
\begin{overpic}[width=5in]{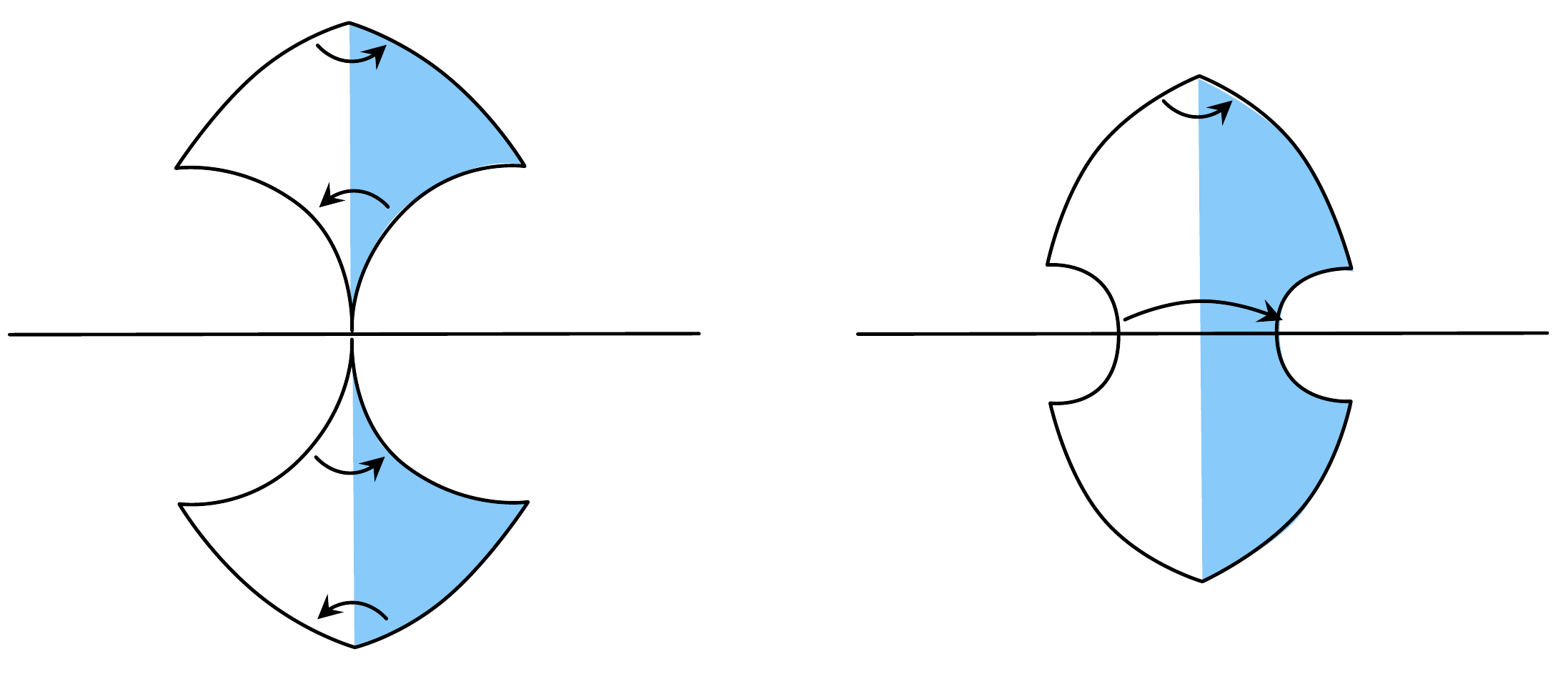}
\put (1,36) {$(a)$}
\put (56,36) {$(b)$}
\end{overpic}
\end{center}
\caption{
$(a)$ Boundary of the Maldacena-Maoz wormhole with three defects, in the limit where one operator is taken to the black hole threshold $\Delta_2 \to \frac{c}{12}$. In this limit $\gamma_2$ is parabolic and the defect approaches the boundary of $\mathbb{H}_2$. $(b)$ The boundary when $\gamma_1$ is elliptic and $\gamma_2$ is hyperbolic has only one connected component, and there are only two defects. Thus the 3-defect Maldacena-Maoz wormhole is related by analytic continuation to a geometry with the topology of a ball and two defects.
 \label{fig:fatman}}
\end{figure}

This leads to the interesting question: what would happen if we were to increase one (or more) of the $\Delta_i$ until its dimension is bigger than $\frac{c}{12}$?  To begin, consider the example above in the limit where $\phi_2\to 0$ so that $\Delta_2\to\frac{c}{12}$.  In doing so, we would like to keep $\phi_1$ and $\phi_3$ fixed.  So from (\ref{phi3}) we must take $\alpha\to\infty$ so that the fixed point of $\gamma_2$ moves to the real axis.  In particular, as $\phi_2\to0$ the generator $\gamma_2$ becomes
\begin{equation}
\gamma_2 \to 
\begin{pmatrix} 1 & 0 \\ 2 {\cos\phi_1-\cos\phi_3\over \sin\phi_1} & 1 \end{pmatrix}
\end{equation}
which is a parabolic rather than an elliptic element of $PSL(2,{\mathbb R})$.
Its fixed point is the origin $z=0$, so $\Sigma$ is now a sphere with two conical defects and a single parabolic cusp.
The boundary geometry is then that depicted in Figure \ref{fig:fatman}a.\footnote{This family of Fuchsian groups includes a familiar example: when $\phi_1=\frac{\pi}{2}$ and $\phi_3=\frac{\pi}{3}$ the group is just $\Gamma=PSL(2,{\mathbb Z})$. The generators are $\gamma_1=S=\left({0~-1\atop 1~0}\right)$ and $\gamma_2=STS=\left({1~0\atop -1~1}\right)$, respectively, and the three fixed points are the usual ones that lie on the boundary of (the $S$-transform of) the fundamental domain.  
}
In this case the wormhole has a boundary which is two copies of $\Sigma$, but we see that -- with respect to the hyperbolic geometry of the bulk -- the two copies of $\Sigma$ are glued together at their cusps.  
This nongeneric situation arises because a cusp is sourced by an operator with $\Delta=\frac{c}{12}$, exactly the boundary between ``light" and ``heavy" states.

We are then led to an important question: what happens if we try to go further, and consider operators of higher dimension?  For example, we can consider the same Fuchsian group, but with $\phi_2= \frac{i \ell}{2}$ where $\ell$ is real.  According to equation (\ref{defect}), this would correspond to an operator of dimension
\begin{equation}\label{heavy}
\Delta_2 = \frac{c}{12}\left(1+\left(\frac{\ell}{2\pi}\right)^2\right)
\end{equation}
which is now in the black hole regime.  In order to keep the matrix $\gamma_2$ real, we will take $\alpha=\beta+\frac{\pi i}{2}$ with $\beta$ real.  The generators are now
\begin{equation}
\gamma_1 = 
\begin{pmatrix} \cos \phi_1 & -\sin\phi_1 \\ \sin\phi_1 & \cos\phi_1 \end{pmatrix}
,~~~~~
\gamma_2 = 
\begin{pmatrix} \cosh \frac{\ell}{2} & e^{-\beta} \sinh \frac{\ell}{2} \\ e^{\beta}\sinh\frac{\ell}{2} & \cosh \frac{\ell}{2} \end{pmatrix}
\end{equation}
which are elliptic and hyperbolic, respectively. 
We wish to keep $\phi_3$ fixed, meaning that the element $\gamma_1\gamma_2$ is still required to be elliptic, so  the parameter $\beta$ is determined just as in equation (\ref{phi3}):
\begin{equation}
\sinh \beta =\frac{\cos \phi_3-\cos \phi_1\cosh \frac{\ell}{2}}{\sinh \frac{\ell}{2} \sin \phi_1}
\end{equation}
The hyperbolic generator $\gamma_2$ now has two fixed points on the real axis, at $\pm e^{-\beta}$. 

The boundary of the resulting quotient  is depicted in Fig \ref{fig:fatman}b.  The important conclusion is that the surface ${\mathbb H}_2/\Gamma$ now has the topology of a sphere with two singularities (conical defects parameterized by $\phi_1$, $\phi_3$) and a single boundary, which is a circle. 
The full wormhole geometry now has a single connected boundary rather than two disconnected boundaries; the boundary is a single sphere with four defects.  The parameter $\ell$ has the interpretation of a geodesic length of a circle which separates the two pairs of operators, as measured with the hyperbolic metric on the boundary.  In the CFT language, it is related to the dimension (according to (\ref{heavy})) of a heavy operator which propagates in the channel which separates the two pairs of operators.

In order to compare with a more standard picture, the natural question is this: since the quotient describes a sphere with four operators at four points, what is the cross ratio of those four points?  Unfortunately, there is no simple closed form expression for the cross ratio in terms of the parameter $\ell$, although it is possible to describe it in a series expansion.

The important implication here is that wormholes and connected geometries are related by the analytic continuation of operator dimensions as the go from ``light" to ``heavy".  When all operators are light, the boundary is the disconnected union of two three-punctured spheres.  When one operator becomes heavy, the spheres are sewn together at one of the punctures, and the boundary is a single four-punctured sphere.  As other operators are taken to be heavy, this procedure will continue.  In particular, if an additional operator is taken to be heavy we obtain the twice-punctured torus, and if all operators are taken to be heavy the geometry is a single smooth surface of genus 2.  

\subsection{Action of the $n$-point wormhole}\label{ss:fuchsianaction}

We will now calculate the action of the Maldacena-Maoz wormhole with genus zero and $n$ conical defects. Similar results can be found in \cite{MR882831,MR889594,MR1953529,Krasnov:2000zq,Takhtajan:2002cc}, but subtleties involving the normalization of the defect operators that are necessary for comparison to the dual CFT have not been treated carefully in the literature. 

The metric is
\begin{equation}\label{fuchsianmetricphi}
    ds^2=d\rho^2+\cosh^2(\rho)e^{\Phi}dzd\overline{z} \ .
\end{equation}
The conical defects are at fixed $z = z_i$, for $i=1\dots n$. The Einstein equations require $e^{\Phi} dz d\bz$ to be hyperbolic, so $\Phi$ must satisfy the Liouville equation sourced by the conical defects,
\begin{equation} \label{5.17}
    \partial\overline{\partial}\Phi=\frac{e^\Phi}{2}-4\pi G \sum m_i\delta^{(2)}(z-z_i)
\end{equation}
where $m_i = \eta_i/(2G)$ is the local mass parameter appearing as the coefficient of the worldline action of particle $i$. For this to have a real solution, we require $n \geq 3$ and $\sum \eta_i > 1$.

 Let us calculate the on-shell action for this geometry,
\begin{equation}  \label{5.19}
 S=S_{\text{bulk}}+S_{\text{bdry}}+S_{\text{defect}}
\end{equation}
with the terms defined as\footnote{Because we are in a hyperbolic slicing, there is an additional GHY term from the boundary at $|z|=R$, which contributes a term
\begin{equation}
    -\frac{1}{8\pi G}\int \sqrt{h}K= \frac{2}{4G}(\log (\frac{2}{\epsilon})+2\log R)
\end{equation}
This will eventually be cancelled by the normalization so we will not keep track of it.}
\begin{equation}
  \begin{split}
     & S_{\text{bulk}}=-\frac{1}{16\pi G}\int_{\Gamma \times \mathbb{R}} \sqrt{g}(R+2)\\
     & S_{\text{bdry}}=-\frac{1}{8\pi G}\bigg(\int_\Gamma\sqrt{h}(K-1)+\sum_i\int_{D_i}\sqrt{h}K\bigg)\\
     & S_{\text{defect}}=\sum_i\bigg ( -\frac{1}{16\pi G}\int_{D_i \times \mathbb{R}}\sqrt{g}R+m_i\int dl_i \bigg)
  \end{split}
\end{equation}
where we define the regions, $\Gamma=\{|z-z_i|>\epsilon_i;|z|<R\}$ and $D_i=\{|z-z_i|<\epsilon_i\}$. The behaviour of the Liouville field as we approach the boundaries of $\Gamma$ is 
\begin{equation}
    \Phi(z,\overline{z}) \sim
    \begin{cases}
      -4\eta_i\log(|z-z_i|) \quad & z\to z_i \\
      -4\log(|z|) \quad & z\to \infty
    \end{cases}
\end{equation}

We eventually want to relate the gravitational action to correlators in the dual CFT calculated with respect to a flat background metric. But in the present hyperbolic slicing, with the boundaries at $\rho \to \pm \infty$, the induced metric on a constant $\rho$ surface near the boundary is hyperbolic. Therefore we choose a $z$-dependent cutoff surface such that the induced metric on the surface is flat to leading order in the cutoff parameter $\epsilon$ (outside small disks excised around the defects),\footnote{We choose a hierarchy between the cutoffs, $\epsilon_i^{2(1-\eta_i)}\ll \epsilon \ll \epsilon_i^{2\eta_i}$. The first inequality serves to suppress the contribution from the GHY term evaluated inside the small disk enclosing the puncture. The second inequality ensures that cutoff surfaces asymptote to $\rho \to \pm \infty$ close to the defects.}
\begin{equation}\label{cutoffs}
    \rho_0(z,\overline{z},\epsilon)=
    \begin{cases}
       \log(\frac{2}{\epsilon})-\frac{\Phi}{2} \quad \quad & |z-z_i|>\epsilon_i\\
       \log(\frac{2}{\epsilon})+2\eta_i\log \epsilon_i-\frac{C_i}{2} \quad \quad & |z-z_i|<\epsilon_i
    \end{cases}
\end{equation}
where $C_i$ are the constant ($O(\epsilon_i^0$)) terms appearing in the expansion of the Liouville field around the defect. Away from the defects, the induced metric on the cutoff surfaces $\rho = \pm \rho_0$ is
\begin{equation}
    ds^2_{\text{bdry}}\approx (\frac{1}{\epsilon^2}+\frac{e^{\Phi}}{2})dzd\overline{z} + \frac{1}{4}(\partial \Phi dz+\overline{\partial}\Phi d\overline{z})^2 \ ,
\end{equation}
as $\epsilon \to 0$. 
Near the defects, we chose a different cutoff in \eqref{cutoffs} because otherwise the constant-$\rho_0$ surface would not be near the AdS boundary. With the cutoff in \eqref{cutoffs}, the boundary metric in the region $D_i$ is not flat. Since this region is taken to zero size at the end, we can simply absorb this into the definition of the local operator dual to the defect.

\subsubsection{Defect action}
The contribution to the action from the region near a defect is
\begin{equation}
    S_{\text{defect}_i}= -\frac{1}{16\pi G}\int_{D_i \times \mathbb{R}}\sqrt{g}R+m_i\int dl_i
\end{equation}
where recall $D_i=\{|z-z_i|<\epsilon_i\}$. The matter stress tensor obtained by varying the worldline action of the defect is
\begin{equation}
    T^{\mu \nu}(y^\alpha)=-m\int ds \frac{\delta^{(3)}(y^\sigma-x^\sigma(s))}{\sqrt{g(y^\sigma)}}\frac{dx^\mu}{ds}\frac{dx^\nu}{ds} \ .
\end{equation}
For defects propagating along fixed $z$, we can do the integral over $\rho$, and the Einstein equation near the defect reads
\begin{equation}
    G_{\mu\nu}=-8\pi G m_i n_{\mu}n_{\nu}\frac{\delta^{(2)}(z-z_i)}{\sqrt{h}}
\end{equation}
where $h$ denotes the determinant of the induced 2-metric on the fixed $\rho$ slice and $n_{\mu}$ is the unit normal to the slice. Contracting both sides with the inverse metric,
\begin{equation}
    \sqrt{h}R=16\pi G m_i \delta^{(2)}(z-z_i) \ .
\end{equation}
Therefore the worldline action from the defect cancels the delta function in the curvature, and as $\epsilon_i \to 0$, 
\begin{align}
    S_{\text{defect}_i} = 0 \ .
\end{align}

\subsubsection{The unnormalised gravitational action}

Away from the conical defect trajectories,  $R=-6$ on-shell, so the bulk term is proportional to the volume of the regularised manifold. This volume is quadratically divergent,\footnote{ $\int d^2 z = \int dxdy$}
\begin{equation}
    S_{\text{bulk}}=\frac{V_\epsilon}{4\pi G}=\frac{1}{4\pi G}\int_{\Gamma} d^2z(\frac{1}{\epsilon^2}+e^\Phi\log(\frac{2}{\epsilon})-\frac{\Phi}{2}e^\Phi)
\end{equation}
We can similarly evaluate the boundary term which is proportional to the area of the two cutoff surfaces and can be calculated using the induced metric. The extrinsic curvature is $K=2+2\epsilon^2(\partial\overline{\partial}\Phi-\frac{e^{\Phi}}{2})+O(\epsilon^3)$, so when the Liouville equation is satisfied, $K=2+O(\epsilon^3)$. Thus,
\begin{equation}
    S_{\text{bdry}}=-\frac{A_\epsilon}{8\pi G}=-\frac{1}{8\pi G}\int_{\Gamma} d^2z(\frac{2}{\epsilon^2}+e^\Phi+\partial \Phi \overline{\partial}\Phi)
\end{equation}
Adding the bulk and boundary actions, the quadratic divergences cancel, leaving behind a logarithmic divergence and terms independent of the cutoff,
\begin{equation}
    \begin{split}
        S_{\text{bulk}}+S_{\text{bdry}}=&-\frac{1}{4\pi G}\int_{\Gamma} d^2z(\frac{1}{2}\partial\Phi\overline{\partial}\Phi+\frac{e^\Phi}{2}(1+\Phi+2\log(\frac{\epsilon}{2})))\\
        & =\frac{1}{2\pi G}\int_{\Gamma} d^2z(\frac{1}{4}(\partial\Phi\overline{\partial}\Phi+e^\Phi)-\frac{1}{2}\overline{\partial}(\Phi\partial\Phi)-\partial\overline{\partial}\Phi(1+\log(\frac{\epsilon}{2})))
    \end{split}
\end{equation}
To arrive at the last equality, we used the Liouville equation 
$\partial\overline{\partial}\Phi=\frac{e^\Phi}{2}$. The terms have also been rearranged such that the first term is the 2d bulk part of the Liouville action, the second term can be simplified using the divergence theorem and, using the Liouville equation, the last term is proportional to the hyperbolic area of the boundary, 
$\int_{\Gamma}d^2z e^\Phi=4\pi(\sum_{i=1}^n \eta_i-1)$.
Following \cite{Zamolodchikov:1995aa}, define the renormalized Liouville action by 
\begin{equation}
    S_L=\frac{1}{4\pi}\int_{\Gamma} d^2z(\partial\Phi\overline{\partial}\Phi+e^\Phi)+\Phi_R+2\log R-\sum(\eta_i\Phi_i+2\eta_i^2\log\epsilon_i)
\end{equation}
where 
\begin{equation}
     \Phi_i=\frac{i}{4\pi \eta_i}\oint_{|z-z_i|=\epsilon_i}dz \Phi\partial\Phi \quad\quad\quad \Phi_R=\frac{i}{4\pi}\oint_{|z|=R}dz \Phi\partial\Phi
\end{equation}
are the boundary terms. The additional cutoff dependent terms are counter terms which make the on-shell Liouville action finite. Using the Brown-Henneaux relation $c = 3/(2G)$, we can therefore express the on-shell action for the wormhole geometry as
\begin{equation}\label{SNprenorm}
    S= \frac{c}{3} \left(S_L-2\log R+2\sum\eta_i^2\log\epsilon_i-2(1+\log(\frac{\epsilon}{2}))(\sum \eta_i-1))\right)
\end{equation}

\subsubsection{Renormalized action}

The action \eqref{SNprenorm} calculates the correlation function of unnormalized defect operators. We can normalize it by adding counterterms for the defects. The counterterms are determined in appendix \ref{s:normalization} by calculating the defect 2-point functions. The result is that the renormalized action of the $n$-point defect wormhole is
\begin{align}\label{SrenDefects}
S_{\rm wormhole} &= S + \sum_{i=1}^n S_{\rm ct}(\eta_i) - (n-2) \frac{c}{3}\log \frac{R}{\epsilon} \ , 
\end{align}
where $S_{\rm ct}$ is given in \eqref{scteta}. The counterterms cancel all the cutoff dependence in $S$, and the result can be expressed in terms of the Liouville action as
\begin{align} \label{Wormaction}
S_{\rm wormhole} &= \frac{c}{3}S_L - \frac{c}{6} \sum_{i=1}^n s(\eta_i)-\frac{c}{3}(1-\log 2)(n-2)
\end{align}
where
\begin{equation} \label{5.57}
  s(\eta_i)=2(1-2\eta_i)(\log (1-2\eta_i)+\log 2-1)
\end{equation}
is related to the semiclassical limit of the Liouville reflection amplitude, analytically continued to the defect regime,
\begin{equation} \label{reflectioncl}
e^{-\frac{c}{6}s(\eta)} \approx S(P) \ ,
\end{equation}
with $\eta \sim \frac{1}{2} + i P \sqrt{\frac{6}{c}}$. The reflection factors define the normalized Liouville operators (see appendix \ref{s:liouville}) so we find simply\footnote{To compare the wormhole amplitude with the semiclassical limit of Liouville correlators, we scale the Liouville cosmological constant in the semiclassical limit according to $\mu=\frac{1}{4\pi b^2}$. With this choice, the last term in the wormhole action (\ref{Wormaction}) can be interpreted in terms of the constant $c_b$ entering the normalisation of Liouville vertex operators in (\ref{Vertexnorm}) with its semiclassical limit given by (\ref{Slimitcb}).}
 \begin{align}
e^{-S_{\rm wormhole}}    &\approx | G_L(z_i, \bz_i)|^2
\end{align}
This agreement holds at the level of the classical action; in section \ref{s:oneloop} we will extend it to include the 1-loop graviton correction.

\subsection{The three-point wormhole}\label{ss:threepointwormhole}
We now specialise to the wormhole with three defects propagating. The on-shell Liouville action for this case becomes the semiclassical limit of the DOZZ formula. The renormalized wormhole action, calculated from \eqref{SrenDefects}, is
\begin{align}
S_{\rm wormhole} &= 
\left((\Delta_1+\Delta_2-\Delta_3)\log(|z_{12}|^2)+\text{cyclic perm}\right)-\log C
\end{align}
where the last term is
\begin{multline}
    \log C=\frac{c}{3}\bigg(\sum F(2\eta_i)-F(\sum \eta_i-1)-(F(\eta_1+\eta_2-\eta_3)+\text{cyc})+F(0)\\+2(\sum \eta_i-1)+\sum (1-2\eta_i)\log(1-2\eta_i)\bigg)
\end{multline}
Here $F(x)=\int_\frac{1}{2}^x dy\log(\frac{\Gamma(y)}{\Gamma(1-y)})$. Note that we are working in the regime: $0<\eta_i<\frac{1}{2}$ and $\eta_1+\eta_2+\eta_3>1$, where the arguments of the function $F$ in the above expression always lie between 0 and 1. In this range, $F(x)$ is unambiguous, and $\log C$ is real. Comparing to the DOZZ formula in appendix \ref{s:liouville}, we see that
\begin{align}
\log C \sim \log \tCdozz^2(\eta_1, \eta_2, \eta_3) \ .
\end{align}
For defect operators, this is equal to $\log C_0(\eta_1, \eta_2, \eta_3)^2$. 
Therefore, the gravity calculation of the average, 
\begin{align}
 \overline{\left| \langle \O_1(z_1,\bz_1) \O_2(z_2,\bz_2) \O_3(z_3,\bz_3) \rangle\right|^2 } \approx e^{-S_{\rm wormhole}} \ ,
\end{align}
agrees with the large-$c$ ensemble of CFTs defined around \eqref{C0dozz}. This is the equation stated pictorially in the introduction in equation \eqref{introC0wormhole}.

This analysis was restricted to defects which are sufficiently heavy, with $\eta_i \in [\frac{1}{4}, \frac{1}{2}]$ and $\sum \eta_i > 1$. For lighter defects, or even quantum field correlators, there is no real classical saddle, but this does not necessarily mean there is no averaging; there may be complex saddles, like those that appear in Liouville CFT in the same range of operator weights \cite{Harlow:2011ny}.\footnote{For $\sum \eta_i <1$, there are also single-boundary solutions studied in \cite{Chang:2016ftb} that were argued to contribute to $\langle \O_1 \O_2 \O_3\rangle$. The bulk solution has three defects meeting at a point in the middle of AdS$_3$, with only one boundary and trivial topology apart from the defects. Interestingly, the action of this solution is also given by the universal OPE coefficient, $S =-\log |C_0|$ \cite{Collier:2019weq}. However, it comes with an unknown coupling constant, and it is unclear how to deal with the singular intersection of the three defects. The geometry studied in \cite{Chang:2016ftb} does not exist for $\sum \eta_i > 1$ (as a real solution).}

\subsection{General Maldacena-Maoz wormholes}

The extension to higher genus boundaries can be pieced together from results in \cite{Krasnov:2000zq,Krasnov:2000ia,Takhtajan:2002cc} plus contributions from the conformal anomaly (see also \cite{Yin:2007gv,Yin:2007at}). The easiest method to find the answer directly is to calculate the boundary stress tensor, then integrate the holographic Ward identity to find the on-shell action. This method, which sidesteps the issue of regulating and renormalizing the defect operators at the expense of being less explicit, also applies to the genus-zero $n$-point wormhole. 

Let $T_F(z)$ be the Brown-York stress tensor on the left boundary, in the metric \eqref{fuchsianmetricphi}. A straightforward calculation \cite{Balasubramanian:1999re} gives
\begin{align}
T_F(z) = \frac{c}{6} T^\Phi(z) \ ,
\end{align}
where $T^\Phi = \frac{1}{2}\p^2 \Phi - \frac{1}{4}(\p \Phi)^2$ is the classical stress tensor of Liouville theory. We see that $T_F$ satisfies the same Ward identities as the quantum Liouville stress tensor, $\frac{c}{6}T^\Phi$. As we vary the moduli, the total variation of the gravitational action gets conjugate contributions from the two ends of the wormhole. Therefore the on-shell agrees with two copies of the Liouville CFT in the semiclassical limit. That is, denoting the moduli of $\Sigma$ by $(\lambda, \bar{\lambda})$, we have
\begin{align}\label{mmaction}
e^{-S_{\rm wormhole}} \approx \left| G_L(\lambda, \bar{\lambda}) \right|^2 \ .
\end{align}
This wormhole contributes to the ensemble CFT observable
\begin{align}
\overline{ G(\lambda, \blambda) G(\blambda, \lambda) } = \overline{|G(\lambda, \blambda)|^2}
\end{align}
so \eqref{mmaction} agrees with the ensemble average calculated in section \ref{s:twocopy}.

Note that in the 3D gravity literature,  the Maldacena-Maoz wormhole is often considered with a hyperbolic metric in the boundary CFT \cite{Maxfield:2016mwh,Belin:2020hea,Schlenker:2022dyo}. In this case, the gravitational action is a topological invariant. However this is not the action that should be compared to the usual expression for the genus-2 partition function in CFT, constructed for example by sewing two flat tori together with a cylinder. The action in \eqref{mmaction} is calculated with the boundary metric suitable for direct comparison to CFT, and clearly it is not a topological invariant.

\section{Two-boundary wormholes with different moduli}\label{s:twoboundary}
\newcommand{\bm}{\bar{m}}
\newcommand{\bmu}{\bar{\mu}}

The Maldacena-Maoz wormholes have equal moduli in the two asymptotic regions and are described by Fuchsian quotients of AdS.  They contribute to CFT averages of the form $\overline{G^2}$. \textit{Quasi-Fuchsian} wormholes connect two asymptotic regions with the same topology, but possibly different moduli, and they contribute to averaged products of two different CFT observables, $\overline{GG'}$.\footnote{This is a slight abuse of terminology. A (quasi-)Fuchsian group is a discrete group so by definition it has no limit points. It follows that any elliptic elements must have finite order. As we discussed in section \ref{s:fuchsian}, we are allowing the masses of the particles to vary continuously and therefore do not require the elliptic identifications to be of finite order. We will nonetheless refer to such solutions as (quasi-)Fuchsian.}

A quasi-Fuchsian group is a discrete subgroup of $PSL(2,\mathbb{C})$ that can be conjugated to a Fuchsian group by a quasiconformal map. A quotient of $\mathbb{H}_3$ by such a group is a quasi-Fuchsian wormhole. The geometry of these manifolds (or more generally, conifolds) can be quite complicated. There is no known explicit expression for the metric. The situation is considerably simpler for the \textit{almost Fuchsian} manifolds. A quasi-Fuchsian manifold is called almost Fuchsian if it has a unique minimal surface that represents the fundamental group and has principal curvatures in the range $(-1,1)$. Uhlenbeck proved that on almost Fuchsian manifolds, the distance from the minimal surface provides a global radial coordinate, so they admit a nice foliation and the metric can be written explicitly \cite{MR795233}. Almost Fuchsian manifolds have been considered in the context of 3D gravity previously in \cite{Krasnov:2005dm,Scarinci:2011np,Kim:2015qoa,Garbarz:2020fky}.

The space of almost Fuchsian manifolds is  an open set in the space of quasi-Fuchsian manifolds, with the same dimension. Therefore, any small deformation of a Fuchsian manifold is almost Fuchsian. This has the remarkable consequence that we can calculate the action of almost Fuchsian wormholes, then analytically continue in the moduli to obtain the action for all quasi-Fuchsian wormholes.  We will follow this strategy to analyze the 4-point wormhole with different cross ratios on the two boundaries. The results of our explicit calculation (to third order in the deformation away from the Fuchsian slice) agree with a less direct but quite general analysis of the action by McMullen \cite{MR1745010} and Teo and Takhtajan \cite{Takhtajan:2002cc}, which applies to a large class of quasi-Fuchsian manifolds. Both methods lead to the conclusion that the gravitational action of a quasi-Fuchsian wormhole, with or without defects, is given by\footnote{Fine print: This equation does not appear in \cite{MR1745010,Takhtajan:2002cc}, but follows without much difficulty from the results therein, plus our results in section \ref{s:fuchsian} above, if we assume the action is sufficiently analytic in the moduli. The theorems in \cite{MR1745010,Takhtajan:2002cc} did not consider quasi-Fuchsian groups with elliptic elements, ie conical defects, but this does not appear to be a serious obstruction --- our third-order calculation in section \ref{ss:almostfuchsian} does allow for conical defects, and leads to the same formula. Our conclusion is that the formula applies quite generally to quasi-Fuchsian wormholes.\label{footnote:fineprint}}
\newcommand{\bl}{\bar{\lambda}}
\begin{align}\label{sgqf}
\exp\left[-S_{\rm grav}(\l, \bl; \l', \bl')\right] \approx
G_L(\l,\l') G_L(\bl', \bl) \ ,
\end{align}
to leading order in the semiclassical expansion. Here $(\l, \bl)$ are the moduli of the surface $\Sigma_{g,n}$ on the left boundary, and $(\l', \bl')$ are the moduli on the right boundary. The right-hand side of \eqref{sgqf} is a product of Liouville observables, which agrees with the connected part (or, in the case where there are multiple contractions, what we defined as the `paired' part) of the CFT average calculated in section \ref{s:twocopy}.

Taking the logarithm of \eqref{sgqf} gives a factorized action of the form
\begin{align}\label{needSgrav1}
S_{\rm grav} = S(\l,\l') + S(\bl', \bl) \ ,
\end{align}
with $S = - \log G_L$. 
On the Fuchsian slice $(\l,\bl) = (\bl', \l')$, we have already confirmed this equation in section \ref{s:fuchsian}. 
Therefore to establish \eqref{sgqf}, it suffices to show that the on-shell gravitational action satisfies 
\begin{align}\label{needSgrav2}
\frac{\p^2 S_{\rm grav}}{\p \l \p \bl} &= 0\\
\frac{\p^2 S_{\rm grav}}{\p \l \p \bl'} &= 0 \ .\label{needSgrav2b}
\end{align}
The holographic Ward identity relates $\frac{\p}{\p \l} S_{\rm grav}$ to the stress tensor. Therefore it is enough to show
\begin{align}\label{needTL}
\frac{\p}{\p\bl} T(z) &= 0 \\
\frac{\p}{\p \bl'} T(z) &= 0  \ ,\label{needTLb}
\end{align}
where $T(z)$ is the Brown-York stress tensor on the left boundary.

\subsection{Geometry of almost Fuchsian wormholes}
\newcommand{\bw}{\bar{w}}
\newcommand{\bt}{\bar{t}}
\newcommand{\balpha}{\bar{\alpha}}
To find the metric of an almost Fuchsian wormhole, we start with the ansatz
\begin{align}
    ds^2=d\rho^2+\cosh^2\rho e^{\Phi}|dz+f(\rho)\overline{t}(\bz)e^{-\Phi}d\overline{z}|^2 \ ,
\end{align}
where $z$ is a coordinate on a genus-$g$ Riemann surface $\Sigma_{g,n}$ with $n$ defects, and $t(z) dz^2$ is a holomorphic quadratic differential on $\Sigma_{g,n}$. The Einstein equations require
\begin{align}
2 \sinh \rho f'(\rho) + \cosh\rho f''(\rho) = 0 \ .
\end{align}
Two convenient solutions of this equation, which are equivalent under a change of coordinates, are $f(\rho) = \tanh \rho$ and $f(\rho) = \frac{1}{2}(1+\tanh \rho)$. In the former case the metric is \cite{Krasnov:2005dm}
\begin{align}\label{fockmetric}
ds^2 = d\rho^2 + \cosh^2 \rho e^{\Phi_f} | dz + \tanh( \rho)  \overline{t}(\bz) e^{-\Phi_f} d\bz|^2 
\end{align}
and the last remaining Einstein equation is 
\begin{align}
\p\bar{\p}\Phi_f = \frac{1}{2}e^{\Phi_f} + \frac{1}{2} |t|^2 e^{-\Phi_f} - 2\pi \sum_i \eta_i \delta^{(2)}(z-z_i) \ .
\end{align}
In the metric \eqref{fockmetric}, the surface $\rho = 0$ is minimal, so this is the foliation established by Uhlenbeck. The restriction that the principal curvatures of the minimal surface lie in the range $[-1,1]$ ensures that the metric is non-degenerate. 
We will use a different radial coordinate, with $f(\rho) = \frac{1}{2}(1+\tanh \rho)$. In this case the metric is
\begin{align}\label{afmetric}
ds^2 = d\rho^2 + \cosh^2 \rho e^{\Phi} |dz + \frac{1}{2}(1+\tanh \rho) \bar{t}(\bz) e^{-\Phi} d\bz|^2 \ ,
\end{align}
and $\Phi$ satisfies the ordinary Liouville equation, 
\begin{align}
\p\bar{\p}\Phi = \frac{1}{2}e^{\Phi} - 2\pi \sum_i \eta_i \delta^{(2)}(z-z_i) \ .
\end{align}
The defects sit at fixed coordinate positions $z_i$, but their cross-ratios change because of the $\rho$-dependent metric deformation in \eqref{afmetric}. In the asymptotic regions, the metric approaches
\begin{align}
\mbox{Left:} \quad &\rho \to -\infty, \quad ds^2  \approx d\rho^2 + \frac{1}{4}e^{-2\rho + \Phi} |dz|^2 \\
\mbox{Right:} \quad &\rho \to +\infty, \quad ds^2 \approx d\rho^2 + \frac{1}{4}e^{2\rho+\Phi}| dz + \mu d\bz|^2 \ .
\end{align}
The complex structure on the right boundary is deformed by the Beltrami coefficient,
\begin{align}
\mu = \bar{t}(\bz) e^{-\Phi} \ .
\end{align}
Thus $\mu$ parameterizes the difference in moduli between the two ends of the wormhole.

Near the left boundary, as in the Fuchsian case (see equation \eqref{cutoffs}), we choose a $z$-dependent cutoff on $\rho$ to cancel the factor of $e^{\Phi}$. Thus the metric on the left boundary (after rescaling by $\epsilon^2$) is 
\begin{align}
ds_{\rm left}^2 = |dz|^2 \ .
\end{align}
The right boundary is more subtle due to the Beltrami coefficient. Let $w(z,\bz)$ be a solution of the Beltrami equation,
\begin{align}
\frac{\bar{\p} w}{\p w} = \mu \ ,
\end{align}
so that
\begin{align}
|dz + \mu d\bz|^2  = \left| \frac{\p w}{\p z}\right|^{-2} |dw|^2 \ .
\end{align}
The coordinate $w$ with canonical complex structure is called `isothermal', and $w(z,\bz)$ is a quasiconformal map.
We want to compare to CFT observables defined in the metric $|dz'|^2$, so we choose the radial cutoff 
$\rho_c = \log \frac{2}{\epsilon} - \frac{1}{2}\Phi +\log |\frac{\p w}{\p z}|$; after rescaling by $\epsilon^2$, the metric on the right boundary is then
\begin{align}
ds_{\rm right}^2 = |dw|^2 \ .
\end{align}
There is also an orientation reversal on the right boundary with respect to the left, so the CFT coordinate is $z' = \bw$.  When we compare the gravitational action to CFT, we must account for the quasiconformal map; for example, it is the coordinate $z'$ that is used to define the cross ratios on the right boundary

\subsection{The almost Fuchsian 4-point wormhole}\label{ss:almostfuchsian}
We now turn to an explicit example, the quasi-Fuchsian wormhole with four defects:
\begin{align}
(x,\bx) \vcenter{\hbox{\includegraphics[width=1.4in]{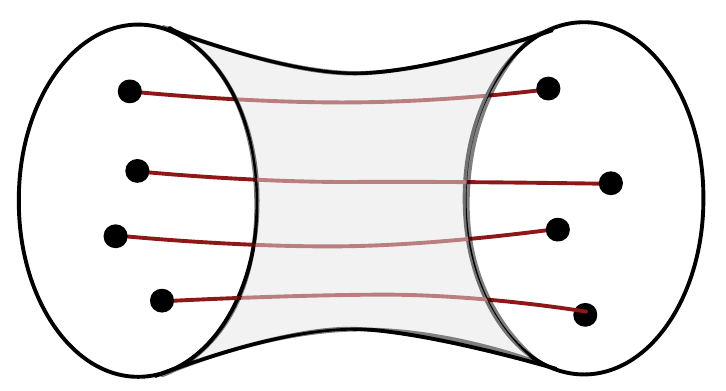}}} (x', \bx') \qquad \ .
\end{align}
The only modulus of $\Sigma_{0,4}$ is the cross ratio of the four insertion points. Denote this cross ratio by $x$ on the left boundary, and $x'$ on the right boundary, and place the defects so  that this wormhole contributes to the averaged product of CFT 4-point functions
\begin{align}
\overline{
\langle \O_1(0)\O_2(x,\bx) \O_3(1) \O_4(\infty) \rangle
\langle \O_1(0) \O_2(x', \bx') \O_3(1) \O_4(\infty) \rangle
} \ .
\end{align}
There is a unique holomorphic quadratic differential on $\Sigma_{0,4}$, up to scaling. With insertions at $0,x,1,\infty$, it is $Q(z,x)dz^2$ with
\begin{align}
Q(z,x) &:= \frac{4x(x-1)}{z(z-1)(z-x)} \ .
\end{align}
The numerator is just a convenient normalization. In the metric ansatz \eqref{afmetric} we therefore take
\begin{align}
t(z) = \alpha Q(z,x)  , \quad \bt(\bz) = \balpha Q(\bz, \bx) \ ,
\end{align}
where $\alpha$ is a complex parameter. It is straightforward to check that the principal curvatures of the minimal slice are bounded near the defects if and only if the defect weights satisfy $\eta > \frac{1}{4}$. We therefore impose this restriction, and choose $\alpha$ small enough so that the wormhole is non-degenerate. This is the almost Fuchsian 4-point wormhole. 

The complex structure on the right boundary is  
\begin{align}
|dz + \mu d\bz|^2 , \qquad
\mu = \balpha Q(\bz, \bx) e^{-\Phi} \ .
\end{align}
The wormhole is labeled by two complex parameters, which are coordinates on the doubled Teichmuller space ${\cal T}(\Sigma_{0,4}) \times {\cal T}(\Sigma_{0,4}) \cong \hat{\mathbb{C}} \times \hat{\mathbb{C}}$, with $\hat{\mathbb{C}} = \mathbb{C} \backslash \{0,1,\infty\}$. The natural coordinates for comparison to CFT are the cross ratios $(x,\bx)$ on the left boundary and $( x', \bx')$ on the right boundary, with these cross ratios defined in isothermal coordinates:
\begin{align}\label{isocross}
x = \frac{(z_1-z_2)(z_3-z_4)}{(z_1-z_3)(z_2-z_4)} , \quad
x' = \frac{( \bw_1 - \bw_2)(\bw_3 - \bw_4)}{(\bw_1 - \bw_3)(\bw_2-\bw_4)} \ , 
\end{align}
where $w_i = w(z_i, \bz_i)$. 
In the metric \eqref{afmetric}, the parameters are  instead $(x, \bx)$ and $(\alpha, \balpha)$. We will see momentarily that $\alpha$ controls the stress tensor on the left boundary. Therefore by writing down the wormhole metric, we have established an implicit relationship between the modulus on the right boundary and the stress tensor on the left boundary. This relationship, in which moduli of a Riemann surface $\Sigma'$ are parameterized by holomorphic quadratic differentials on a quasi-Fuchsian partner $\Sigma$, is known as the Bers embedding.

We now turn to the calculation of the action. As explained around \eqref{needTL}, the important question is how the Brown-York stress tensor on the left boundary, $T(z)$, depends on the moduli. A standard gravity calculation \cite{Balasubramanian:1999re,Scarinci:2011np} gives
\begin{align}\label{tlby}
T(z) = \frac{c}{6}\left[- \frac{1}{4}\alpha Q(z,x) + T^{\Phi}(z) \right]\ ,
\end{align}
where $T^\Phi(z) = \frac{1}{2}\p^2 \Phi - \frac{1}{4}(\p \Phi)^2$ is the classical Liouville stress tensor associated to the Liouville field that appears in the metric \eqref{afmetric}. The Liouville term takes the usual meromorphic form of a CFT stress tensor in the presence of operator insertions,
\begin{align}
T^\Phi(z)  = \sum_{i=1}^4\left[ \frac{\eta_i(1-\eta_i)}{(z-z_i)^2} - \frac{c_i^F}{z-z_i}\right] \ ,
\end{align}
where the accessory parameters $c_2^F = c_i^F(z_i; \bz_i) $ are independent of $(z,\bz)$ but depend on the moduli. Three of the accessory parameters are fixed by regularity,\footnote{$T(z) \sim 1/z^4$ as $z\to \infty$ before moving any operators to infinity.} leaving just the one associated to $\O_2$, which we denote $c_2^F$. With insertions at $0,x,1,\infty$, the full stress tensor is 
\begin{equation}
\begin{aligned}
\label{fullT}
\frac{6T(z)}{c} &=\frac{6h_1/c}{z^2} + \frac{6h_2/c}{(z-x)^2} + \frac{6h_3/c}{(z-1)^2} + \frac{6(h_1+h_2+h_3-h_4)/c}{z(1-z)}  \\
&\qquad -\frac{1}{4} [\alpha+  c_2^F(x,\bx) ] Q(z,x) \ .
\end{aligned}
\end{equation}
Polyakov conjectured, and Zograf and Takhtajan proved \cite{MR806856,MR882831,MR889594}, that the residue $c_2^F$ of the semiclassical Liouville stress tensor is the accessory parameter for Fuchsian uniformization. In the language of AdS/CFT, Fuchsian uniformization corresponds to the Maldacena-Maoz wormhole so this is the statement that the holographic stress tensor of that wormhole is proportional to $T^{\Phi}$. Define the accessory parameter for the almost-Fuchsian wormhole by
\begin{align}
c_2^{AF} &:= -\mbox{res}_{z \sim x} \frac{6T(z)}{c} \ .
\end{align} 
From \eqref{tlby} we have
\begin{align}
c_2^{AF} = c_2^F + \alpha \ .
\end{align}
Note that $c_2^F = c_2^F(x,\bx)$, i.e., $c_2^F$ is not holomorphic in the moduli.

So far, we have determined the stress tensor in terms of $(x,\bx, \alpha,\balpha)$. 
To complete the calculation, we must find the change of coordinates from 
\begin{align}
(x, \bx, \alpha, \balpha) \qquad \mbox{to} \qquad (x, \bx, x', \bx') \ .
\end{align}
We claim the change of coordinates is
\begin{align}\label{c2rels}
\alpha &= c_2^F(x, x') - c_2^F(x, \bx) \\
\balpha &= c_2^F(\bx, \bx') - c_2^F(\bx, x) \ .\notag
\end{align}
That is,
\begin{align}\label{c2afrel}
c_2^{AF}(x,\bx, x', \bx') = c_2^F(x,x') \ .
\end{align}
These equations describe how to choose the stress tensor on the left boundary in order to have cross-ratio $(x', \bx')$ on the right boundary.
In appendix \ref{s:beltrami} we check this by explicitly solving the Beltrami equation to 3rd order in $|x' - \bx|$, and plugging into formula  \eqref{isocross} to find $x'$. We will see below that the general statement is implied by a theorem of Teo and Takhtajan (modulo the subtleties in footnote \ref{footnote:fineprint}). 
This establishes \eqref{c2afrel}, from which follows \eqref{needTL} and therefore  the Liouville formula \eqref{sgqf} for the almost Fuchsian 4-point wormhole.

\subsection{General quasi-Fuchsian wormholes}\label{ss:TT}

McMullen \cite{MR1745010} has studied the holographic stress tensor for  general quasi-Fuchsian wormholes, though not in that language. Takhtajan and Teo \cite{Takhtajan:2002cc} built upon this result, and earlier work of Zograf and Takhtajan \cite{MR806856,MR882831,MR889594,MR1204463}, to write a Liouville-like expression for the on-shell gravitational action. Their results apply to purely loxodromic groups, i.e. those without punctures or conical defects; they were extended to allow for punctures ($\eta = \frac{1}{2}$) in \cite{MR3570149}. There does not appear to be an analogous theorem for groups with elliptic elements, which produce conical defects in the bulk, but we consider it very likely that the final results apply without modification because they agree with our results for the elliptic 4-point wormhole in the previous subsection (see footnote \ref{footnote:fineprint}). 

We will now explain (without proof) the main results of \cite{MR1745010,Takhtajan:2002cc,MR3570149}. Consider a quasi-Fuchsian 3-manifold with boundary given by the disjoint union $\Sigma \sqcup \Sigma'$ of two Riemann surfaces, each with genus $g$ and $n$ punctures. The gravitational action of this solution depends on the moduli at each end, and it depends on the boundary metrics due to the conformal anomaly (or in bulk language, due to the dependence on the cutoff). Define the canonical hyperbolic metrics $ds^2_{\rm hyper}(\Sigma)$ and $ds^2_{\rm hyper}(\Sigma')$ on each boundary by Fuchsian uniformization. With the cutoff chosen so that the induced metrics on the two boundary components are
\begin{align}\label{ds2hypers}
\frac{1}{\epsilon^2} ds^2_{\rm hyper}(\Sigma) \qquad \mbox{and} \qquad
\frac{1}{\epsilon^2} ds^2_{\rm hyper}(\Sigma') \ ,
\end{align}
the gravitational on-shell action was expressed as an integral over the boundaries in \cite{Takhtajan:2002cc,MR3570149}. With our conventions for the gravitational action,\footnote{Our $S_{TT}$ is equal to the action called $S$ in \cite{Takhtajan:2002cc,MR3570149} up to an additive constant.} 
\begin{align}
S_{\rm grav}^{\rm hyperbolic} = -\frac{c}{24\pi} S_{TT} \ .
\end{align}
The expression for $S_{TT}$ is quite complicated; it is a Liouville-like action that depends on both $\Sigma$ and $\Sigma'$. It is unclear how to directly relate it to observables in the Liouville CFT. It would be interesting to do so --- especially for the generalization to Kleinian wormholes with $k>2$ boundaries as discussed below --- but we will instead rely on properties of the stress tensor.

Consider the Brown-York stress tensor on the boundary component $\Sigma$. In the slicing with hyperbolic boundary metrics \eqref{ds2hypers}, denote the holomorphic component by $T_{\rm hyper}(z)$; then $T_{\rm hyper}(z) dz^2$ is a holomorphic quadratic differential  on $\Sigma$ (this will be justified momentarily). Choose a basis $Q^i(z)dz^2$ for such differentials, with $i = 1$ to $3g-3+n$ (which is the complex dimension of the space of quadratic differentials).  Expanding in this basis, 
\begin{align}
T_{\rm hyper}(z) = \sum_{i=1}^{3g-3+n} c_i^{\rm hyper} Q^i(z) \ ,
\end{align}
where the coefficients $c_i^{\rm hyper}$ define the accessory parameters. Now let us consider the case where the cutoff for the metric on $\Sigma$ is instead flat,  $\frac{1}{\epsilon^2}|dz|^2$. The boundary stress tensor in this slicing, denoted $T_{QF}(z)$,  is shifted by the conformal anomaly, so
\begin{align}\label{trels}
T_{ QF}(z) = T_{\rm hyper}(z) + T_F(z) \ ,
\end{align}
where $T_F$ comes from the anomaly. The anomaly action is itself a Liouville action that is associated only to the boundary component $\Sigma$. It has two pieces: one that is fixed by the Ward identities, like the first line in \eqref{fullT}, and a remainder given by a quadratic differential, $\sum_i c_i^F Q^i$. The coefficients $c_i^F$ of the quadratic differentials are the accessory parameters for Fuchsian uniformization.

Equation \eqref{trels} expresses $T_{\rm hyper}$ as the difference of two projective connections on $\Sigma$. This justifies the claim that $T_{\rm hyper}dz^2$ is a quadratic differential, because as we have noted, the portion of a projective connection that is \textit{not} a quadratic differential is fixed uniquely by the Ward identities.

We can view $S_{TT}$ as a function on two copies of Teichmuller space, ${\cal T}(\Sigma) \times {\cal T}(\Sigma')$, whose coordinates are the moduli on the two boundaries. For now, we will hold fixed $\Sigma'$ and vary $\Sigma$, treating $S_{TT}$ as a function on ${\cal T}(\Sigma)$. The variation defines $T_{\rm hyper}$, so using \eqref{trels}, it is 
\begin{align}
\delta S_{TT} = \frac{12\pi}{c} \int_{\Sigma} d^2 z \sqrt{|g|} (T^{QF}_{\mu\nu}  - T^{F}_{\mu\nu} ) \delta g^{\mu\nu} \ .
\end{align}
Take the metric variation $\delta g^{\mu\nu}$ to be an holomorphic deformation of the complex structure. Then $\delta g^{zz}$ can be expanded in a basis of harmonic (with respect to the Liouville metric on $\Sigma$) Beltrami differentials, $\sqrt{|g|}\delta g^{zz} = -2 \sum_i \mu_i \delta \lambda^i$,  and plugging in the expansion of $T^{QF} - T^F$ in quadratic differentials $Q_i$ we have  
\begin{align}
\delta S_{TT} &=  -\sum_{i,j}(c_i^{QF} - c_i^F) \delta \lambda^j \int d^2 z Q^i \mu_j  \ .
\end{align}
Since the vector space of harmonic Beltrami differentials is dual to the vector space of holomorphic quadratic differentials, we may choose dual bases for the two vector spaces so that $\int d^2 z Q^i \mu_j= \delta^i_j$. Then
\begin{align}\label{diffptt}
\delta S_{TT} = -\sum_i (c_i^{QF} - c_i^F) \delta \lambda^i \ .
\end{align}
This equation is Theorem 4.1 in \cite{Takhtajan:2002cc}.\footnote{
In \cite{Takhtajan:2002cc},  the holomorphic variation of the complex structure moduli is written as $\p$, and $\p S_{\rm TT}$ is viewed as a holomorphic $(1,0)$-form on  the deformation space. They define $P_{QF}$ as the quasi-fuchsian projective connection, and depending on how $P_{QF}$ is used, it can refer either to our $T_{QF}$ or to our $\sum_i c_i^{QF} \delta \lambda^i$. When the projective connection is viewed as a holomorphic function on $\Sigma$, the relation to our notation is 
$P_{QF}(z) =\frac{3}{c} T_{QF}(z)$. When it is viewed as a $(1,0)$ form on the deformation space, it can be evaluated on a harmonic Beltrami differential variation, $\delta\mu = \mu_i \delta \lambda^i$ and the relation to our notation is  $P_{QF}(\delta\mu) =\frac{3}{c} \int d^2 z T_{QF} (z) \delta \mu =-\frac{1}{2} \sum_i c_i^{QF}\delta \lambda^i$.
} Now that we have translated the notation to our language, we appeal to their Lemma 4.2, which states that $c_i^{ QF}$ is holomorphic on the moduli space of $\Sigma \sqcup \Sigma'$. This gives the two equations \eqref{needTL} and \eqref{needTLb}, and therefore establishes our formula \eqref{sgqf} relating the gravitational action to Liouville CFT.

\section{Handles and random bulk couplings}\label{s:coleman}

Coleman \cite{Coleman:1988cy} and Giddings and Strominger \cite{Giddings:1988cx,Giddings:1988wv} suggested that microscopic wormholes would lead to a theory with random coupling constants. They supposed that these wormholes, stabilized by some unknown physics at the Planck or string scale (or arising as classical solutions supported by axions \cite{Giddings:1987cg,Giddings:1989bq}), could have their endpoints anywhere in spacetime. This would induce a bilocal coupling in the low energy theory, 
\begin{align}
Z = \int D\phi \, e^{-\int dx {\cal L}(x) +\frac{1}{2} \sum_i \int dx dy \O_i(x) \O_i(y) } \ ,
\end{align}
where $\phi$ denotes all the fields, and $\O_i$ are operators produced at the wormhole endpoints whose specific form depends on the microscopic details. The bilocal coupling can be reinterpreted by integrating in a random coupling constant,
\begin{align}
Z = \int d\alpha^i P(\alpha^i) \int D\phi \, e^{-\int dx [ {\cal L}(x) - \alpha^i \O_i(x) ] }
\end{align}
where $P(\alpha^i)$ is a Gaussian probability distribution. Thus the effects of wormholes are indistinguishable from random couplings. This cannot be detected experimentally, because once the couplings are measured, they are completely fixed; the universe decoheres into a superselection sector. 

Recently, this idea has been revived and extended to show that wormholes in JT gravity and related toy models can be interpreted in terms of superselection sectors (or `$\alpha$-states') \cite{Marolf:2020xie,Marolf:2020rpm,Saad:2021uzi}. The mechanism is similar to Coleman's, except that in the recent discussions there is no apparent role for random bulk coupling constants. 

We will show that in 3D gravity coupled to sufficiently massive point particles, an on-shell wormhole induces random bulk 3-point couplings among the point particles. The bulk 3-point couplings obey the same statistics as the boundary OPE coefficients $c_{ijk}$. Throughout this section we assume that the operators $\O_1, \O_2, \O_3$ satisfy $\eta_i \in [\frac{1}{4}, \frac{1}{2}]$ and $\sum \eta_i > 1$, so they are sufficiently heavy to produce a real on-shell 3-point wormhole.

This effect closely resembles that studied by Coleman, Giddings, and Strominger, but there are some differences, the main one being that our analysis is entirely on-shell. The wormholes that we integrate out are classical solutions. To argue that they can be interpreted as random bulk couplings, we will calculate their contribution to boundary correlation functions and compare to a bulk EFT without wormholes, but with random couplings among the massive point particles. Also, our wormholes have large, AdS-scale mouths. This could lead to additional non-localities in more complicated observables.

\subsection{Construction of the handle wormhole}

Consider the boundary 4-point function of sub-threshold operators,
\begin{align}
G_{1221} = \langle O_1 O_2 O_2 O_1 \rangle \ .
\end{align}
On the gravity side, the leading contribution comes from AdS$_3$ with two conical defects \cite{Faulkner:2013yia,Hartman:2013mia}, 
\begin{align}
\includegraphics[width=2in]{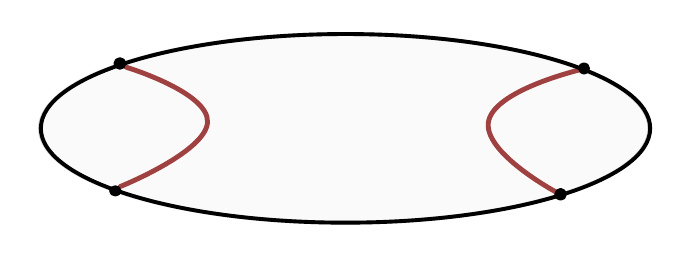}
\end{align}
We will now construct an additional saddlepoint with the topology
\begin{align}\label{handleNice}
\begin{overpic}[width=2.5in]{figures/handle.pdf}
\put (8,24) {$1$}
\put (11,3) {$2$}
\put (80,2) {$2$}
\put (85,26) {$1$}
\put (47,15) {$3$}
\end{overpic}
\end{align}
The blue line running around the handle is a conical defect corresponding to another sub-threshold operator, $\O_3$. The starting point to construct this solution is the 2-boundary wormhole with three defects discussed in section \ref{ss:threepointwormhole}. Let's place the three operators $\O_1, \O_2, \O_3$ at $z=1,\infty,0$ respectively on both ends of the wormhole. Now we cut open this wormhole along two identical extremal surfaces and glue these surfaces together:
\begin{align}\label{cutandglue}
\vcenter{\hbox{
\begin{overpic}{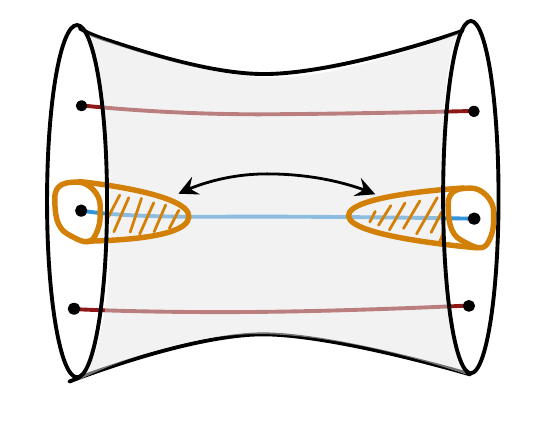}
\put(40,48) {identify}
\end{overpic}
}}
\end{align}
This is obviously still a saddlepoint, and it is straightforward to see that it has the same topology as the picture in \eqref{handleNice}. In gluing the two extremal surfaces together we have also glued together the boundaries, so the geometry now has just one asymptotic region. To see that there are really extremal surfaces like this, we refer to the Fuchsian representation in figure \ref{fig:batman}. We are adding to the Fuchsian group an additional generator, not in $PSL(2,\mathbb{R})$, which identities a circle in the lower half-plane of this figure to its reflection in the upper half plane.\footnote{This procedure may be possible only if the regions cut out around the $\O_3$ insertions are sufficiently small.}

\subsection{Gravitational action}\label{s:handleaction}

We will now show that the action of this wormhole is
\begin{align}\label{handleaction}
e^{-S_{\rm wormhole}} \ \approx  \   \overline{c^2_{123}} |{\cal F}_{1221}(h_3, x)|^2  \ , 
\end{align}
to leading order, where $x$ is the cross-ratio of the insertions $\O_1 \O_2 \O_2 \O_1$ on the glued spherical boundary. Let $w$ be a coordinate on the full boundary; it covers two copies of the $z$ plane, minus the parts that were removed by the cutting and gluing procedure. To calculate the action we will integrate the Ward identity for the boundary stress tensor $T(w)$. By construction, this stress tensor satisfies the Ward identities of a four-point function with operators $\O_1 \O_2 \O_2 \O_1$. Therefore it takes the form
\begin{align}
T(w) &= \frac{c}{6}\sum_{i=1}^4 \left( \frac{6h_i/c}{(w-w_i)^2} - \frac{c_i}{w-w_i}\right) \ .
\end{align}
Three of the accessory parameters $c_i$ are determined by regularity as $w \to \infty$. The final piece of information that specifies $T(w)$ can be characterized as a monodromy of the auxiliary equation
\begin{align}
\psi''(w) + \frac{6}{c}T(w) \psi(w) = 0 \ .
\end{align}
Around an operator insertion of weight $h = \frac{c}{6}\eta(1-\eta)$, the monodromy $M$ of the two solutions to this differential equation satisfies
\begin{align}
\mbox{Tr\ } M = -2 \cos( \pi (1-2\eta)) \ .
\end{align}
In our case, since the classical boundary stress tensor is unaffected by cutting and gluing, we conclude that the monodromy around points $w_1$ and $w_2$ is set by the $h_3$, the weight of the operator that is running around the handle:
\begin{align}
\mbox{Tr\ }M_{w_1w_2} = -2 \cos (\pi(1-2\eta_3)) \ .
\end{align}
In principle, this fixes the final accessory parameter $c_2$. 

The procedure that we have just followed --- write down a semiclassical stress tensor, and determine $c_2$ by setting the monodromy around $w_1$ and $w_2$ appropriate to an operator of weight $h_3$ --- is exactly the prescription used by Zamolodchikov \cite{Belavin:1984vu,ZamoRecursion} to calculate Virasoro blocks in the semiclassical limit.\footnote{This prescription is related to 3D gravity in \cite{Hartman:2013mia}, in situations without wormholes.} In this limit the blocks exponentiate,
\begin{align}
\F_{ijkl}(h; x_{a}) \approx \exp\left[-\frac{c}{6}f_{ijkl}(\frac{h}{c}; x_a) \right] \ ,
\end{align}
where $x_a = (x_1,x_2,x_3,x_4)$ and we are including all of the position dependence in the block.
The final step in Zamolodchikov's prescription is to integrate the Ward identity
\begin{align}
\frac{\p f}{\p x_2} = c_2(x_a)
\end{align}
to find the semiclassical block $f$. The action of the wormhole that we just constructed is also obtained by integrating the Ward identity, with the same $c_2$. Therefore 
\begin{align}
\frac{\p}{\p x_2} S_{\rm wormhole} &= \frac{c}{6} \frac{\p}{\p x_2} f_{1221}(h_3; x_a)\\
\frac{\p}{\p \bx_2} S_{\rm wormhole} &= \frac{c}{6} \frac{\p}{\p \bx_2} f_{1221}(\bh_3; \bx_a)\ .
\end{align}
To fix the integration constant, we take the OPE limit in which the two operators that started on the left of the wormhole are very far away from the two operators that started on the right. This is the limit where the cutouts in \eqref{cutandglue} are small, with only a tiny dome near each boundary removed.\footnote{Call the left boundary of the original 2-boundary wormhole  in \eqref{cutandglue}  $M_L$ and the right boundary $M_R$. Denote the cutout regions inside the orange circles by $C_L$ and $C_R$. The coordinate $w$ covers $(M_L \backslash C_L) \sqcup (M_R \backslash C_R)$. As we shrink $C_{L,R}$ to zero size, we see that $w$ covers two copies of $ \{ z \in \mathbb{C} | |z|>L\}$ for some radius $L$. To cover this region, we rescale $L \to 1$, do an inversion on one copy, and glue them together along the unit circle. As $L\to 0$ this procedure results in a configuration where two operators $\O_1,\O_2$ are moved near the origin and the other two are near infinity.
} In this limit the cutting and gluing leaves the action unchanged; therefore $S_{\rm wormhole}$ in the OPE limit is the action of the 2-boundary 3-defect wormhole that we started with. This action, calculated in section \ref{ss:threepointwormhole}, is $-\log \overline{c^2_{123}}$ at leading order. Therefore we find
\begin{align}
S_{\rm wormhole} = \frac{c}{6}f_{1221}(h_3, x_a) + \frac{c}{6}f_{1221}(\bh_3, \bx_a) -\log \overline{c^2_{123}}  + o(c) \ .
\end{align}
Exponentiating gives \eqref{handleaction}.

\subsection{Bulk interpretation}
Combining this with the trivial bulk solution, we have found from the gravity calculation
\begin{align}\label{handleexpansion}
\langle \O_1 \O_2 \O_2 \O_1 \rangle \quad \approx  \quad
\vcenter{\hbox{\includegraphics[width=1.2in]{figures/witten-diagram-decoupled.pdf}}}
+ \sum_{\rm defects} 
\vcenter{\hbox{\includegraphics[width=1.2in]{figures/handle.pdf}}}
\end{align}
The first term is the identity block, discussed in section \ref{ss:ensemblecrossing}. The wormhole contribution is exactly what we would obtain from a Witten diagram with a random coupling constant,
\begin{align}
e^{-S_{\rm wormhole}} \quad \approx  \quad \overline{c^2_{123}} |{\cal F}_{1221}(h_3, x)|^2  \quad  \approx \quad
\vcenter{\hbox{
\begin{overpic}[width=1.6in]{figures/handle-shrunk.pdf}
\put (11,30) {$1$}
\put (11,0) {$2$}
\put (88,30) {$1$}
\put (83,0) {$2$}
\put (48,22) {$3$}
\end{overpic}
}} \ .
\end{align}
The operators are heavy enough to backreact, so this is not quite a regular Witten diagram. It is a network of conical defects, which has classical action $\frac{c}{6}(f_{1221}(h_3,x_a)+\mbox{barred})$ \cite{Hartman:2013mia,Faulkner:2013yia,Hijano:2015qja}, so it reproduces the conformal block. The 3-point vertices come with bulk coupling constants $c_{ijk}$. The dashed line indicates that the couplings are treated as random variables, with $\overline{c_{ijk}} = 0$ and $\overline{c^2_{ijk}}$ given by the universal OPE formula in the dual CFT.

As we described in the introduction, the handle wormhole is essential for the ensemble interpretation of the bulk theory. In the boundary CFT, the exchange $\O_1 \O_2 \to \O_3 \to \O_1 \O_2$ contributes to the 4-point function, and it is the handle wormhole that accounts for this exchange on the gravity side.

\subsection{Another diagram in the EFT}\label{ss:anotherDiagram}
For the bulk EFT interpretation to be consistent, any other diagrams that exist in the effective theory should match with the wormhole picture.  In particular, since we have argued that the EFT has a random  3-point coupling $c_{123}\phi_1 \phi_2\phi_3$, where $\phi_i$ is the bulk field dual to $\O_i$ and $c_{123}$ is drawn from a Guassian distribution, the bulk effective theory also predicts corrections to the 6-point function:
\begin{align}
\overline{\langle \O_1 \O_2 \O_3 \O_3 \O_2 \O_1 \rangle} =
\vcenter{\hbox{\includegraphics[width=1.2in]{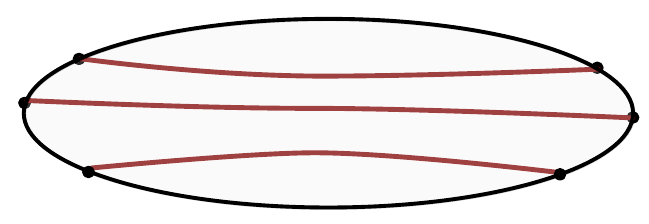}}}
+ \vcenter{\hbox{\includegraphics[width=1.5in]{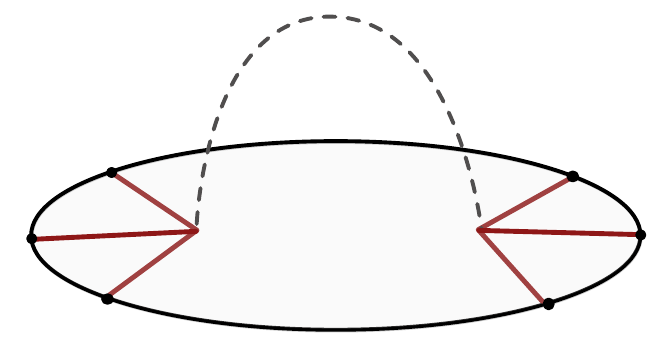}}} + \cdots
\end{align}
The first term is the Witten diagram for the identity block, which already appears without wormholes. The second term is the correction from the random bulk 3-point coupling. Note that in this diagram, the first three particles are coupled to the other three particles not just by the random coupling, but by gravitational backreaction. 

This should match a wormhole, and indeed it does. The wormhole picture is
\begin{align}
\vcenter{\hbox{\includegraphics[width=1.5in]{figures/handleG6-shrunk.pdf}}}
=
\vcenter{\hbox{\includegraphics[width=1.5in]{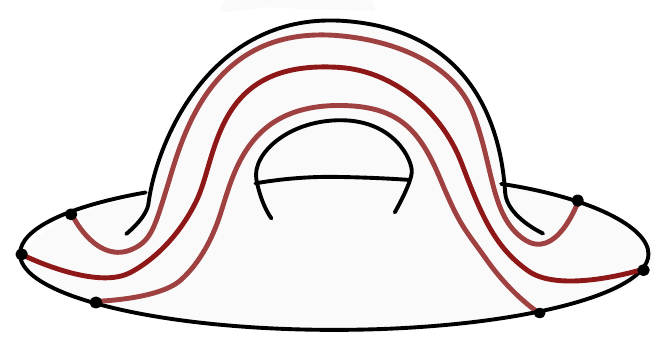}}}
\end{align}
The wormhole is constructed by a cutting and gluing procedure similar to \eqref{cutandglue}, except that now the identification is done in a way that avoids all three defects. Thus they all travel through the handle. 
Following similar steps to the calculation above, it is straightforward to see that this wormhole produces the correction
\begin{align}
\overline{c^2_{123}} \left| {\cal F}^{\rm comb}_{123321}(h_3, \mathbb{1}, h_3, x_i) \right|^2 \ .
\end{align}
This is exactly the correction expected from the bulk EFT with random couplings, and from the CFT ensemble.

\subsection{Non-Gaussianities}\label{ss:nongaussianities}

As we emphasized in section \ref{s:ensemble}, our definition of the large-$c$ ensemble is not exact. We have only provided the CFT data at leading order in the semiclassical expansion. At subleading orders, the statistics of the OPE coefficients must be non-Gaussian due to the structure of the correlators and crossing relations \cite{Foini:2018sdb,Belin:2021ryy}. In the bulk, non-Gaussian statistics are also expected to arise from multi-wormholes \cite{Banks:1988je}. The contribution of the handle topology to the four-point function provides a simple example of this effect by introducing correlations between different OPE coefficients. In the channel $\O_1 \O_2 \to p \to \O_1 \O_2$, we have interpreted this handle as the contribution of a sub-threshold state in the OPE. In the dual channel, 
\begin{align}
\overline{ \langle \O_1 \O_2 \O_2 \O_1 \rangle }  = \sum_p \overline{c_{11p} c_{22p}} \left| \F_{1122}(h,1-x) \right|^2 \ .
\end{align}
In the approximation where the OPE coefficients are independent Gaussian random variables, $\overline{c_{11p} c_{22p}}$ is non-vanishing only for $\O_p = \id$. Therefore to reproduce the handle we must add a correction at subleading order,
\begin{align}
\overline{c_{11p}c_{22p}} = \delta_{p\id} + W_{12p} + \cdots \ .
\end{align}
The correction term $W_{12p}$ is determined by comparison to the other channel, where the expansion is \eqref{handleexpansion}. Note that $\overline{c^2_{123}}$ is exponentially suppressed at large $c$, so the corrections are non-perturbatively small. We leave a detailed analysis for the future.

\subsection{Handle in a higher genus example}

This story appears to generalize: handles in the bulk, supported by defects propagating through them, are dual to defect contributions in the boundary conformal block expansion. We will not try to address this systematically here but we will describe one more example. Let's start with the 2-boundary wormhole with torus boundary and one conical defect. From this 2-boundary wormhole we can construct a 1-boundary wormhole by the same cutting-and-gluing procedure as above:
\begin{align}\label{glued-torus}
\begin{overpic}[width=2in]{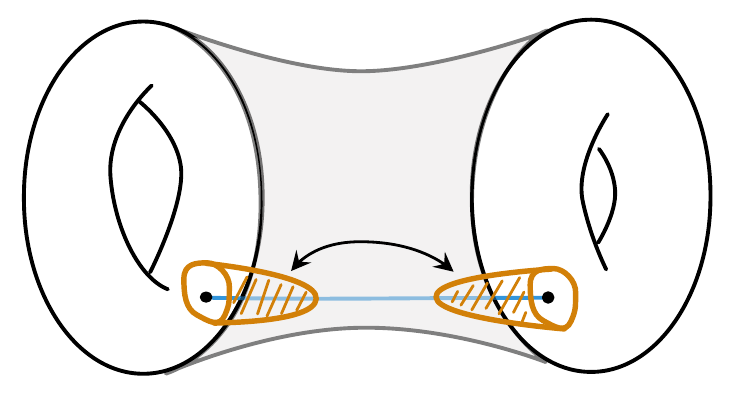}
\put (40,24) {\footnotesize identify}
\end{overpic}
\end{align}
The boundary is now genus-2. This is therefore a non-handlebody contribution to the genus-2 partition function $Z_{g=2}$ of the CFT, $Z_{g=2}$, with a defect running around a non-contractible loop.

Now consider the dual CFT. The conformal block expansion of the genus-2 partition function in the dumbbell channel is
\begin{align}\label{dumbg2B}
Z_{g=2} &= \sum_{p,q,r} c_{ppq} c_{rrq} 
\left|
\vcenter{\hbox{
\begin{tikzpicture}[scale=0.5]
\draw (1,2) circle (1);
\draw (1,0) -- (1,1);
\draw (1,-1) circle (1);
\node at (-0.25,2) {$p$};
\node at (0.75, 0.5) {$q$};
\node at (-0.25, -1) {$r$};
\end{tikzpicture}
}}
\right|^2
\end{align}
The ensemble-averaged contributions to this sum from heavy operators can be written as a sum of Virasoro identity blocks in various channels. This will be discussed in section \ref{ss:genus2average} below. Here we are interested in the terms where $q$ is sub-threshold, and $p$, $r$ are above threshold.
For a given $q$, this contribution is 
\begin{align}\label{genus2handle}
\sum_{p,r \in H} 
c_{ppq} c_{rrq} 
\left|\vcenter{\hbox{
\begin{tikzpicture}[scale=0.5]
\draw (1,2) circle (1);
\draw (1,0) -- (1,1);
\draw (1,-1) circle (1);
\node at (-0.25,2) {$p$};
\node at (0.75, 0.5) {$q$};
\node at (-0.25, -1) {$r$};
\end{tikzpicture}
}}  \right|^2 = \langle O_q\rangle_{\tau_1} \langle O_q\rangle_{\tau_2} \ell^{h_q} \bar{\ell}^{\bh_q} \ ,
\end{align}
where $H$ denotes the set of above-threshold primaries, $\tau_1$, $\tau_2$ and $\ell$ are the moduli of the genus-2 surface, and $\langle O_q \rangle_\tau$ is a thermal 1-point function. The average of this quantity is therefore determined by $\overline{\langle O_q\rangle_{\tau_1} \langle O_q\rangle_{\tau_2}}$, which is computed by the 2-boundary torus wormhole with 1 defect that we started with in \eqref{glued-torus}. Following the same steps as in section \ref{s:handleaction}, the gravitational action of the non-handlebody constructed by the cutting and gluing procedure in \eqref{glued-torus} reproduces the contribution in \eqref{genus2handle}. We omit the details.

\section{Many-boundary wormholes}\label{s:kboundary}

There are solutions of 3D gravity, with or without defects, with any number of boundary components. They arise as quotients of $\mathbb{H}_3$ by Kleinian groups that are not necessarily quasi-Fuchsian. There is a Liouville-like expression for the gravitational action of a large class of Kleinian wormholes derived by Takhtajan and Teo \cite{Takhtajan:2002cc}, but it is not straightforward to evaluate it or compare to CFT. We will not attempt a general analysis of many-boundary wormholes but we will check one example in a particular limit and show that it agrees with the CFT ensemble.

\begin{figure}
\begin{center}
\includegraphics{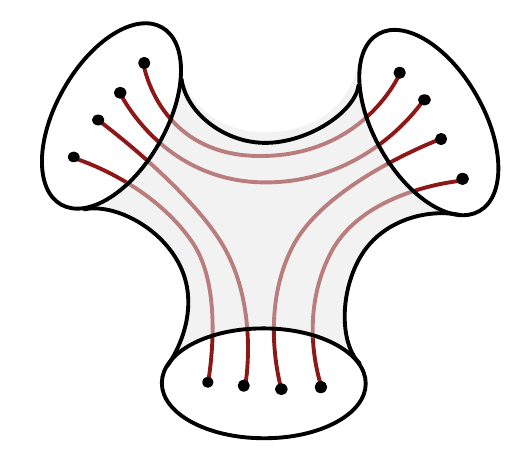}
\end{center}
\caption{3-boundary wormhole that contributes to a product of 4-point functions, $\overline{G_{1234}G_{3456}G_{5612}}$.\label{fig:threeboundary}}
\end{figure}

On the gravity side, we consider wormholes with $k \geq 2$ boundaries and four sub-threshold operators inserted on each boundary. The case $k=3$ is illustrated in figure \ref{fig:threeboundary}. We can either choose the operator species so this is the only OPE contraction, or allow some operators to be identical but consider only this term. If there is a saddle, it contributes to the ensemble average of a product of 4-point functions,
\begin{align}
\overline{G^k} &:= \overline{G_{1234}G_{3456} G_{5678} \cdots G_{(2k-1)(2k)12}}
\end{align}
where $G_{ijkl} = \langle \O_i \O_j \O_k \O_l\rangle$ and $\O_1,\O_2,\dots \O_{2k}$ are sub-threshold operators.

\subsection{CFT calculation of $\overline{G^k}$}
The CFT calculation proceeds in the usual way. We expand all of the $G$'s in conformal blocks,
\begin{align}
G_{1234}(x^{(1)}, \bx^{(1)})
&G_{3456}(x^{(2)}, \bx^{(2)}) 
\cdots G_{(2k-1)(2k)12}(x^{(k)}, \bx^{(k)})
\\
&= \sum_{p_1, p_2, \dots, p_k} c_{12p_1} c_{34p_1} c_{34p_2} c_{56p_2} \cdots \left|
\vcenter{\hbox{
\begin{tikzpicture}[scale=0.5]
\draw (0,1) -- (1, 0);
\draw (0, -1) -- (1,0);
\draw (1,0) -- (3,0);
\draw (3,0) -- (4,1);
\draw (3,0) -- (4,-1);
\node at (-0.25,1) {$1$};
\node at (-0.25,-1) {$2$};
\node at (4.25,1) {$3$};
\node at (4.25,-1) {$4$};
\node at (2,0.4) {$p_1$};
\end{tikzpicture}
}}
\vcenter{\hbox{
\begin{tikzpicture}[scale=0.5]
\draw (0,1) -- (1, 0);
\draw (0, -1) -- (1,0);
\draw (1,0) -- (3,0);
\draw (3,0) -- (4,1);
\draw (3,0) -- (4,-1);
\node at (-0.25,1) {$3$};
\node at (-0.25,-1) {$4$};
\node at (4.25,1) {$5$};
\node at (4.25,-1) {$6$};
\node at (2,0.4) {$p_2$};
\end{tikzpicture}
}}
\cdots
\right|^2
\notag
\end{align}
where $x^{(i)}$ is the cross-ratio on the $i$th boundary.
Taking the ensemble average sets all of the internal weights $p_i$ equal. Thus 
\begin{align}\notag
\overline{G^k} &= \sum_p \overline{|c^2_{12p}|}\ \overline{|c^2_{34p}|}\ \cdots \overline{|c^2_{(2k-1)(2k)p}|} 
 \left|
\vcenter{\hbox{
\begin{tikzpicture}[scale=0.5]
\draw (0,1) -- (1, 0);
\draw (0, -1) -- (1,0);
\draw (1,0) -- (3,0);
\draw (3,0) -- (4,1);
\draw (3,0) -- (4,-1);
\node at (-0.25,1) {$1$};
\node at (-0.25,-1) {$2$};
\node at (4.25,1) {$3$};
\node at (4.25,-1) {$4$};
\node at (2,0.4) {$p$};
\end{tikzpicture}
}}
\vcenter{\hbox{
\begin{tikzpicture}[scale=0.5]
\draw (0,1) -- (1, 0);
\draw (0, -1) -- (1,0);
\draw (1,0) -- (3,0);
\draw (3,0) -- (4,1);
\draw (3,0) -- (4,-1);
\node at (-0.25,1) {$3$};
\node at (-0.25,-1) {$4$};
\node at (4.25,1) {$5$};
\node at (4.25,-1) {$6$};
\node at (2,0.4) {$p$};
\end{tikzpicture}
}}
\cdots
\right|^2\\
&\sim \left| \int dh \rho_0(h)^{1-k/2} I_{1234}(h,x^{(1)}) I_{3456}(h, x^{(2)}) \cdots I_{(2k-1)(2k)12}(h,x^{(k)})
\right|^2 \label{gkpred}
\end{align}
where
\begin{align}
I_{ijkl}(h,x) &= \sqrt{|\tCdozz(P_i,P_j, P)\tCdozz(P_k,P_l,P)|}\F_{ijkl}(h,x)
\end{align}
with the $h$'s and $P$'s related by \eqref{hprelation}. To be more explicit about the leading term, write the holomorphic and antiholomorphic factors in \eqref{gkpred} as
\begin{align}\label{gkpred2}
\overline{G^k} &\approx \exp\left[ - \frac{c}{6}(S_k + \overline{S_k}) \right] \ .
\end{align}
Parameterize the heavy weight over which we integrate by $h = \frac{c}{24}(1+\gamma^2)$. 
Then using the Cardy formula $\rho_0(h) \approx e^{\pi c \gamma/6}$, the semiclassical DOZZ formula $|\tCdozz(P_1, P_2, P)| \approx e^{-\frac{c}{6}w_{12}(\gamma)}$ and the semiclassical Virasoro blocks ${\cal F}(h,x) \approx e^{-\frac{c}{6}f(\gamma,x)}$, the result in the CFT ensemble is
\begin{align}\label{skcft}
S_k = \mbox{ext}_{\gamma}\big[
\left(\frac{k}{2} -1\right)\pi \gamma &+ 
w_{12}(\gamma) + w_{34}(\gamma) + \cdots + w_{(2k-1)(2k)}(\gamma) \\
&+ f_{1234}(\gamma, x^{(1)}) + f_{3456}(\gamma, x^{(2)}) + \cdots + f_{(2k-1)(2k)12}(\gamma, x^{(k)}) 
\big]\notag
\end{align}
and similarly for the antiholomorphic term.
For $k=2$, the Cardy factors cancel, and we recover the Liouville correlator as in section \ref{s:twocopy}.  For $k>2$ there is no apparent way to relate this to any standard observable in Liouville CFT, because of the first term in \eqref{skcft}. We have checked numerically for $k=3$ that there is an extremum (which for real cross ratios is a minimum with $\gamma \in \mathbb{R}$), so there is a large semiclassical contribution that should correspond to a bulk solution.

\subsection{Comparison to gravity for $k \approx 2$}

Our prediction is that $\frac{c}{6}(S_k + \overline{S_k})$ equals the gravitational action of the $k$-boundary wormhole. We will do one simple consistency check of this prediction. Let us take all the operators to be identical, and all the cross-ratios real and equal, $x^{(1)} = x^{(2)} = \cdots = x^{(k)}$. Then the wormhole has a $\mathbb{Z}/k\mathbb{Z}$ cyclic symmetry, rotating the $k$ boundaries. It is quite similar to a replica wormhole \cite{Almheiri:2019qdq,Penington:2019kki}, or to related wormholes involving operator insertions considered in \cite{Stanford:2020wkf}. We can therefore follow the derivation of the Ryu-Takayanagi formula by the gravitational replica method \cite{Lewkowycz:2013nqa} in order to compare gravity to CFT in the vicinity of $k=2$. We will borrow the following result from \cite{Lewkowycz:2013nqa}. Denote by $I(n)$ the gravitational action of the solution replicated $n$ times by branching around a minimal surface. Then to first order in $n-1$, the action is related to the area of the minimal surface by
\begin{align}\label{lmvar}
I(n) = n I(1)  + (n-1) \frac{\mbox{Area}}{4G}  + O((n-1)^2)\ .
\end{align}

\begin{figure}
\begin{center}
\begin{overpic}{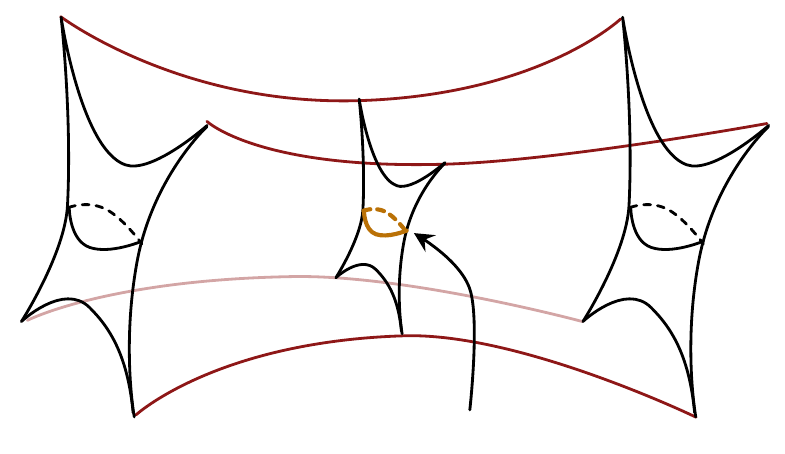}
\put (57, 0) {$\Gamma$}
\end{overpic}
\end{center}
\caption{Two-boundary wormhole with four conical defects, drawn in the hyperbolic metric. The $\rho=0$ slice in the middle is a minimal surface, and the orange curve $\Gamma$ around its waist is a geodesic.\label{fig:RTsurface}}
\end{figure}

For our purposes, the $n=1$ solution is the 2-boundary, 4-defect Fuchsian wormhole discussed in section \ref{s:fuchsian}. This solution has a minimal 2-surface at the center of the wormhole, $\rho=0$, and on this minimal 2-surface there is a minimal geodesic $\Gamma$ pictured in figure \ref{fig:RTsurface}. Consider $n$ copies of the 2-boundary solution, with $k = 2n$ boundaries in total, branched around the minimal surface (and backreacted). Applying \eqref{lmvar} we have
\begin{align}\label{Ivargrav}
I(n=1+\epsilon) = I(1) + \epsilon I(1) + \epsilon \frac{c}{6} \mbox{length}(\Gamma) +O(\epsilon^2) \ .
\end{align}
Let's compare to the CFT. For the general (non-symmetric) wormhole with $k$ boundaries, the action is \eqref{skcft}. Assuming a $Z_k$ symmetry and identical real cross-ratios, it simplifies to
\begin{align}
\frac{c}{6}(S_k + \overline{S}_k) &= \frac{ck}{3}\left[ \frac{k-2}{2k} \pi \gamma + w_{11}(\gamma) + f_{1111}(\gamma,x) \right] \ .
\end{align}
At $k=2$, this is the exponent in the Liouville 4-point function, for which the saddlepoint weight is denoted $\gamma_*$. Expanding in $k=2(1+\epsilon)$, 
\begin{align}\label{Ivarcft}
\frac{c}{6}(S_k + \overline{S}_k) = \frac{c}{6}(S_2 + \bar{S}_2)(1+\epsilon) + 
\epsilon \frac{c}{6} 2\pi \gamma_* +O(\epsilon^2) \ ,
\end{align}
where we have used the saddlepoint equation to drop the variation of $w_{11} + f_{1111}$. 

This is to be compared with the gravity result, \eqref{Ivargrav}. To relate the geodesic length to the saddlepoint weight we note that the bulk metric on the $\rho=0$ slice is the Liouville metric, 
$ds^2|_{\rho=0} = e^{\Phi}|dz|^2$.
Therefore $\Gamma$ is a geodesic in the hyperbolic metric on the sphere with four insertions. It turns out that this geodesic length is related to the saddlepoint momentum in the semiclassical Liouville 4-point function \cite{Hadasz:2005gk}. The relation is\footnote{This is the relation mentioned in \eqref{heavy}. It is equation (1.4) in \cite{Hadasz:2005gk}. Our conventions are $\mu_{\rm there} = 1$, $\delta_{\rm there} = \frac{1}{4}(1+\gamma^2)$. } 
\begin{align}
\mbox{length}(\Gamma) = 2 \pi \gamma_* \ .
\end{align}
Therefore we find perfect agreement between the CFT \eqref{Ivarcft} and gravity \eqref{Ivargrav} at this order.

\section{Ensemble interpretation of simple topologies}\label{s:singleboundary}

Throughout this paper we have shown in concrete and explicit terms how wormhole amplitudes in semiclassical gravity admit an interpretation in terms of averaged solutions to the bootstrap in the semiclassical limit. Here we will show that similar considerations apply to configurations in semi-classical gravity that involve only a single boundary, \emph{without} wormholes. This expands upon our discussion of crossing invariance in the large-$c$ ensemble in section \ref{ss:ensemblecrossing}.

The geometries that we discuss in this section include defects propagating in a trivial topology as well as handlebody 3-manifolds, the higher-genus cousins of the BTZ black hole. These geometries are saddlepoints whose on-shell action is given by the semiclassical Virasoro identity block in some channel; this was demonstrated for partition functions, i.e. handlebodies in the bulk, in \cite{Yin:2007gv,Maloney:2007ud}, and for correlation functions involving bulk defects in \cite{Hartman:2013mia}. We will show in various examples that the identity block can be reinterpreted as an average over heavy states in a dual channel. This is complementary to recent results in e.g. \cite{Cardy:2017qhl,Collier:2019weq,Belin:2021ryy,Belin:2021ibv,Anous:2021caj} where it was argued that crossing can be used to infer the statistics of heavy-operator OPE coefficients, averaged in the microcanonical sense. It is natural to suppose that these statistics are self-averaging, so that ensemble averages and microcanonical averages agree.

 There is a close connection between the wormholes discussed in the rest of this paper and the single-boundary identity blocks discussed in this section: Single-boundary observables that correspond to an identity block may be viewed in terms of an analytic continuation of the wormhole amplitudes, where the weight of an external defect state is taken above the black hole threshold, as discussed in section \ref{ss:quotient}.
 
 \subsection{Sphere four-point function}
 
 The simplest example of this phenomenon involves the two-boundary sphere three-point wormhole discussed in section \ref{ss:threepointwormhole}. There we found (with insertions at $0,1,\infty$)
 \begin{equation}\label{eq:3ptWormholeRecap}
	e^{-S_{\rm wormhole}} \approx \overline{|c_{123}|^2} = |C_0(h_1,h_2,h_3)|^2,
\end{equation}
where the $\approx$ indicates that this is a classical equivalence, and the weights satisfy $\eta_1+\eta_2+\eta_3 > 1$ with $\eta_i \in (0, \frac{1}{2})$. The right-hand side is, of course, the universal formula for the averaged structure constants, analytically continued to the regime where all three weights correspond to those of defects.

Consider what happens to this geometry, and its CFT interpretation, if we take one or more of the external operators above the black hole threshold. As illustrated in fig.~\ref{fig:fatman}, taking one operator above threshold merges the two boundary components, and eliminates the propagating defect. The result is a four-point function on the sphere, calculated by two defects propagating in trivial topology:
\begin{align}\label{g3g3merge}
\vcenter{\hbox{\includegraphics[width=2in]{figures/G3G3.pdf}}} \qquad \Longrightarrow \qquad \vcenter{\hbox{\includegraphics[width=1.4in]{figures/G4.pdf}}} \ .
\end{align}
We find therefore
\begin{align}\label{g4cc}
 \vcenter{\hbox{\includegraphics[width=1.4in]{figures/G4.pdf}}}  \qquad \approx \qquad |C_0(h_1,h_2,h_3)|^2 \ ,
\end{align}
where the left-hand side of this equation represents $e^{-S_{\rm grav}}$ for this solution, and now one of the operators, say $\O_3$, is above threshold, $h_3 > \frac{c}{24}$. Note, however, that the boundary metric on the sphere in \eqref{g4cc} is inherited from the analytic continuation in \eqref{g3g3merge}, and it is not the usual flat metric.  The dependence of $e^{-S_{\rm grav}}$ on $h_3$ enters through the nontrivial metric. Transforming to the flat metric on the boundary $S^2$ gives an additional anomaly factor, so that in the flat metric, with the four points at $0,x,1,\infty$, the answer must be
\begin{align}\label{g4ccFlat}
 \vcenter{\hbox{\includegraphics[width=1.4in]{figures/G4.pdf}}}  \qquad \approx \qquad  |\rho_0(h_3(x)) C_0(h_1,h_2,h_3(x))|^2 \left| \F_{1221}(h_3(x), x)\right|^2 \ .
\end{align}
This equation was derived in \cite{Faulkner:2013yia,Hartman:2013mia} by calculating the bulk action with a flat boundary metric.  It should also be possible to derive it directly from \eqref{g4cc} by adding the contribution of the conformal anomaly, but we will not attempt to do so. 
In this expression, $x$ and $h_3$ are not independent parameters. (The relation between them is explained in detail and worked out numerically in \cite{Faulkner:2013yia}.) The original 3-point wormhole was parameterized by $h_3$, but it is more natural to parameterize the 4-point function by the cross ratio, so we have written $h_3$ as a function of $x$ (it also depends implicitly on $h_1$ and $h_2$). The right-hand side of \eqref{g4ccFlat} is now interpreted as the semiclassical four-point function, calculated by a saddlepoint in the conformal block expansion,
\begin{align}
G_4 &\approx |\rho_0(h_3(x)) C_0(h_1,h_2,h_3(x))|^2 \left| \F_{1221}(h_3(x), x)\right|^2 \\
&\approx \sum_p \overline{|c_{12p}|^2} \left| \F_{1221}(h_p, x)\right|^2 \notag \\
&\approx 
\left|\vcenter{\hbox{
\begin{tikzpicture}[scale=0.5]
\draw (0,1) -- (1, 0);
\draw (0, -1) -- (1,0);
\draw (1,0) -- (3,0);
\draw (3,0) -- (4,1);
\draw (3,0) -- (4,-1);
\node at (2,0.4) {$\mathbb{1}$};
\end{tikzpicture}
}}\right|^2\notag
\end{align}
In the last line we have related this to the identity block in the dual channel (see \eqref{G4identity}). The saddlepoint in the sum over $p$ defines the function $h_3(x)$ implicitly. 
The conclusion is that in the CFT, when $h_3$ is below threshold it labels an external operator in $\langle \O_1 \O_2 \O_3\rangle^2$, but when it is taken above threshold, it labels the saddlepoint weight in the sum over conformal blocks appearing in $\langle \O_1 \O_2 \O_1 \O_2\rangle$. 

If a second defect in the 3-point wormhole is taken above the black hole threshold, it disappears and is replaced by a handle on the boundary sphere, so we now have a geometry with a single defect propagating inside the solid torus. This contributes to the torus 2-point function with a single boundary. Finally, if the last defect is taken heavy, this adds another handle, so the boundary is now a single component with genus two, with no defects. This leads to an averaged interpretation of the genus-2 identity block that we will explain in section \ref{ss:genus2average}.  The process of analytically continuing defect masses above the black hole threshold is depicted in figure \ref{fig:continuation}.

\ \\

\subsection{Sphere six-point function}

We now consider the sphere six-point function of pairwise identical operators expanded in ``comb-channel'' conformal blocks 
\begin{equation}
	G_{123321}(x_i,\bar x_i) = \sum_{a,4,b}c_{12a}c_{34a}c_{43b}c_{b21}|\mathcal{F}_{123321}^{\rm comb}(h_a,h_4,h_b;x_i)|^2,
\end{equation}
where we have suggestively named one of the internal primaries $\mathcal{O}_4$ in anticipation of the connection with the sphere four-point wormhole. 
The averaged six-point function in our large-$c$ ensemble 
is then given by\footnote{It will sometimes be convenient to neglect the contribution of exchanges of the identity in studying the conformal block expansions of averaged observables. We will denote this by $\overline{(\cdots)}_{\rm heavy}$. Obviously, this is a channel-dependent definition. In particular, below we are neglecting the contribution of the exchange $a=3$, $4=\id$, which we saw in section \ref{ss:anotherDiagram} has a holographic interpretation in terms of the addition of a bulk handle.} 
\begin{equation}\label{eq:averagedSixPoint}
\begin{aligned}
	\left.\overline{G_{123321}(x_i,\bar x_i)}\right|_{\rm heavy} &= \sum_{a,4,c}\overline{c_{12a}c_{34a}c_{43b}c_{b21}}|\mathcal{F}_{123321}^{\rm comb}(h_a,h_4,h_b;x_i)|^2\\
	&= \sum_{a,4} \overline{|c_{12a}|^2}\,\overline{|c_{34a}|^2}|\mathcal{F}_{123321}^{\rm comb}(h_a,h_4,h_a;x_i)|^2\\
	&\approx \left|\int_{c-1\over 24}^\infty dh_a dh_4\, \rho_0(h_a)\rho_0(h_4)C_0(h_1,h_2,h_a)C_0(h_3,h_4,h_a)\mathcal{F}_{123321}^{\rm comb}(h_a,h_4,h_a;x_i)\right|^2,
\end{aligned}
\end{equation}
where we have again suppressed the averaged contributions of internal sub-threshold states relative to the collective contributions of the states in the black hole regime. So we see that the averaged six-point function is given by a linear combination of comb-channel Virasoro blocks with two of the internal weights correlated.

In fact, the semiclassical approximation of the averaged six-point function (\ref{eq:averagedSixPoint}) can also be realized as a vacuum block. To see this, consider the vacuum Virasoro block in the ``star channel'' $\mathcal{F}^{\rm star}_{112233}(\id,\id,\id;q_i)$. By a series of crossing moves, it can be related to the right-hand side of (\ref{eq:averagedSixPoint}) as follows:
\begin{align}
	&\vcenter{\hbox{
	\begin{tikzpicture}[scale=0.75]
	\draw[thick] (0,0) -- (0,1);
	\draw[thick](0,1) -- (-0.866*0.8,1+0.8*0.5);
	\draw[thick] (0,1) -- (0.866*0.8,1+0.8*0.5);
	\draw[thick] (0,0) -- (0.866,-1/2);
	\draw[thick] (0,0) -- (-0.866,-1/2);
	\draw[thick] (0.866,-1/2) -- (0.866+0.8*0.866,-1/2+0.8*1/2);
	\draw[thick] (0.866,-1/2) -- (0.866,-1/2-0.8);
	\draw[thick] (-0.866,-1/2) -- (-0.866-0.8*0.866,-1/2+0.8*1/2);
	\draw[thick] (-0.866,-1/2) -- (-0.866,-1/2-0.8);
	\node[left,scale=0.75] at (-0.866*0.8,1+0.8*0.5) {$2$};
	\node[right,scale=0.75] at (0.866*0.8,1+0.8*0.5) {$2$};
	\node[right,scale=0.75] at (0.866+0.8*0.866,-1/2+0.8*1/2) {$3$};
	\node[below,scale=0.75] at (0.866,-1/2-0.8) {$3$};
	\node[left, scale=0.75] at (-0.866-0.8*0.866,-1/2+0.8*1/2) {$1$};
	\node[below, scale=0.75] at (-0.866,-1/2-0.8) {$1$};
	\node[left, scale=0.75] at (0,1/2) {$\id$};
	\node[above, scale=0.75] at (1/2*0.866,-1/2*1/2) {$\id$};
	\node[above, scale=0.75] at (-1/2*0.866,-1/2*1/2) {$\id$};
	\end{tikzpicture}
	}}  
	= 
	\vcenter{\hbox{
	\begin{tikzpicture}[scale=.75]
	\draw[thick] (-1/2,0) -- (1/2,0);
	\draw[thick] (-1/2,0) -- (-1/2, 0.8);
	\draw[thick] (1/2,0) -- (1/2, 0.8);
	\draw[thick] (-1/2,0) -- (-3/2,0);
	\draw[thick] (-3/2,0) -- (-3/2-0.8*1/2,0.8*0.866);
	\draw[thick] (-3/2,0) -- (-3/2-0.8*1/2,-0.8*0.866);
	\draw[thick] (1/2,0) -- (3/2,0);
	\draw[thick] (3/2,0) -- (3/2+0.8*1/2,0.8*0.866);
	\draw[thick] (3/2,0) -- (3/2+0.8*1/2,-0.8*0.866);
	\node[left,scale=0.75] at (-3/2-0.8*1/2,0.8*0.866) {$1$};
	\node[left,scale=0.75] at (-3/2-0.8*1/2,-0.8*0.866) {$1$};
	\node[right,scale=0.75] at (3/2+0.8*1/2,0.8*0.866) {$3$};
	\node[right,scale=0.75] at (3/2+0.8*1/2,-0.8*0.866) {$3$};
	\node[above,scale=0.75] at (-1/2, 0.8) {$2$};
	\node[above,scale=0.75] at (1/2,0.8) {$2$};
	\node[below,scale=0.75] at (0,0) {$2$};
	\node[above,scale=0.75] at (-1,0) {$\id$};
	\node[above,scale=0.75] at (1,0) {$\id$};
	\end{tikzpicture}
	}}
\\
	&\qquad \qquad = \int_{c-1\over 24}^\infty dh_a\, 
	\rho_0(h_a)C_0(h_1,h_2,h_a)
	\vcenter{\hbox{
	\begin{tikzpicture}[scale=.75]
	\draw[thick] (-1/2,0) -- (1/2,0);
	\draw[thick] (-1/2,0) -- (-1/2,1);
	\draw[thick] (-1/2,1) -- (-1/2-0.8*0.866,1+0.8*1/2);
	\draw[thick] (-1/2,1) -- (-1/2+0.8*0.866,1+0.8*1/2);
	\draw[thick] (-1/2,0) -- (-3/2,0);
	\node[left,scale=0.75] at (-3/2,0) {$1$};
	\node[above,scale=0.75] at (-1/2-0.8*0.866,1+0.8*1/2) {$1$};
	\node[above,scale=0.75] at (-1/2+0.8*0.866,1+0.8*1/2) {$2$};
	\node[left,scale=0.75] at (-1/2,1/2) {$a$};
	\node[below,scale=0.75] at (0,0) {$2$};
	\draw[thick] (1/2,0) -- (1/2,0.8);
	\node[above,scale=0.75] at (1/2,0.8) {$2$};
	\draw[thick] (1/2,0) -- (3/2,0);
	\node[below,scale=0.75] at (1,0) {$\id$};
	\draw[thick] (3/2,0) -- (3/2+0.8*1/2,0.8*0.866);
	\draw[thick] (3/2,0) -- (3/2+0.8*1/2,-0.8*0.866);
	\node[right,scale=0.75] at (3/2+0.8*1/2,0.8*0.866) {$3$};
	\node[right,scale=0.75] at (3/2+0.8*1/2,-0.8*0.866) {$3$};
	\end{tikzpicture}
	}}\notag
	\\
	&\qquad \qquad = \int_{c-1\over 24}^\infty dh_a\, 
	\rho_0(h_a)C_0(h_1,h_2,h_a)
	\vcenter{\hbox{
	\begin{tikzpicture}[scale=.75]
	\draw[thick] (0,0) -- (1,0);
	\node[above,scale=0.75] at (1/2,0) {$\id$};
	\draw[thick] (0,0) -- (0,1);
	\node[left,scale=0.75] at (0,1/2) {$a$};
	\draw[thick] (0,0) -- (0,-1);
	\node[left,scale=0.75] at (0,-1/2) {$a$};
	\draw[thick] (0,1) -- (0.8*0.866,1+0.8*1/2);
	\draw[thick] (0,1) -- (-0.8*0.866,1+0.8*1/2);
	\node[above,scale=0.75] at (-0.8*0.866,1+0.8*1/2) {$1$};
	\node[above,scale=0.75] at (0.8*0.866,1+0.8*1/2) {$2$};
	\draw[thick] (0,-1) -- (0.8*0.866,-1-0.8*1/2);
	\draw[thick] (0,-1) -- (-0.8*0.866,-1-0.8*1/2);
	\node[below,scale=0.75] at (-0.8*0.866,-1-0.8*1/2) {$1$};
	\node[below,scale=0.75] at (0.8*0.866,-1-0.8*1/2) {$2$};
	\draw[thick] (1,0) -- (1+0.8*1/2,0.8*0.866);
	\draw[thick] (1,0) -- (1+0.8*1/2,-0.8*0.866);
	\node[right,scale=0.75] at (1+0.8*1/2,0.8*0.866) {$3$};
	\node[right,scale=0.75] at (1+0.8*1/2,-0.8*0.866) {$3$};
	\end{tikzpicture}
	}}\notag
	\\
	&\qquad \qquad = \int_{c-1\over 24}^\infty dh_4 dh_a\, 
	\rho_0(h_a)\rho_0(h_4) C_0(h_1,h_2,h_a)C_0(h_3,h_4,h_a)
	\vcenter{\hbox{
	\begin{tikzpicture}[scale=.75]
	\draw[thick] (-1/2,0) -- (1/2,0);
	\node[below,scale=0.75] at (0,0) {$4$};
	\draw[thick] (-1/2,0) -- (-3/2,0);
	\node[above,scale=0.75] at (-1,0) {$a$};
	\draw[thick] (-1/2,0) -- (-1/2,1*0.8);
	\node[above,scale=0.75] at (-1/2,0.8) {$3$};
	\draw[thick] (-3/2,0) -- (-3/2,0.8);
	\node[above,scale=0.75] at (-3/2,0.8) {$2$};
	\draw[thick] (-3/2,0) -- (-3/2-0.8,0);
	\node[left,scale=0.75] at (-3/2-0.8,0) {$1$};
	\draw[thick] (1/2,0) -- (1/2,0.8);
	\node[above,scale=0.75] at (1/2,0.8) {$3$};
	\draw[thick] (1/2,0) -- (3/2,0);
	\node[above,scale=0.75] at (1,0) {$a$};
	\draw[thick] (3/2,0) -- (3/2,0.8);
	\node[above,scale=0.75] at (3/2,0.8) {$2$};
	\draw[thick] (3/2,0) -- (3/2+0.8,0);
	\node[right,scale=0.75] at (3/2+0.8,0) {$1$};
	\end{tikzpicture}
	}}\notag
\end{align}
The first and the third crossing moves were trivial.  Note the presence of the correlated exchanges of the primaries $\mathcal{O}_a$. This is exactly the averaged six-point function in the semiclassical limit given in (\ref{eq:averagedSixPoint}):\footnote{The map between the cross-ratios $x_i$ in the comb channel and the plumbing parameters $q_i$ that are natural in the star channel is nontrivial but unimportant for our present purposes.} 
\begin{equation}
	\left.\overline{G_{123321}(x_i,\bar x_i)}\right|_{\rm heavy} \approx \left|\mathcal{F}^{\rm star}_{112233}(\id,\id,\id;q_i)\right|^2.
\end{equation}
Similarly to the case of the averaged four-point function, one could have arrived at this conclusion by starting directly in the star channel; the only term in the OPE sum that survives the averaging in the semiclassical limit is that for which all internal operators are the identity. 

The result for the averaged six-point function can also be understood by analytic continuation of the amplitude for the sphere four-point wormhole (or equivalently of the connected part of the averaged product of sphere four-point functions) to the regime where one of the four external operators is taken above the black hole threshold. That is, we start with the wormhole that contributes to 
\begin{align}
\overline{| \langle \O_1 \O_2 \O_3 \O_4\rangle|^2} 
\end{align}
when all four operators are sub-threshold, then increase the weight of $\O_4$ above the threshold. 
As in the case of the three-point wormholes, when one of the external operators is taken to the black hole regime, the formerly disconnected boundaries are glued, leading to a configuration with three defects propagating through the bulk of a hyperbolic ball:
\begin{align}
\hspace{-0.5in}\vcenter{\hbox{\includegraphics[width=1.5in]{figures/G4G4.pdf}}} \qquad \Rightarrow \qquad
\vcenter{\hbox{\includegraphics[width=1in]{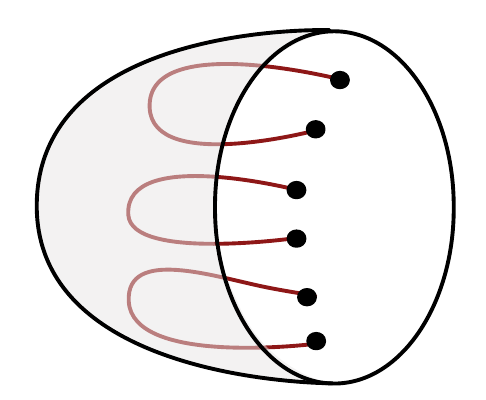}}}
\end{align}
 In the CFT, the gluing of the two boundaries corresponds to a sum over $\O_4$, which is now a black hole state propagating in the intermediate channel. The saddlepoint weight in the sum over black holes now takes the place of $h_4$ in the original wormhole.

\subsection{Genus-two partition function}\label{ss:genus2average}

This procedure is slightly more elaborate when the observable has many intermediate channels in which the identity operator can propagate. To illustrate the point, consider the averaged genus-two partition function. It can be computed by applying the OPE for instance in the sunset channel 
\begin{equation}
	Z_{g=2} = \sum_{p,q,r}c_{pqr}c^*_{pqr}\left|\vcenter{\hbox{
	\begin{tikzpicture}[scale=0.5]
	\draw[thick] (0,0) circle (3/2);
	\draw[thick] (-3/2,0) -- (3/2,0);
	\node[above] at (0,3/2) {$p$};
	\node[above] at (0,0) {$q$};
	\node[above] at (0,-3/2) {$r$};
	\end{tikzpicture}
	}}\right|^2. 
\end{equation}
Sub-threshold operators are subleading, so we will ignore their contributions to averaged quantities in the following discussion. 
When computing the average, one must take care to account for the contributions of the identity:
\begin{equation}\label{genus2sunset}
	\hspace{-0.6cm}\overline{Z_{g=2}} = \left|\vcenter{\hbox{
	\begin{tikzpicture}[scale=0.5]
	\draw[thick] (0,0) circle (3/2);
	\draw[thick] (-3/2,0) -- (3/2,0);
	\node[above] at (0,3/2) {$\id$};
	\node[above] at (0,0) {$\id$};
	\node[above] at (0,-3/2) {$\id$};
	\end{tikzpicture}
	}}\right|^2
	+ \sum_{p\ne \id}\left(
	\left|\vcenter{\hbox{
	\begin{tikzpicture}[scale=0.5]
	\draw[thick] (0,0) circle (3/2);
	\draw[thick] (-3/2,0) -- (3/2,0);
	\node[above] at (0,3/2) {$p$};
	\node[above] at (0,0) {$p$};
	\node[above] at (0,-3/2) {$\id$};
	\end{tikzpicture}
	}}\right|^2+
	\left|\vcenter{\hbox{
	\begin{tikzpicture}[scale=0.5]
	\draw[thick] (0,0) circle (3/2);
	\draw[thick] (-3/2,0) -- (3/2,0);
	\node[above] at (0,3/2) {$\id$};
	\node[above] at (0,0) {$p$};
	\node[above] at (0,-3/2) {$p$};
	\end{tikzpicture}
	}}\right|^2+
	\left|\vcenter{\hbox{
	\begin{tikzpicture}[scale=0.5]
	\draw[thick] (0,0) circle (3/2);
	\draw[thick] (-3/2,0) -- (3/2,0);
	\node[above] at (0,3/2) {$p$};
	\node[above] at (0,0) {$\id$};
	\node[above] at (0,-3/2) {$p$};
	\end{tikzpicture}
	}}\right|^2
	\right) + \sum_{p,q,r\ne \id}\overline{|c_{pqr}|^2}
	\left|\vcenter{\hbox{
	\begin{tikzpicture}[scale=0.5]
	\draw[thick] (0,0) circle (3/2);
	\draw[thick] (-3/2,0) -- (3/2,0);
	\node[above] at (0,3/2) {$p$};
	\node[above] at (0,0) {$q$};
	\node[above] at (0,-3/2) {$r$};
	\end{tikzpicture}
	}}\right|^2
\end{equation}
In fact, all of these terms can be realized as various identity blocks. For example, the second term above is well-approximated in our ensemble by
\begin{equation}
	\sum_{p\ne \id}
	\left|\vcenter{\hbox{
	\begin{tikzpicture}[scale=0.5]
	\draw[thick] (0,0) circle (3/2);
	\draw[thick] (-3/2,0) -- (3/2,0);
	\node[above] at (0,3/2) {$p$};
	\node[above] at (0,0) {$p$};
	\node[above] at (0,-3/2) {$\id$};
	\end{tikzpicture}
	}}\right|^2 \approx
	\left|\int dh_p\, \rho_0(h_p) \vcenter{\hbox{
	\begin{tikzpicture}[scale=0.5]
	\draw[thick] (0,3/2) circle (1);
	\draw[thick] (0,1/2) -- (0,-1/2);
	\draw[thick] (0,-3/2) circle (1);
	\node[left] at (-1,3/2) {$p$};
	\node[left] at (0,0) {$\id$};
	\node[left] at (-1,-3/2) {$\id$}; 
	\end{tikzpicture}
	}}\right|^2 =
	\left|\vcenter{\hbox{
	\begin{tikzpicture}[scale=0.5]
	\draw[thick] (0,3/2) circle (1);
	\draw[thick] (0,1/2) -- (0,-1/2);
	\draw[thick] (0,-3/2) circle (1);
	\node[left] at (-1,3/2) {$\id$};
	\node[left] at (0,0) {$\id$};
	\node[left] at (-1,-3/2) {$\id$}; 
	\node[right] at (1,3/2) {$'$};
	\end{tikzpicture}
	}}\right|^2,
\end{equation} 
where the prime denotes a modular S-transformation on the corresponding torus factor.
Similarly, the fourth term can be expressed as an identity block by first exchanging the bottom line and middle line, then following the same procedure. This mixes up the two sub-tori, so to distinguish this channel from the others we will draw this block horizontally, as
\begin{equation}
	\sum_{p\ne\id} \left|\vcenter{\hbox{
	\begin{tikzpicture}[scale=0.5]
	\draw[thick] (0,0) circle (3/2);
	\draw[thick] (-3/2,0) -- (3/2,0);
	\node[above] at (0,3/2) {$p$};
	\node[above] at (0,0) {$\id$};
	\node[above] at (0,-3/2) {$p$};
	\end{tikzpicture}
	}} \right|^2 \approx 
	\left| \vcenter{\hbox{
	\begin{tikzpicture}[scale=0.5]
	\draw[thick] (0,0) circle (1);
	\draw[thick] (3,0) circle (1);
	\draw[thick] (1,0) -- (2,0);
	\node[above] at (0,-2) {$\id$};
	\node[above] at (3,-2) {$\id$};
	\node[above] at (3/2,-1) {$\id$};	
	\end{tikzpicture}
	}}\right|^2
\end{equation}
Finally, the last term is computed in terms of a dumbbell channel vacuum block using the crossing relation (\ref{eq:genus2IdentityCrossing}) as
\begin{equation}
\hspace{-0.6cm}
	\sum_{p,q,r\ne \id}\overline{|c_{pqr}|^2}
	\left|\vcenter{\hbox{
	\begin{tikzpicture}[scale=0.5]
	\draw[thick] (0,0) circle (3/2);
	\draw[thick] (-3/2,0) -- (3/2,0);
	\node[above] at (0,3/2) {$p$};
	\node[above] at (0,0) {$q$};
	\node[above] at (0,-3/2) {$r$};
	\end{tikzpicture}
	}}\right|^2 \approx
	\left|\int dh_p dh_q dh_r \, \rho_0(h_p)\rho_0(h_q)\rho_0(h_r)C_0(h_p,h_q,h_r) \vcenter{\hbox{
	\begin{tikzpicture}[scale=0.5]
	\draw[thick] (0,0) circle (3/2);
	\draw[thick] (-3/2,0) -- (3/2,0);
	\node[above] at (0,3/2) {$p$};
	\node[above] at (0,0) {$q$};
	\node[above] at (0,-3/2) {$r$};
	\end{tikzpicture}
	}}\right|^2
	= \left|\vcenter{\hbox{
	\begin{tikzpicture}[scale=0.5]
	\draw[thick] (0,3/2) circle (1);
	\draw[thick] (0,1/2) -- (0,-1/2);
	\draw[thick] (0,-3/2) circle (1);
	\node[left] at (-1,3/2) {$\id$};
	\node[left] at (0,0) {$\id$};
	\node[left] at (-1,-3/2) {$\id$}; 
	\node[right] at (1,3/2) {$'$};
	\node[right] at (1,-3/2) {$'$};
	\end{tikzpicture}
	}}\right|^2
\end{equation}
We have dropped the extra contractions that appear when some of $p,q,r$ are identical because they have fewer $\rho_0$ factors and are therefore subleading. Assembling the pieces, this gives the following for the averaged genus-two partition function in the semiclassical limit
\begin{equation}\label{eq:averagedGenusTwoIdentityBlocks}
\begin{aligned}
	\overline{Z_{g=2}} =& \, \left|\vcenter{\hbox{
	\begin{tikzpicture}[scale=0.5]
	\draw[thick] (0,3/2) circle (1);
	\draw[thick] (0,1/2) -- (0,-1/2);
	\draw[thick] (0,-3/2) circle (1);
	\node[left] at (-1,3/2) {$\id$};
	\node[left] at (0,0) {$\id$};
	\node[left] at (-1,-3/2) {$\id$}; 
	\end{tikzpicture}
	}}\right|^2
	+
	\left|\vcenter{\hbox{
	\begin{tikzpicture}[scale=0.5]
	\draw[thick] (0,3/2) circle (1);
	\draw[thick] (0,1/2) -- (0,-1/2);
	\draw[thick] (0,-3/2) circle (1);
	\node[left] at (-1,3/2) {$\id$};
	\node[left] at (0,0) {$\id$};
	\node[left] at (-1,-3/2) {$\id$}; 
	\node[right] at (1,3/2) {$'$};
	\end{tikzpicture}
	}}\right|^2+
	\left|\vcenter{\hbox{
	\begin{tikzpicture}[scale=0.5]
	\draw[thick] (0,3/2) circle (1);
	\draw[thick] (0,1/2) -- (0,-1/2);
	\draw[thick] (0,-3/2) circle (1);
	\node[left] at (-1,3/2) {$\id$};
	\node[left] at (0,0) {$\id$};
	\node[left] at (-1,-3/2) {$\id$}; 
	\node[right] at (1,-3/2) {$'$};
	\end{tikzpicture}
	}}\right|^2+
	\left|\vcenter{\hbox{
	\begin{tikzpicture}[scale=0.5]
	\draw[thick] (0,3/2) circle (1);
	\draw[thick] (0,1/2) -- (0,-1/2);
	\draw[thick] (0,-3/2) circle (1);
	\node[left] at (-1,3/2) {$\id$};
	\node[left] at (0,0) {$\id$};
	\node[left] at (-1,-3/2) {$\id$}; 
	\node[right] at (1,3/2) {$'$};
	\node[right] at (1,-3/2) {$'$};
	\end{tikzpicture}
	}}\right|^2\\
	& \, + 
		\left| \vcenter{\hbox{
	\begin{tikzpicture}[scale=0.5]
	\draw[thick] (0,0) circle (1);
	\draw[thick] (3,0) circle (1);
	\draw[thick] (1,0) -- (2,0);
	\node[above] at (0,-2) {$\id$};
	\node[above] at (3,-2) {$\id$};
	\node[above] at (3/2,-1) {$\id$};	
	\end{tikzpicture}
	}}\right|^2
	\end{aligned}
\end{equation}

It is straightforward to check that one gets the same answer by instead performing the ensemble average in the dumbbell channel. Separating out the identity, 
\begin{equation}\hspace{-0.8cm}
	\overline{Z_{g=2}} = 
	\left|\vcenter{\hbox{
	\begin{tikzpicture}[scale=0.5]
	\draw[thick] (0,3/2) circle (1);
	\draw[thick] (0,1/2) -- (0,-1/2);
	\draw[thick] (0,-3/2) circle (1);
	\node[left] at (-1,3/2) {$\id$};
	\node[left] at (0,0) {$\id$};
	\node[left] at (-1,-3/2) {$\id$}; 
	\end{tikzpicture}
	}}\right|^2
	+ \sum_{p\ne\id}\left(
	\left|\vcenter{\hbox{
	\begin{tikzpicture}[scale=0.5]
	\draw[thick] (0,3/2) circle (1);
	\draw[thick] (0,1/2) -- (0,-1/2);
	\draw[thick] (0,-3/2) circle (1);
	\node[left] at (-1,3/2) {$p$};
	\node[left] at (0,0) {$\id$};
	\node[left] at (-1,-3/2) {$\id$}; 
	\end{tikzpicture}
	}}\right|^2+
	\left|\vcenter{\hbox{
	\begin{tikzpicture}[scale=0.5]
	\draw[thick] (0,3/2) circle (1);
	\draw[thick] (0,1/2) -- (0,-1/2);
	\draw[thick] (0,-3/2) circle (1);
	\node[left] at (-1,3/2) {$\id$};
	\node[left] at (0,0) {$\id$};
	\node[left] at (-1,-3/2) {$p$}; 
	\end{tikzpicture}
	}}\right|^2\right) + \sum_{p,q\ne\id}
	\left|\vcenter{\hbox{
	\begin{tikzpicture}[scale=0.5]
	\draw[thick] (0,3/2) circle (1);
	\draw[thick] (0,1/2) -- (0,-1/2);
	\draw[thick] (0,-3/2) circle (1);
	\node[left] at (-1,3/2) {$p$};
	\node[left] at (0,0) {$\id$};
	\node[left] at (-1,-3/2) {$q$}; 
	\end{tikzpicture}
	}}\right|^2 + \sum_{p,q,r\ne \id}\overline{c_{ppq}c_{rrq}}
	\left|\vcenter{\hbox{
	\begin{tikzpicture}[scale=0.5]
	\draw[thick] (0,3/2) circle (1);
	\draw[thick] (0,1/2) -- (0,-1/2);
	\draw[thick] (0,-3/2) circle (1);
	\node[left] at (-1,3/2) {$p$};
	\node[left] at (0,0) {$q$};
	\node[left] at (-1,-3/2) {$r$}; 
	\end{tikzpicture}
	}}\right|^2.
\end{equation}
The first four terms are the dumbbell channel vacuum block summed over the action of modular S-transformations on the sub-tori. These are also the first four terms in (\ref{eq:averagedGenusTwoIdentityBlocks}). For the last term, the ensemble average gives 
\begin{align}\label{eq:dumbbellNonIdAverage}
	\sum_{p,q,r\ne \id}\overline{c_{ppq}c_{rrq}}
	\left|\vcenter{\hbox{
	\begin{tikzpicture}[scale=0.5]
	\draw[thick] (0,3/2) circle (1);
	\draw[thick] (0,1/2) -- (0,-1/2);
	\draw[thick] (0,-3/2) circle (1);
	\node[left] at (-1,3/2) {$p$};
	\node[left] at (0,0) {$q$};
	\node[left] at (-1,-3/2) {$r$}; 
	\end{tikzpicture}
	}}\right|^2 &\approx \left|\int dh_pdh_q\, \rho_0(h_p)\rho_0(h_q)C_0(h_p,h_p,h_q) \vcenter{\hbox{
	\begin{tikzpicture}[scale=0.5]
	\draw[thick] (0,3/2) circle (1);
	\draw[thick] (0,1/2) -- (0,-1/2);
	\draw[thick] (0,-3/2) circle (1);
	\node[left] at (-1,3/2) {$p$};
	\node[left] at (0,0) {$q$};
	\node[left] at (-1,-3/2) {$p$}; 
	\end{tikzpicture}
	}}\right|^2\\
	&=
		\sum_{p\ne\id} \left|\vcenter{\hbox{
	\begin{tikzpicture}[scale=0.5]
	\draw[thick] (0,0) circle (3/2);
	\draw[thick] (-3/2,0) -- (3/2,0);
	\node[above] at (0,3/2) {$p$};
	\node[above] at (0,0) {$\id$};
	\node[above] at (0,-3/2) {$p$};
	\end{tikzpicture}
	}} \right|^2
\end{align}
This exactly reproduces the last term of (\ref{eq:averagedGenusTwoIdentityBlocks}). Thus we see that the averaged genus-two partition function is approximately crossing invariant in the semiclassical limit, in the sense that the average acts the same way in either the sunset or the dumbbell channel.

Similarly to the case of the other single-boundary observables we've considered, the contribution (\ref{eq:dumbbellNonIdAverage}) to the averaged genus-two partition function can be understood via analytic continuation of a two-boundary wormhole amplitude. Note in particular the pairing between internal operators propagating along the sub-tori induced by the average in (\ref{eq:dumbbellNonIdAverage}). In particular, consider the averaged product of torus one-point functions, dual to a two-boundary wormhole whose boundaries are tori with a single operator insertion, studied in section \ref{sss:thermalOnePointWormhole}. When the dimension of the external operator is continued above the black hole threshold the disparate boundaries are glued, leading to a genus-two handlebody. 

Each genus-2 identity block dominates in a different region of moduli space, and each corresponds to a different handlebody geometry in the bulk. It was already known that the genus-2 partition function of 3D gravity is approximated by a sum over identity blocks; what we have shown is that this sum has a natural interpretation as an ensemble average.

\subsection{Higher-genus partition function}

The statement that the averaged genus-$g$ partition function is given by a suitable vacuum block to leading order in the semiclassical limit holds for all genera $g$. To see this, consider the genus-$g$ partition function, computed for example in the ``cascade channel'' (which generalizes the genus-two sunset channel):\footnote{Below we are suppressing all dependence on the moduli of the Riemann surface, which is left implicit in the blocks.}
\begin{equation}
\begin{aligned}
	Z_{\text{genus-}g} =& \, \sum_{\substack{1,2,\ldots,g+1 \\ p_1,p_2,\ldots,p_{g-2}\\ q_1,q_2,\ldots,q_{g-2}}}c_{12p_1}c^*_{12q_1}c_{p_1 3 p_2}c^*_{q_1 3 q_2}\cdots c_{p_{g-3}(g-1)p_{g-2}}c^*_{q_{g-3}(g-1)q_{g-2}}c_{p_{g-2}(g)(g+1)}c^*_{q_{g-2}(g)(g+1)}\\
	& \, \times 
	\left|\vcenter{\hbox{
	\begin{tikzpicture}[scale=0.75]
	\draw[thick] (5/4,0) arc (0:180:5/4 and 3/2);
	\node at (0,-3/8) {$\ldots$};
	\draw[thick] (5/4,-3/4) arc (0:-180:5/4 and 3/2);
	\draw[thick] (-.93,3/2-1/2) -- (.93,3/2-1/2);
	\draw[thick] (-1.185,3/2-1/2-1/2) -- (1.185,3/2-1/2-1/2);
	\draw[thick] (-1.185,-3/4-1/2) -- (1.185,-3/4-1/2);
	\draw[thick] (-.93,-3/4-1/2-1/2) -- (.93,-3/4-1/2-1/2);
	\node[scale=0.75, above] at (0,3/2) {$1$};
	\node[scale=0.75, left] at (-1.0575,3/4) {$p_1$};
	\node[scale=0.75, left] at (-5/4,1/4) {$p_2$};
	\node[scale=0.75, above] at (0,1) {$2$};
	\node[scale=0.75, above] at (0,1/2) {$3$};
	\node[scale=0.75, right] at (1.0575,3/4) {$q_1$};
	\node[scale=0.75, right] at (5/4,1/4) {$q_2$};
	\node[scale=0.75, above] at (0,-3/4-1/2) {$g-1$};
	\node[scale=0.75, above] at (0,-3/4-1) {$g$};
	\node[scale=0.75, above] at (0,-3/4-3/2) {$g+1$};
	\node[scale=0.75, left] at (-5/4,-3/4-1/4) {$p_{g-3}$};
	\node[scale=0.75, left] at (-1.0575,-3/4-3/4) {$p_{g-2}$};
	\node[scale=0.75, right] at (5/4,-3/4-1/4) {$q_{g-3}$};
	\node[scale=0.75, right] at (1.0575,-3/4-3/4) {$q_{g-2}$};
	\end{tikzpicture}
	}}\right|^2.
\end{aligned}
\end{equation}
In the semiclassical limit, the Gaussian contribution to the ensemble average of the genus-$g$ partition function is given by the Wick contraction\footnote{As the previous example demonstrated, already at genus-two it is becoming unwieldy to keep track of all of the distinct exchanges of the identity operator. For this reason, in what follows we will focus on the terms in the average for which all exchanged internal primaries are nontrivial, which we denote by $\left.\overline{(\cdots)}\right|_{\text{heavy}}$ as before.}
\begin{align}\label{eq:genusgAverage}
	&\left.\overline{Z_{\text{genus-}g}}\right|_{\text{heavy}}\\
	&\qquad \approx  \, \sum_{\substack{1,2,\ldots,g+1 \\ p_1,p_2,\ldots,p_{g-2}}}\overline{|c_{12p_1}|^2}\, \overline{|c_{p_1 3 p_2}|^2}\cdots\overline{|c_{p_{g-3}(g-1)p_{g-2}}|^2}\, \overline{|c_{p_{g-2}(g)(g+1)}|^2} \, \left|\vcenter{\hbox{
	\begin{tikzpicture}[scale=0.75]
	\draw[thick] (5/4,0) arc (0:180:5/4 and 3/2);
	\node at (0,-3/8) {$\ldots$};
	\draw[thick] (5/4,-3/4) arc (0:-180:5/4 and 3/2);
	\draw[thick] (-.93,3/2-1/2) -- (.93,3/2-1/2);
	\draw[thick] (-1.185,3/2-1/2-1/2) -- (1.185,3/2-1/2-1/2);
	\draw[thick] (-1.185,-3/4-1/2) -- (1.185,-3/4-1/2);
	\draw[thick] (-.93,-3/4-1/2-1/2) -- (.93,-3/4-1/2-1/2);
	\node[scale=0.75, above] at (0,3/2) {$1$};
	\node[scale=0.75, left] at (-1.0575,3/4) {$p_1$};
	\node[scale=0.75, left] at (-5/4,1/4) {$p_2$};
	\node[scale=0.75, above] at (0,1) {$2$};
	\node[scale=0.75, above] at (0,1/2) {$3$};
	\node[scale=0.75, right] at (1.0575,3/4) {$p_1$};
	\node[scale=0.75, right] at (5/4,1/4) {$p_2$};
	\node[scale=0.75, above] at (0,-3/4-1/2) {$g-1$};
	\node[scale=0.75, above] at (0,-3/4-1) {$g$};
	\node[scale=0.75, above] at (0,-3/4-3/2) {$g+1$};
	\node[scale=0.75, left] at (-5/4,-3/4-1/4) {$p_{g-3}$};
	\node[scale=0.75, left] at (-1.0575,-3/4-3/4) {$p_{g-2}$};
	\node[scale=0.75, right] at (5/4,-3/4-1/4) {$p_{g-3}$};
	\node[scale=0.75, right] at (1.0575,-3/4-3/4) {$p_{g-2}$};
	\end{tikzpicture}
	}}\right|^2 \notag\\
	&\qquad \approx\, \Bigg|\int_{c-1\over 24}^\infty dh_1 \rho_0(h_1)\cdots dh_{g+1} \rho_0(h_{g+1}) dh_{p_1}\rho_0(h_{p_1}) \cdots dh_{p_{g-2}}\rho_0(h_{p_{g-2}}) C_0(h_1,h_2,h_{p_1})\notag\\
	&\qquad \quad  \times C_0(h_{p_1},h_3,h_{p_2})\cdots C_0(h_{p_{g-3}},h_{g-1},h_{p_{g-2}})C_0(h_{p_{g-2}},h_g,h_{g+1})\vcenter{\hbox{
	\begin{tikzpicture}[scale=0.75]
	\draw[thick] (5/4,0) arc (0:180:5/4 and 3/2);
	\node at (0,-3/8) {$\ldots$};
	\draw[thick] (5/4,-3/4) arc (0:-180:5/4 and 3/2);
	\draw[thick] (-.93,3/2-1/2) -- (.93,3/2-1/2);
	\draw[thick] (-1.185,3/2-1/2-1/2) -- (1.185,3/2-1/2-1/2);
	\draw[thick] (-1.185,-3/4-1/2) -- (1.185,-3/4-1/2);
	\draw[thick] (-.93,-3/4-1/2-1/2) -- (.93,-3/4-1/2-1/2);
	\node[scale=0.75, above] at (0,3/2) {$1$};
	\node[scale=0.75, left] at (-1.0575,3/4) {$p_1$};
	\node[scale=0.75, left] at (-5/4,1/4) {$p_2$};
	\node[scale=0.75, above] at (0,1) {$2$};
	\node[scale=0.75, above] at (0,1/2) {$3$};
	\node[scale=0.75, right] at (1.0575,3/4) {$p_1$};
	\node[scale=0.75, right] at (5/4,1/4) {$p_2$};
	\node[scale=0.75, above] at (0,-3/4-1/2) {$g-1$};
	\node[scale=0.75, above] at (0,-3/4-1) {$g$};
	\node[scale=0.75, above] at (0,-3/4-3/2) {$g+1$};
	\node[scale=0.75, left] at (-5/4,-3/4-1/4) {$p_{g-3}$};
	\node[scale=0.75, left] at (-1.0575,-3/4-3/4) {$p_{g-2}$};
	\node[scale=0.75, right] at (5/4,-3/4-1/4) {$p_{g-3}$};
	\node[scale=0.75, right] at (1.0575,-3/4-3/4) {$p_{g-2}$};
	\end{tikzpicture}
	}}\Bigg|^2\notag
\end{align}
All other Wick contractions lead to fewer factors of $\rho_0$ and thus are suppressed relative to the above contribution in the semiclassical limit.
This in turn is equivalent to a genus-$g$ Virasoro vacuum block in the ``bead channel'' (which generalizes the genus-two dumbbell channel). To see this, we start with the bead channel vacuum block and apply $g$ modular S transformations on the $g$ sub-tori\footnote{As before, we have added primes to the dumbbell-channel vacuum block on the left-hand side to denote the relative modular S-transformations compared to the block on the right-hand side.}
\begin{equation}
\begin{aligned}
	& \, \vcenter{\hbox{
	\begin{tikzpicture}[scale=.75]
	\draw[thick] (0,0) circle (1);
	\draw[thick] (1,0) -- (2,0);
	\draw[thick] (3,0) circle (1); 
	\draw[thick] (4,0) -- (5,0);
	\node at (5.5,0) {$\ldots$};
	\draw[thick] (6,0) -- (7,0);
	\draw[thick] (8,0) circle (1);
	\draw[thick] (9,0) -- (10,0);
	\draw[thick] (11,0) circle (1);
	\node[below] at (0,-1) {$\id$};
	\node[below] at (3,-1) {$\id$};
	\node[above] at (3,1) {$\id$};
	\node[below] at (8,-1) {$\id$};
	\node[above] at (8,1) {$\id$};
	\node[below] at (11,-1) {$\id$};
	\node[below] at (1.5,0) {$\id$};
	\node[below] at (4.5,0) {$\id$};
	\node[below] at (6.5,0) {$\id$};
	\node[below] at (9.5,0) {$\id$};
	\node[above] at (1,1/2) {$'$};
	\node[above] at (4,1/2) {$'$};
	\node[above] at (9,1/2) {$'$};
	\node[above] at (12,1/2) {$'$};
	\end{tikzpicture}
	}}\\
	=& \, \int_{c-1\over 24}^\infty dh_1 \rho_0(h_1) dh_{p_1}\rho_0(h_{p_1})\cdots dh_{p_{g-2}}\rho_0(h_{p_{g-2}})dh_{g+1}\rho_0(h_{g+1})\\
	&\, \times 
	\vcenter{\hbox{
	\begin{tikzpicture}[scale=.75]
	\draw[thick] (0,0) circle (1);
	\draw[thick] (1,0) -- (2,0);
	\draw[thick] (3,0) circle (1); 
	\draw[thick] (4,0) -- (5,0);
	\node at (5.5,0) {$\ldots$};
	\draw[thick] (6,0) -- (7,0);
	\draw[thick] (8,0) circle (1);
	\draw[thick] (9,0) -- (10,0);
	\draw[thick] (11,0) circle (1);
	\node[below] at (0,-1) {$1$};
	\node[below] at (3,-1) {$p_1$};
	\node[above] at (3,1) {$p_1$};
	\node[below] at (8,-1) {$p_{g-2}$};
	\node[above] at (8,1) {$p_{g-2}$};
	\node[below] at (11,-1) {$g+1$};
	\node[below] at (1.5,0) {$\id$};
	\node[below] at (4.5,0) {$\id$};
	\node[below] at (6.5,0) {$\id$};
	\node[below] at (9.5,0) {$\id$};
	\end{tikzpicture}
	}}
\end{aligned}
\end{equation}
One then applies fusion transformations to the remaining $g-1$ internal legs involving the exchange of the identity, which introduces integrals over the internal weights $h_2$ through $h_{g}$, to arrive at the right-hand side of (\ref{eq:genusgAverage}). So we can express heavy contributions to the averaged genus-$g$ partition function as a Virasoro vacuum block in the bead channel to leading order in the semiclassical limit
\begin{equation}
	\left.\overline{Z_{\text{genus-}g}}\right|_{\text{heavy}} \approx  \left|\vcenter{\hbox{
	\begin{tikzpicture}[scale=.75]
	\draw[thick] (0,0) circle (1);
	\draw[thick] (1,0) -- (2,0);
	\draw[thick] (3,0) circle (1); 
	\draw[thick] (4,0) -- (5,0);
	\node at (5.5,0) {$\ldots$};
	\draw[thick] (6,0) -- (7,0);
	\draw[thick] (8,0) circle (1);
	\draw[thick] (9,0) -- (10,0);
	\draw[thick] (11,0) circle (1);
	\node[below] at (0,-1) {$\id$};
	\node[below] at (3,-1) {$\id$};
	\node[above] at (3,1) {$\id$};
	\node[below] at (8,-1) {$\id$};
	\node[above] at (8,1) {$\id$};
	\node[below] at (11,-1) {$\id$};
	\node[below] at (1.5,0) {$\id$};
	\node[below] at (4.5,0) {$\id$};
	\node[below] at (6.5,0) {$\id$};
	\node[below] at (9.5,0) {$\id$};
	\node[above] at (1,1/2) {$'$};
	\node[above] at (4,1/2) {$'$};
	\node[above] at (9,1/2) {$'$};
	\node[above] at (12,1/2) {$'$};
	\end{tikzpicture}
	}}\right|^2.
\end{equation}

The averaged genus-$g$ partition function is what one gets starting from the two-boundary sphere $(g+1)$-point wormhole and analytically continuing all $g+1$ external operators to the black hole regime. The two sphere boundaries are glued at their $g+1$ pairs of insertion points, leading to the genus-$g$ handlebody.

\section{One Loop Corrections}
\label{s:oneloop}

We now describe the computation of one-loop corrections to the partition function of 3D gravity, and show that, at least in the examples we check, these one-loop contributions match the result of our CFT ensemble computations.  We will restrict to two-boundary wormholes without conical defects, where the computations are reasonably straightforward.  On the other hand, as we have seen in section \ref{s:singleboundary}, all of the wormholes with conical defects that we have studied can be related to smooth saddles, with higher genus but a smaller number of boundary components, by analytic continuation in the defect mass. We therefore expect the one-loop results to carry over to these cases. Spectral analysis on orbifolds has additional complications, and working this out explicitly would require carefully regulating the defects at one loop (see \cite{MR3946487} for mathematical results in this direction).

In section \ref{s:twocopy} we studied the ensemble average of products of two similar observables, and showed that it is related to the product of Liouville correlators at the level of the classical action. We will now show that the formula extends to one-loop order, i.e.,
\begin{equation}
\left.\overline{G(\lambda, {\bar \lambda}) G(\lambda', {\bar \lambda}')}\right|_{\rm paired} \sim G_L(\lambda, \lambda') G_L( {\bar \lambda}',{\bar \lambda}) ~,
\end{equation}
where recall that $x\sim y$ means that $\lim_{c\to\infty}\frac{x}{y}=1$.
Each Liouville correlation function in this expression has a large central charge expansion
\begin{equation}
G_L(\lambda, \lambda')\sim e^{-\frac{c}{6} S_{L}} Z^{L}_{\rm 1-loop}
\end{equation}
which includes both a classical action as well as a one-loop prefactor.  
We will now show that this one-loop correction matches the corresponding one-loop term in Einstein gravity.

We will consider wormholes with two boundaries, each of which is a smooth surface $\Sigma$ of genus $g$.
Since we are assuming there are no conical defects,   $(\lambda, {\bar \lambda})$ and $(\lambda', {\bar \lambda}')$ are the period matrices of the two boundaries, with orientation conventions such that the Maldacena-Maoz wormhole has $\lambda' = \blambda$.
Writing the bulk geometry as a quotient $M={\mathbb H}_3/\Gamma$, with $\Gamma$ a discrete subgroup of $PSL(2,{\mathbb C})$ acting on hyperbolic space ${\mathbb H}_3$, the one-loop determinant of three dimensional gravity was computed in \cite{Giombi:2008vd}.\footnote{It is important to remember that even though three dimensional general relativity has no {\it local} degrees of freedom, the loop expansion is still non-trivial due to non-local degrees of freedom, such as those which arise in the presence of boundaries or non-trivial bulk topology.}  The result is
\begin{equation}\label{gr1loop}
Z^{\rm gravity}_{\rm 1-loop}= 
\prod_{\gamma \in {\cal P}} \left(\prod_{n=2}^\infty {1\over |1-q_\gamma^n|^2}\right) 
\end{equation}
where $q_{\gamma}=e^{2\pi i \tau_\gamma}$ with  ${\rm Tr} \left(\gamma\right) = 2 \cos \pi \tau_\gamma$. 
Here ${\cal P}$ is the set of primitive conjugacy classes of $\Gamma$; an element $\gamma\in\Gamma$ is primitive if it cannot be written as the power of any other element of $\Gamma$.  Each primitive element generates a ${\mathbb Z}$ subgroup of $\Gamma$.\footnote{In our definition of ${\cal P}$ we are not counting both $\gamma$ and $\gamma^{-1}$ separately; in the literature these elements are often counted separately.}  We also exclude from ${\cal P}$ the identity element, which would lead to a divergence which can be absorbed by a local bulk counterterm (i.e. the cosmological constant).  

\subsection{The Maldacena-Maoz Wormhole}

We begin by considering the Maldacena-Maoz wormhole, with metric $d\rho^2 + \cosh^2 \rho d\Sigma^2$, where $d\Sigma^2$ is the constant negative curvature metric on $\Sigma$.  The boundary is a pair of Riemann surfaces with conjugate moduli. In terms of the moduli space coordinates described above, we have  $\lambda'={\bar \lambda}$.
The gravitational one-loop determinant is 
\begin{equation}
Z^{\rm gravity}_{\rm 1-loop}= 
\prod_{\gamma \in {\cal P}} \left(\prod_{n=2}^\infty {1\over |1-e^{-n\ell(\gamma)}|^2}\right) = Z(2)^{-1}
\end{equation}
where $\ell(\gamma) = 2 \cosh^{-1}(\frac{1}{2} \tr \gamma)$ is the length of the geodesic associated with the element $\gamma$, and
\begin{equation}
Z(s)\equiv \prod_{\gamma \in {\cal P}} \prod_{n=0}^\infty |1-e^{-(n+s)\ell(\gamma)}|^2
\end{equation}
is the Selberg zeta function. Although we have not indicated it explicitly, $Z(s)$ is a complicated function of the moduli coordinates $(\lambda, {\bar \lambda})$.
Famously (and most importantly for our purposes), $Z(s)$ is a meromorphic function of $s$ with zeros when $\frac{1}{4}s(1-s)$ is a zero of the scalar Laplacian on $\Sigma$.\footnote{
With conventions such that $\Delta_0 = -\frac{1}{4}y^{2} (\p_x^2+\p_y^2)$ on $\mathbb{H}_2$.
}  So the Selberg zeta function also has the interpretation as a (regularized) one-loop determinant 
\cite{DHoker:1986eaw,MR885573}
\begin{equation}
\det(\Delta_0-\frac{1}{4}s(1-s)) = c_g Z(s)
\end{equation}
The result is that, up to an overall constant, the gravity one-loop determinant for the Maldacena-Maoz wormhole is
\begin{equation}\label{mmloop}
Z^{\rm gravity}_{\rm 1-loop}= \frac{1}{\det(\Delta_0+\frac{1}{2})}
\end{equation}
where $\Delta_0$ is proportional to the scalar Laplacian on $\Sigma$. In the metric $e^{\Phi} |dz|^2$, 
\begin{equation}\label{scalarLap}
\Delta_0 = -e^{-\Phi} \partial {\bar \partial}  \ .
\end{equation}

We now wish to compare this to the result one obtains for Liouville theory
on the surface $\Sigma$.  A classical solution $\Phi_{o}$ of the Liouville equation
\begin{equation}
\partial {\bar \partial} \Phi_{o}=\frac{1}{2}e^{\Phi_{o}}
\end{equation}
gives a constant negative curvature metric $ds^2 = e^{\Phi_0}dz d{\bar z}$ on $\Sigma$.
Expanding the Liouville action to quadratic order around $\Phi_{o}$ gives: 
\begin{align}
S_L[\Phi] &= \frac{1}{4\pi} \int d^2z \left( \partial \Phi {\bar \partial}\Phi  + e^{\Phi}\right)
\cr
&= S_L[\Phi_{o}]+\frac{1}{4\pi} \int d^2 z e^{\Phi_{o}}\chi\left(\Delta_0 + \frac{1}{2}\right)\chi+\dots
\end{align}
where $\chi=\Phi-\Phi_o$ is a small perturbation of the Liouville field, $\dots$ denotes boundary terms 
as well as terms which are higher order in $\chi$, and $\Delta_0$ is the scalar Laplacian associated to the background Liouville field. 
Thus at quadratic order the action is just that of a massive scalar field
in a negatively curved background, and the one loop contribution to the Liouville partition function is the functional determinant
\begin{equation}
Z^L_{\rm 1-loop} = \frac{1}{\sqrt{\det \left(\Delta_0+\frac{1}{2}\right)}}
\end{equation}
Putting this together, we conclude that for the Maldacena-Maoz wormhole
\begin{equation}\label{MM1loop}
Z^{\rm gravity}_{\rm 1-loop}(\lambda, {\bar \lambda}) = Z^L_{\rm 1-loop}(\lambda, {\bar \lambda}) Z_{\rm 1-loop}^L({\bar \lambda}, {\lambda})\ .
\end{equation}
Here we have indicated explicitly the dependence on moduli.
We find a match between the average CFT partition function and the gravitational one-loop determinant.

\subsection{Quasi-Fuchsian Case}

We now turn to the general quasi-Fuchsian case, where the boundary is $\Sigma \sqcup \Sigma'$ for a pair Riemann surfaces $\Sigma$ and $\Sigma'$ with moduli  $(\lambda, {\bar \lambda})$ and $(\lambda', {\bar \lambda}')$.  
To match with the CFT ensemble, one must show that the gravity one loop determinant obeys 
\begin{equation}
Z^{\rm gravity}_{\rm 1-loop}(\lambda, {\bar \lambda}; \lambda', {\bar \lambda}')= Z^L_{\rm 1-loop}(\lambda, \lambda')Z^L_{\rm 1-loop}({\bar \lambda}', {\bar \lambda})~.
\end{equation}
We have already shown this in the case of the Maldacena-Maoz wormhole, for which $\lambda'={\bar \lambda}$.   So, just as in section \ref{s:twoboundary}, the general case follows if
\begin{equation}\label{mt1}
\frac{\partial^2}{\partial { \lambda} \partial {\bar \lambda}}\log Z^{\rm gravity}_{\rm 1-loop} = \frac{\partial^2}{\partial \lambda \partial \blambda'}\log Z^{\rm gravity}_{\rm 1-loop} = 0  \ . 
\end{equation}
Remarkably, this exact result was proven by McIntyre and Teo \cite{mcintyre2008holomorphic}, albeit in a somewhat different language.

In the remainder of the section, we will explain briefly the results of \cite{mcintyre2008holomorphic}.  The essential result of \cite{mcintyre2008holomorphic} is that for any quasi-Fuchsian group $\Gamma$, the general expression (\ref{gr1loop}) for the gravitational one-loop determinant can be written as a product of functional determinants:\footnote{In fact, \cite{mcintyre2008holomorphic} does not consider the gravitational determinant directly, but instead proves a more general result involving Laplacians acting on tensors of general rank.} 
\begin{equation}\label{mt}
\prod_{\gamma \in {\cal P}} \left(\prod_{n=2}^\infty {|1-q_\gamma^n|^2}\right) = e^{\frac{13}{12\pi}S_{TT}}\frac{\det \Delta_2(\Sigma) \det \Delta_2(\Sigma')}{\det N_2(\Sigma) \det N_2(\Sigma')} \ .
\end{equation}
Here $\Delta_2(\Sigma)$ is a Laplacian acting on quadratic differentials on the Riemann surface $\Sigma$, and $S_{TT}$ is the Liouville action defined by Takhtajan and Teo \cite{Takhtajan:2002cc}. In the denominator, $N_2(\Sigma)$ is the matrix of inner products for a basis of quadratic differentials on $\Sigma$. The bases for $\Sigma$ and $\Sigma'$ must be chosen to be dual to each other in a certain sense in the expression \eqref{mt}, which introduces additional moduli dependence; see \cite{mcintyre2008holomorphic} for details. On the Fuchsian slice, the denominator is equal to one, and the spectrum of $\Delta_2$ is related to that of $\Delta_0$ (see e.g.~\cite{MR885573}), so this reduces to the expression in terms of $\det (\Delta_0 + \frac{1}{2})$ in \eqref{mmloop}.

Now let us consider the dependence on moduli. It is proved in \cite[Section 3]{mcintyre2008holomorphic} that this combination satisfies
\begin{align}
\frac{\p^2}{\p Z \p \bar{Z}}
\log \left(\frac{\det \Delta_2(\Sigma) \det \Delta_2(\Sigma')}{\det N_2(\Sigma) \det N_2(\Sigma')}\right)
=
-
\frac{\p^2}{\p Z \p \bar{Z}}
 {\frac{13}{12 \pi}S_{TT}}
\ ,
\end{align}
where $(Z,\bar{Z})$ are complex coordinates on the moduli space for the union $\Sigma \sqcup \Sigma'$. The holomorphic coordinates on this moduli space are $Z = (\lambda, \lambda')$, and the anti-holomorphic coordinates are $\bar{Z} = (\blambda, \blambda')$. Therefore this implies both equations in (\ref{mt1}) and establishes the equivalence between gravity and the CFT ensemble at one-loop in the quasi-Fuchsian case.

\ \\
\bigskip

\noindent\textbf{Acknowledgments} \\
We thank Ahmed Almheiri, Alexandre Belin, Nathan Benjamin, Wan Zhen Chua, Chao-Ming Jian, Yikun Jiang, Henry Lin, Eric Perlmutter, Julian Sonner, Phil Saad, Edward Witten, Mengyang Zhang, and Sasha Zhiboedov for helpful conversations. TH and JC are supported by the Simons Foundation and NSF grant PHY-2014071. Research of AM is supported in part by the Simons Foundation Grant No. 385602 and the Natural Sciences and Engineering Research Council of Canada (NSERC), funding reference number
SAPIN/00047-2020.  
Some of this work was done at the Aspen Center of Physics, which is supported by NSF grant PHY-1607611. 

\appendix

\section{Liouville review and subtleties in $C_0$}\label{s:liouville}

\subsection{Liouville basics}\label{ss:liouvilleBasics}
We will often find it convenient to adopt the Liouville parameterizations of various CFT quantities. For example, we will write the conformal weights in terms of the ``Liouville momentum'' $P$ as
\begin{align}
h = \frac{c-1}{24} + P^2 \ ,
\end{align}
and the central charge $c$ in terms of the ``Liouville background charge'' $Q$ or $b$ as
\begin{equation}
    Q = \sqrt{c-1\over 6} = b+b^{-1}
\end{equation}
so that the semiclassical limit corresponds to the $b\to 0$ limit.

The Liouville CFT is a 
solution to the crossing equations in which the only Virasoro primary operators are scalars. It is noncompact, meaning that the spectrum of local primary operators is continuous and the vacuum is not a normalizable state in the Hilbert space of the theory on a circle. In particular, the identity operator never appears as an intermediate state in the OPE decomposition of any CFT observable.
The spectrum of primaries in the Liouville CFT is a continuum of scalars with $P \in \mathbb{R}$, and total scaling dimension $\Delta = 2h$. 
The vertex operators of Liouille theory are normalized so that
\begin{align}
\langle V_1(0) V_2(1)\rangle_L = 2 \pi \left[\delta(P_1+P_2) + S(P_1) \delta(P_1-P_2)\right] 
\end{align}
where $S(P)$ is a quantity referred to as the reflection coefficient given by
\begin{equation} \label{Refcoeff}
  S(P)=\frac{1}{b^2}(\pi \mu \gamma(b^2))^{\frac{(Q-2\alpha)}{b}}\gamma\left(\frac{2\alpha}{b}-1-\frac{1}{b^2}\right)\gamma(2b\alpha-b^2)
\end{equation}
where $\alpha=\frac{Q}{2}+iP$ and $\gamma(x)=\frac{\Gamma(x)}{\Gamma(1-x)}$. The structure constants are given by the DOZZ formula \cite{Dorn:1994xn,Zamolodchikov:1995aa}
\begin{equation}
    \langle V_1(0)V_2(1)V_3(\infty)\rangle_L = \Cdozz(P_1,P_2,P_3).
\end{equation}
They are symmetric under permutations of the operators, but they are not invariant under the reflection $P \to -P$ that leaves the conformal weights invariant:
\begin{align}
\Cdozz(P_1, P_2, P_3) &= S(P_1)\Cdozz(-P_1, P_2, P_3)  \ .
\end{align}
With this normalization, the structure constants reproduce the two-point function upon analytic continuation of one of the weights to that of the identity
\begin{equation}
    \Cdozz(P_1,P_2,\id) = 2\pi\left[\delta(P_1+P_2) + S(P_1)\delta(P_1-P_2)\right].
\end{equation} The 4-point function is given by integrating the DOZZ structure constants against the sphere four-point Virasoro conformal blocks 
\begin{equation}
\begin{aligned}
\langle V_1(0) V_2(z,\bz) &V_3(1) V_4(\infty) \rangle_{L} \\
&=
\frac{1}{2}\int_{\mathbb{R}} {dP\over 2\pi} \, \Cdozz(P_1, P_2, P) \Cdozz(-P, P_3, P_4) \left|{\cal F}_{1234}(h_P;z) \right|^2 \ .
\end{aligned}
\end{equation}
There is a simple relation between the DOZZ structure constants of Liouville theory $\Cdozz$ and the universal averaged structure constants $C_0$ that defines our ensemble. The exact, finite-$c$ relation is given by \cite{Collier:2019weq}
\begin{equation}\label{eq:C0CDOZZ}
\begin{aligned}
    C_0(P_1,P_2,P_3) &= {\left(\pi \mu\gamma(b^2)b^{2-2b^2}\right)^{{Q\over 2b}}\over 2^{3\over 4}\pi}{\Gamma_b(2Q)\over \Gamma_b(Q)}{\Cdozz(P_1,P_2,P_3)\over\sqrt{\prod_{k=1}^3S(P_k)\rho_0(P_k)}}\\
    & \equiv c_b {\Cdozz(P_1,P_2,P_3)\over\sqrt{\prod_{k=1}^3S(P_k)\rho_0(P_k)}}
\end{aligned}
\end{equation}
The constant $c_b$ is independent of the conformal weights and cancels the dependence of $\Cdozz$ on the Liouville cosmological constant $\mu$.
The universal OPE function $C_0$ admits the following explicit expression \cite{Collier:2019weq}
\begin{equation}
    C_0(P_1,P_2,P_3) = {\Gamma_b(2Q)\over\sqrt{2}\Gamma_b(Q)^3}{\prod_{\pm_{1,2,3}}\Gamma_b\left({Q\over 2} \pm_1 iP_1 \pm_2 iP_2 \pm_3 iP_3\right)\over\prod_{k=1}^3\Gamma_b(Q+2iP_k)\Gamma_b(Q-2iP_k)}
\end{equation}
where $\Gamma_b(x)$ is a special function with simple poles at $x = -mb-nb^{-1}$ for $m,n\in\mathbb{Z}_{\geq 0}$. Note that this expression is manifestly invariant under reflections $P_i\to -P_i$. See (\ref{semiclassicalC0}) for the expansion of $C_0$ in the semiclassical limit, in which the external weights scale linearly with the central charge.
The limit of $C_0$ in which one of the operator weights is analytically continued to that of the identity operator is given by
\begin{equation}
    C_0(P_1,P_2,\id) = {1\over\rho_0(P_1)}\left[\delta(P_1-P_2) + \delta(P_1+P_2)\right].
\end{equation}

Operator normalization is inherently ambiguous in noncompact CFT. We find it convenient to use a different normalization that we refer to as $\widehat{V}$, with
\begin{equation} \label{Vertexnorm}
    \langle \widehat{V}_1(0)\widehat{V}_2(1)\widehat{V}_3(\infty)\rangle = \tCdozz(P_1,P_2,P_3)
\end{equation}
defined such that 
\begin{equation}\label{eq:ChatDOZZ}
    \tCdozz(P_1,P_2,P_3) = {c_b \Cdozz(P_1,P_2,P_3)\over \sqrt{\prod_{j\in\text{light}}\rho_0(P_j)S(P_j)\prod_{k\in\text{heavy}}S(P_k)}}.
\end{equation}
Here, the distinction ``heavy'' refers to states with weights $h>{c-1\over 24}$ while ``light'' refers to sub-threshold states.  We are normalizing the light vs heavy operators slightly differently; this is natural because in our ensemble, light states appear individually without a density-of-states factor, whereas heavy states are weighted against the Cardy density of states, $\rho_0$. Note that for light states, $\rho_0(P) = \O(1)$ in the semiclassical limit, so the difference in normalization only affects the 1-loop prefactors.

With this normalization, the Liouville OPE coefficients $\tCdozz$ are invariant under reflections. To see this, we write the following explicit expression for the normalized structure constants
\begin{equation}
\begin{aligned}
    &\, \tCdozz(P_1,P_2,P_3)\\
    =& \, {2^{1\over 4}\Gamma_b(2Q)\over\Gamma_b(Q)^3}{\prod_{\pm_{1,2,3}}\Gamma_b\left({Q\over 2} \pm_1 iP_1 \pm_2 iP_2 \pm_3 iP_3\right)\over\left(\prod_{j\in\text{light}}\Gamma_b(Q+2iP_j)\Gamma_b(Q-2iP_j)\right)\left(\prod_{k\in\text{heavy}}|\Gamma_b(2iP_k)\Gamma_b(Q+2iP_k)|\right)}.
\end{aligned}
\end{equation}
For heavy operator weights $P_1,P_2$, the limit in which the third operator weight is taken to zero gives the following canonically normalized two-point function
\begin{equation}
    \tCdozz(P_1,P_2,\id) = \delta(P_1+P_2) + \delta(P_1-P_2).
\end{equation}  
The Liouville 4-point function of sub-threshold operators with this normalization then takes the following form
\begin{equation}
\begin{aligned}\label{GLhat4}
\langle \widehat{V}_1(0) \widehat{V}_2(z,\bar z) &\widehat{V}_3(1) \widehat{V}_4(\infty) \rangle_{L} \\
&= \frac{1}{2} \int_{\mathbb{R}}{dP} \, \tCdozz(P_1, P_2, P)\tCdozz(P_3, P_4,P)  \left|{\cal F}_{1234}(h_P;z) \right|^2 \ .
\end{aligned}
\end{equation}
All of the comparisons to Liouville in the body of the paper use the hatted normalization for the external operators. Thus for the $n$-point function on a genus-$g$ Riemann surface with period matrix $\Omega$, we define
\begin{align}
G_L(x_i, \bx_i; \Omega, \bar{\Omega}) = \langle \widehat{V}_1(x_1, \bx_1) \cdots \widehat{V}_n(x_n, \bx_n)\rangle_{L}^{\Sigma(\Omega, \bar{\Omega)}} \ .
\end{align}
In terms of the semiclassical parameterization \eqref{largecweights}, normalizable states in Liouville have $\gamma \in \mathbb{R}$ and Re $\eta = \frac{1}{2}$.
 In the semiclassical limit, the Virasoro conformal blocks and structure constants exponentiate,
\begin{align}
\F_{1234}(h) &\approx \exp\left[-\frac{c}{6} f_{1234}(\eta) \right] \\
\tCdozz(P_1, P_2, P) &\approx \exp\left[ -\frac{c}{6}w_{12}(\eta) \right]  \ .
\end{align}
It is also useful to note the semiclassical limit of the proportionality constant $c_b$ defined in (\ref{eq:C0CDOZZ}),
\begin{equation} \label{Slimitcb}
   c_b \approx  \exp\left[\frac{c}{6}\left(1+\frac{1}{2}\log (\pi \mu b^2)\right)\right]
\end{equation}
and the semiclassical limit of the reflection coefficient given in (\ref{Refcoeff}) for $\alpha=\frac{\eta}{b}$ with $\eta$ fixed to lie in the range $0<\eta<\frac{1}{2}$,
\begin{equation}
   S(P) \approx \exp\left[-\frac{c}{3}\left((1-2\eta)\left(\log(1-2\eta)-\frac{1}{2}\log (\pi \mu b^2)-1\right)\right)\right]\csc \left(\frac{\pi (1-2\eta)}{b^2}\right)
\end{equation}
The semiclassical 4-point function is given by a saddlepoint in the integral over $P$, so in this limit, crossing invariance reduces to an equality between two saddlepoint expressions. Finally, the semiclassical limit of the universal formula with external operators in the defect regime as in (\ref{largecweights}), with $0<\eta_i < \half$, is given by
\begin{equation}
\begin{aligned}
    \log C_0(\eta_1,\eta_2,\eta_3) \approx& \, {c\over 6} \bigg[\left(F(2\eta_1) - F(\eta_2+\eta_3-\eta_1) + (1-2\eta_1)\log(1-2\eta_1) + \text{2 cyclic permutations}\right)\\
    & \, -F(\eta_1+\eta_2+\eta_3) - 2\left(1-\eta_1-\eta_2-\eta_3\right)\log\left(1-\eta_1-\eta_2-\eta_3\right) + F(0)\\
    & \, + \pi i\left(\eta_1+\eta_2+\eta_3-1\right)\bigg],
\end{aligned}
\end{equation}
where $F(x) = \int_{\half}^x dy\, \log\left({\Gamma(y)\over \Gamma(1-y)}\right)$.

\subsection{On the analytic continuation of the Liouville four-point function and of $C_0$}
Although the spectrum of primary operators in Liouville theory is spanned by scalars with dimensions $\Delta = {c-1\over 12}+P^2$ for real $P$, one may consider the analytic continuation of the Liouville four-point function (\ref{GLhat4}) to the regime where the external operators correspond to scalar defects with conformal weights below the black hole threshold, $0<h_i<{c-1\over 24}$. 
To proceed, it will be convenient to define
\begin{equation}
    \alpha_i = {Q\over 2} + iP_i.
\end{equation}
The conformal weights are given in terms of $\alpha_i$ as
\begin{equation}
    h_i = \alpha_i(Q-\alpha_i).
\end{equation}
The defect regime corresponds to real $\alpha_i\in[0,{Q\over 2}]$.

We start by noting that the structure constants (\ref{eq:ChatDOZZ}) appearing in (\ref{GLhat4}) have simple poles in the complex $P$ plane at the following values of $P$\footnote{Here we assume that $b^2$ is not rational so that none of the poles are overlapping.}
\begin{equation}\label{eq:PPoles}
\begin{aligned}
   {Q\over 2}+iP &= \left({Q\over 2}+iP_1\right) + \left({Q\over 2}+i P_2\right) + mb+nb^{-1},\quad m,n\in\mathbb{Z}_{\geq 0}\\
   {Q\over 2}+iP &= \left({Q\over 2}+iP_3\right) + \left({Q\over 2}+i P_4\right) + mb+nb^{-1},\quad m,n\in\mathbb{Z}_{\geq 0}\\
\end{aligned}
\end{equation}
and those corresponding to all possible combinations of reflections of the three Liouville momenta involved in each expression. If all of the $P_i$ are real, then these poles are well-separated from the contour of integration in the Liouville four-point function (\ref{GLhat4}) and the analytic continuation is straightforward. In the defect regime, provided
\begin{equation}\label{eq:alphaCondition}
    \re(\alpha_1 + \alpha_2) > {Q\over 2},\quad \re(\alpha_3 + \alpha_4) > {Q\over 2}
\end{equation}
then none of the poles of the DOZZ structure constants cross the contour of integration in the Liouville four-point function and the expression (\ref{GLhat4}) is still a valid solution to the crossing equation. On the other hand, if (\ref{eq:alphaCondition}) is violated, then some of the poles in (\ref{eq:PPoles}) cross the contour of integration in (\ref{GLhat4}) which then must be deformed, corresponding to a finite number of additional discrete contributions to the four-point function. These are needed in order to preserve crossing symmetry.  Note that this can happen for observables involving external operators with $\alpha_i < {Q\over 4}$, corresponding to weight $h_i < {c-1\over 32}$. In particular, we have
\begin{align}
     \langle \widehat{V}_1(0) &\widehat{V}_2(z,\bz)\widehat{V}_3(1)\widehat{V}_4(\infty)\rangle_L \\
     &= \, -2\pi i\sum_{m,n}\Res_{P = P_{m,n}}\left(\tCdozz(P_1,P_2,P)\tCdozz(P_3,P_4,P)|\mathcal{F}_{1234}(h_P;z)|^2\right)\notag\\
     &\qquad \,  + \frac{1}{2 } \int_{\mathbb{R}}dP \, \tCdozz(P_1, P_2, P)\tCdozz(P_3, P_4, P)  \left|{\cal F}_{1234}(h_P;z) \right|^2,\notag
\end{align}
where $P_{m,n}$ collectively denotes the set of poles in (\ref{eq:PPoles}) such that 
\begin{equation}
    {Q\over 2}+i P_{m,n} < {Q\over 2}. 
\end{equation}
Note that in the case of pairwise identical operators $P_3 = P_2,\, P_4 = P_1$, these are enhanced to double poles and the additional contributions to the four-point function involve derivatives of the conformal blocks with respect to the internal weight. 

Similar considerations extend to the universal structure constant $C_0$. The defining feature of $C_0$ is that it branches the sphere four-point Virasoro vacuum block into a complete basis of Virasoro blocks in the dual channel
\begin{equation}\label{eq:vacFusion}
    \mathcal{F}_{1122}(\id;1-z) = \frac{1}{2}\int_{\mathbb{R}}dP\, \rho_0(P)C_0(P_1,P_2,P)\mathcal{F}_{1221}(h_P;z).
\end{equation}
Viewed as a function of the internal Liouville momentum $P$, $C_0(P_1,P_2,P)$ has the same set of poles as the DOZZ OPE coefficient listed in (\ref{eq:PPoles}). Thus $C_0$ may straightforwardly be continued to the defect regime where $\alpha_i \in [0,{Q\over 2}]$ provided $\re(\alpha_1 + \alpha_2) > {Q\over 2}$. If on the other hand $\re(\alpha_1 + \alpha_2) < {Q\over 2}$, then (\ref{eq:vacFusion}) receives a finite number of additional discrete contributions from the poles of $C_0$ that have crossed the contour in the decomposition of the vacuum block in the dual channel
\begin{equation}
\begin{aligned}
    \mathcal{F}_{1122}(\id;1-z) =& \, -2\pi i\sum_{m,n}\left(\Res_{P = P_{m,n}}C_0(P_1,P_2,P)\right)\rho_0(P_{m,n})\mathcal{F}_{1221}(h_{P_{m,n}};z)\\
    & \, + \frac{1}{2}\int_{\mathbb{R}}dP\, \rho_0(P)C_0(P_1,P_2,P)\mathcal{F}_{1221}(h_P;z),
\end{aligned}
\end{equation}
where again the sum over $m,n$ is over the poles of $C_0$ such that ${Q\over 2}+i P_{m,n}< {Q\over 2}$. In \cite{Kusuki:2018wpa,Collier:2018exn}, the presence of the additional contributions to the cross-channel decomposition of the T-channel vacuum block was interpreted as signaling accumulation points in the spectrum of twists in the large-spin spectrum of $c>1$ CFTs with no continuous symmetries beyond Virasoro. These infinite towers of states with asymptotic twist below threshold were referred to as multi-twist trajectories, in analogy to the composite states that are known to appear in higher-dimensional CFT \cite{Komargodski:2012ek,Fitzpatrick:2012yx}.

As described in section \ref{s:ensemble}, throughout the paper we  make use of a large-$c$ ensemble whose spectrum includes a discrete set of scalar defect states in addition to a continuum of black hole states. In some cases, we assume that the scalars have
\begin{equation}
    h \geq {c-1\over 32}
\end{equation}
in order to avoid the multi-twist trajectories.  For these defects, the universal OPE coefficient $C_0$ does not acquire any additional delta-function distributional support away from the black hole spectrum.

\section{Renormalization of the defect action}\label{s:normalization}
 In order to compare gravity calculations with conical defects to CFT correlation functions with unit-normalized operators, we need to renormalize the gravitational action by subtracting local terms associated to each defect. In this appendix we find the appropriate counterterm.
 
To this end, we will evaluate the two point functions of the defect operators holographically.  Since the $n$-point wormhole saddle does not exist when the Gauss Bonnet constraint $\sum \eta_i>1$ is violated, we cannot calculate the two point functions by taking an $\eta \to 0$ limit of the action for the $n$-point wormhole. Instead, we must use the spherical slicing of AdS,
\begin{equation}
    ds^2=dr^2+\sinh^2(r)e^\Phi dzd\overline{z} \ .
\end{equation}
We place the defects at $z= \pm 1$.  The Einstein equations reduce to the positive-curvature Liouville equation,
\begin{equation}
    \partial\overline{\partial}\Phi=-\frac{e^\Phi}{2}-2\pi\eta\delta^{(2)}(z+1)-2\pi \eta\delta^{(2)}(z-1)
\end{equation}
with
\begin{equation}
    \Phi(z,\overline{z}) \sim 
    \begin{cases}
      -4\eta\log(|z\pm 1|) \quad & z\to \pm 1 \\
     -4\log(|z|) \quad & z\to \infty
    \end{cases}
\end{equation}
The solution is \footnote{The easiest way to arrive at this is to first put the punctures at $z=0,\infty$, use the map $w=z^{1-2\eta}$  to obtain $e^\Phi=4(1-2\eta)^2\frac{|z|^{-4\eta}}{(1+|z|^{2(1-2\eta)})^2}$ and then do a conformal transformation $z\to\frac{z+1}{z-1}$ to place the punctures at $z=\pm1$.}
\begin{equation}
    e^\Phi= 16(1-2\eta)^2\frac{|z^2-1|^{-4\eta}}{(|z-1|^{2(1-2\eta)}+|z+1|^{2(1-2\eta)})^2}
\end{equation}
Let us again introduce a wiggly cutoff,
\begin{equation}
    r_0=\log(\frac{2}{\epsilon})-\frac{\Phi}{2}
\end{equation}
giving the induced metric on the cutoff surface,
\begin{equation}
    ds^2_{\text{bdry}}=\frac{1}{4}(\partial \Phi dz+\overline{\partial}\Phi d\overline{z})^2+(\frac{1}{\epsilon^2}-\frac{e^{\Phi}}{2})dzd\overline{z}
\end{equation}
Now we will calculcate the on-shell action. We define the regularised twice-punctured sphere to be the region $\Gamma=\{\epsilon_i<|z\pm 1|; |z|<R \}$. We can evaluate the EH+cc term away from the conical defect trajectory,
\begin{equation}
    S_{\text{bulk}}=\frac{V_\epsilon}{4\pi G}=\frac{1}{4\pi G}\int_{\Gamma} d^2z(\frac{1}{2\epsilon^2}-\frac{e^\Phi}{2}\log(\frac{2}{\epsilon})+\frac{\Phi}{4}e^\Phi)
\end{equation}
Next, we evaluate the GHY+ct away from the defect on the cutoff boundary, where we have $K=2+2\epsilon^2(\partial\overline{\partial}\Phi+\frac{e^{\Phi}}{2})+O(\epsilon^3)$,
\begin{equation}
     S_{\text{bdry}}=-\frac{A_\epsilon}{8\pi G}=-\frac{1}{8\pi G}\int_{\Gamma} d^2z(\frac{1}{\epsilon^2}-\frac{e^\Phi}{2}+\frac{1}{2}\partial \Phi \overline{\partial}\Phi)
\end{equation}
The sum is
\begin{equation}
    S_{\text{bulk}}+S_{\text{bdry}}= \frac{1}{4\pi G}\int_{\Gamma} d^2z(\frac{1}{4}(\partial\Phi\overline{\partial}\Phi-e^\Phi)-\frac{1}{2}\overline{\partial}(\Phi\partial\Phi)-\partial\overline{\partial}\Phi(1+\log(\frac{\epsilon}{2})))
\end{equation}
To evaluate this, we use a trick similar to the one used in \cite{Zamolodchikov:1995aa} to evaluate the Liouville action for the thrice punctured sphere. Expand the Liouville field about the defects,
\begin{equation}
    \Phi \to -4\eta \log (|z\pm1|)+4\eta \log 2+ 2\log (1-2\eta) \quad \quad z\to \pm 1
\end{equation}
We define the renormalised Liouville action with positive cosmological constant, whose solutions have positive curvature, as
\begin{equation}
    \widetilde{S}_L=\frac{1}{4\pi}\int_{\Gamma} d^2z(\partial\Phi\overline{\partial}\Phi-e^\Phi)+\widetilde{\Phi}_R+2\log R-\sum(\eta_i\widetilde{\Phi}_i+2\eta_i^2\log\epsilon_i)
\end{equation}
where 
\begin{equation}
    \widetilde{ \Phi}_i=\frac{i}{4\pi \eta_i}\oint_{|z-z_i|=\epsilon_i}dz \Phi\partial\Phi \quad\quad\quad \widetilde{\Phi}_R=\frac{i}{4\pi}\oint_{|z|=R}dz \Phi\partial\Phi
\end{equation}
are the boundary terms. Treating the on-shell action as a function of defect strengths, it satisfies the differential equation,
\begin{equation}
    \frac{d\widetilde{S}_L}{d\eta_i}=-\widetilde{\Phi}_i-4\eta_i\log \epsilon_i
\end{equation}
On-shell, the RHS has contributions only from the boundary and counter terms.  Thus, evaluating the RHS in the limit $\epsilon_i\to 0$, we see that only the constant term in the expansion of the solution around the puncture survives,
\begin{equation}
    \frac{d\widetilde{S}_L}{d\eta}= -4\eta \log 2 - 2\log (1-2\eta) 
\end{equation}
In the present case, there are two defects of equal strengths, so integrating the above equation,
\begin{equation}
   \widetilde{S}_L= -4\eta^2 \log 2+ 2(1-2\eta)\log (1-2\eta)-2(1-2\eta)+k
\end{equation}
where $k$ is the integration constant. Now, adding to this, the term proportional to the area which is independent of the positions of the defects which we evaluated to be $2(1-2\eta)(1+\log (\frac{\epsilon}{2}))$ gives the on-shell action,
\begin{equation}
    S_2=\frac{c}{6}\bigg(4\eta(1-\eta)\log 2+ 2(1-2\eta)\log (1-2\eta)+2(1-2\eta)\log \epsilon+k-2\log 2-2\log R+4\eta^2 \log \epsilon_0\bigg)
\end{equation}
For $\eta = 0$, there is no defect, so the action must agree with the action on the sphere calculated from the conformal anomaly,
\begin{equation}
    S_{\text{anom}}(S^2)=-\frac{c}{3}\log \frac{R}{\epsilon} \ .
\end{equation}
This sets the constant to  $k=2\log2$. We now have the gravity calculation of the unnormalized 2-point function,
\begin{align}
\langle O(-1) \O(1) \rangle = e^{-S_2} \ .
\end{align}
The nomalized 2-point function is $|z_1-z_2|^{-2\Delta} = 2^{-2\Delta}$. Therefore to normalize the defects, we must add to the action a term
\begin{align}
S_{\rm ct} (\eta_i)  &=  2\Delta \log 2- S \notag\\
 &=\frac{c}{6}\bigg(-2(1-2\eta_i)\log (1-2\eta_i)-2(1-2\eta_i)\log \epsilon+2\log R-4\eta_i^2 \log \epsilon_i\bigg) \ , \label{scteta}
\end{align}
for each defect. The final term in \eqref{SrenDefects} comes from normalizing by the vacuum partition function.

\section{Series solution of the Beltrami equation}\label{s:beltrami}
In this appendix, we derive equation \eqref{c2rels}. This equation describes what parameters $(\alpha,\balpha)$ to choose in the almost Fuchsian metric \eqref{afmetric} such that the cross-ratio on the right boundary, in isothermal coordinates, is $x'$. Let
\begin{align}
x' = \bx + \bar{a} \ .
\end{align}
The bar is due to the orientation reversal in going from the left to right boundary (i.e., the Fuchsian wormhole has $x' = \bx$).
Copying the essential equations from section \ref{ss:almostfuchsian}, we choose
\begin{align}
\alpha = c_2^F(x,\bx+\bar{a}) - c_2^F(x,\bx) \ ,
\end{align}
and we need to solve the Beltrami equation 
\begin{align}
\frac{\p \bw}{\bar{\p} w} = \bmu  , \qquad
\bmu = \alpha Q(z,x) e^{-\Phi} \ ,
\end{align}
normalizing the solution to fix $0,1$, and $\infty$. 
The cross ratio on the right boundary is
\begin{align}
x' = \bw(x,\bx) \ .
\end{align}
Therefore our goal is to solve the Beltrami equation, with this $\alpha$, and check that
\begin{align}
\bw(x,\bx) = \bx + \bar{a} \ .
\end{align}
Expand in $\bar{a}$,
\begin{align}
\bmu &= Q(z,x) e^{-\Phi}( \bar{a} c_2' + \frac{\bar{a}^2}{2}c_2'' + \cdots)\\
\bw(z,\bz) &= \bz + \bar{a} \bw_{(1)} + \bar{a}^2 \bw_{(2)} + \cdots
\end{align}
where  we use the shorthand $c_2' = \frac{\p}{\p \bx}c_2^F$, and similarly for higher derivatives. Before proceeding to solve the Beltrami equation, it is useful to note the following two identities obeyed by the Liouville stress tensor,
\begin{equation} \label{6.16}
    \begin{split}
        & \partial_{\overline{x}}T^\Phi(z)=-\frac{c_2'Q(z;x)}{4}-\pi c_2 \delta(z-x)\\
        & -2e^{-\Phi}\partial_{\overline{x}}T^\Phi(z)=\partial_z(-e^{-\Phi}\partial_{\overline{x}}\partial_z \Phi)
    \end{split}
\end{equation}
Define $F(z,\overline{z})\equiv -2e^{-\Phi}\partial_{\overline{x}}\partial_z \Phi$. The Beltrami equation at $O(\overline{a})$ reads,
\begin{equation}
    \partial \bw_{(1)}=Q(z;x)c'_2e^{-\Phi}
\end{equation}
Using the two identities presented in (\ref{6.16}), we see that the Beltrami equation can be written as $\partial \bw_{(1)}=\partial F$ . (The delta function doesn't contribute since $e^{-\Phi}\to 0$ as $z\to x$). Integrating the equation, we get
\begin{align}
\bw_{(1)} &= F(z,\bz) \ .
\end{align}
Using the Liouville equation, we can check that $F(x,\overline{x})=1$. Thus, we see that the relation between the moduli on the two boundaries at $O(\overline{a})$ is $x'=\overline{x}+\overline{a}$.

The Beltrami equation at $O(\overline{a}^2)$ takes the form,
\begin{equation} \label{Bel2}
    \partial \overline{w}_{(2)}= Qe^{-\Phi}(c'_2\overline{\partial}F+\frac{c_2''}{2})
\end{equation}
Just like at $O(\overline{a})$, we show that the RHS is a total $\partial$ derivative. To this end, it is convenient to express the RHS in terms of the Liouville field $\Phi$ alone. The first term on the RHS can be written as $ Qe^{-\Phi}c_2'\overline{\partial}F=\partial F \overline{\partial}F$. To express this in terms of the Liouville field, observe that,
\begin{equation}
    \begin{split}
        & \partial F=2e^{-\Phi}(\partial \Phi \partial_{\overline{x}}\partial\Phi-\partial_{\overline{x}}\partial^2\Phi)\\
        & \overline{\partial}F=2e^{-\Phi}\overline{\partial}\Phi\partial_{\overline{x}}\partial\Phi-\partial_{\overline{x}}\Phi
    \end{split}
\end{equation}
To arrive at the second expression, we used the Liouville equation (Again, note that the delta function drops out). Now, we evaluate $\partial F \overline{\partial}F$ and observe after a few lines of algebra that it can be expressed as a total $\partial$ derivative,
\begin{equation}
    \partial F \overline{\partial}F=\frac{1}{2}\partial \left(e^{-\Phi}\partial (\partial_{\overline{x}}\Phi)^2\right) -2\partial \left(e^{-2\Phi}\overline{\partial}\Phi (\partial_{\overline{x}}\Phi)^2\right)+\partial \left(\partial_{\overline{x}}F-\frac{c_2''}{2c_2'}F\right)
\end{equation}
Since the second term on the RHS of (\ref{Bel2}) is also a total derivative, $c_2''Qe^{-\Phi}=\frac{c_2''}{c_2'}\partial F$, we can integrate (\ref{Bel2}) trivially to get,
\begin{equation}
    \overline{w}_{(2)}=\partial_{\overline{x}}F+\frac{1}{2}e^{-\Phi}\partial (\partial_{\overline{x}}\Phi)^2-2e^{-2\Phi}\overline{\partial}\Phi (\partial_{\overline{x}}\Phi)^2
\end{equation}
Using the Liouville equation and noting that the delta function terms drop out everywhere since they are always accompanied by a $e^{-\Phi}$ prefactor, we see that $\overline{w}_{(2)}(x,\overline{x})=0$. So, the relation between moduli on the two boundaries calculated upto $O(\overline{a}^2)$ reads,
\begin{equation}
    x'=\overline{x}+\overline{a}
\end{equation}
To proceed to $O(\bar{a}^3)$ we use Mathematica. We construct a general ansatz with the correct scaling and up to three powers of $e^{-\Phi}$, plug it into the 3rd order Beltrami equation, and simplify using the Liouville equation and same tricks described at lower orders. This fixes the coefficients in the ansatz and leads to
\begin{multline}
    \overline{w}_{(3)}= \frac{4}{3}e^{-3\Phi}\Phi^3_{z\overline{x}}(-2\Phi^2_{\overline{z}}+\Phi_{\overline{z}\overline{z}})+2e^{-2\Phi}\Phi_{z\overline{x}}(2\Phi_{\overline{x}}\Phi_{z\overline{x}}\Phi_{\overline{z}}-\Phi_{z\overline{x}\overline{x}}\Phi_{\overline{z}}-\Phi_{z\overline{x}}\Phi_{\overline{z}\overline{x}})\\
     +e^{-\Phi}(-\Phi^2_{\overline{x}}\Phi_{z\overline{x}}+\Phi_{\overline{x}\overline{x}}\Phi_{z\overline{x}}+\Phi_{\overline{x}}\Phi_{z\overline{x}\overline{x}}-\frac{1}{3}\Phi_{z\overline{x}\overline{x}\overline{x}})
\end{multline}
where the subscripts are partial derivatives. Using the Liouville equation to evaluate this solution at $(z,\bz)=(x,\bx)$, we find $\overline{w}_{(3)}(x,\overline{x})=0$. Thus, we have verified that 
\begin{equation}
    x'=\overline{x}+\overline{a} + \O(\bar{a}^4) \ .
\end{equation}
As discussed in section \ref{ss:TT}, the equality is extended to all orders by  \cite[Lemma 4.2]{Takhtajan:2002cc}.

\renewcommand{\baselinestretch}{1}\small
\bibliographystyle{ourbst}
\bibliography{biblio}
\end{document}